\documentclass[aps,prl,twocolumn,floatfix,citeautoscript,nofootinbib]{revtex4-2}
\usepackage{amsfonts}
\usepackage{amsbsy}
\usepackage{latexsym,epsfig,graphicx}
\usepackage{dcolumn}
\usepackage{subfigure}
\usepackage{color}
\usepackage[colorlinks,linkcolor=blue,urlcolor=blue,citecolor=blue]{hyperref}
\usepackage{amstext}
\usepackage{amssymb}
\usepackage{setspace}
\usepackage{amsmath}
\usepackage{makeidx}
\usepackage{bm}
\usepackage{ulem}
\usepackage{multirow}
\usepackage{mathrsfs}
\usepackage{array}
\usepackage[section]{placeins}
\usepackage{booktabs}

\DeclareMathOperator{\sech}{sech}

\begin{document}

\title{Asymptotic breathers in the Manakov system}
\author{Liming Ling$^{1}$}
\author{Boris A. Malomed$^{2,3}$}
\author{Yan-Hong Qin$^{4}$}
\email{yhqin@xju.edu.cn}
\author{Jin-Peng Yang$^{5}$}
\author{Hao-Jun Yin$^{1}$}
\author{Li-Chen Zhao$^{6}$}

\begin{abstract}
We construct vector asymptotic Akhmediev breathers in the Manakov system, by matching generalized modulational-instability (gMI) eigenmodes to the preformation tails of the breather states. This branch-resolved matching provides analytical results linking the gMI-perturbation envelopes to the positions of the emerging breathers. Numerical simulations of the system with the focusing and defocusing nonlinearities corroborate the analytical predictions for each MI branch. We further develop the asymptotic theory for two-breather solutions generated by two gMI modes, linking their temporal separation to the input envelopes and interaction-induced shifts. Numerical simulations also reveal that breather patterns exhibit robustness in the presence of weak nonintegrable perturbations. These findings establish a framework relating experimentally controllable input modulations to the branch-selective formation and positioning of breathers in multicomponent nonlinear systems.
\end{abstract}

\address{$^{1}$School of Mathematics, South China University of Technology,
Guangzhou 510640, China}
\address{$^{2}$Department of Physical Electronics,
School of Electrical Engineering, Faculty of Engineering, Tel Aviv
University, Tel Aviv 69978, Israel}
\address{$^{3}$Instituto de Alta
Investigaci\'{o}n, Universidad de Tarapac\'{a}, Casilla 7D, Arica, Chile}
\address{$^{4}$School of Physical Science and Technology, Xinjiang
University, Urumqi 830046, China}
\address{$^{5}$Department of Mathematics,
China University of Mining and Technology, Beijing 100083, China}
\address{$^{6}$School of Physics,
Northwest University, Xi'an 710127, China}

\maketitle

\textit{Introduction}---Akhmediev breathers (ABs) are transient nonlinear structures associated with MI \cite{MI1,MI2,MI3,MI4,MI5,Zhao2016,Sun2023,Sun2025}. They occur in nonlinear optics, hydrodynamics, and Bose-Einstein condensates \cite{Dudley1,Dudley2,Kevrekidis-09}. The controlled generation of ABs is relevant to studies of pulse trains, supercontinuum generation, and extreme events in these fields \cite{fiber,fiber2,fiber3,water,water2,BEC}. Therefore, selective excitation of ABs at prescribed positions is a problem of considerable physical interest. This calls for a predictive framework linking a physically realistic initial perturbation to the exact location, amplitude, and structure of the resulting breathers. Despite vast advances in constructing exact breather solutions in integrable nonlinear systems \cite{Vishnu,Guo2011,Ling2019,Chen2021,Gelash,Kraus,DSWang,Wright2000}, there remains a gap between idealized mathematical considerations and physically realistic excitations. A fundamental issue is that exact solutions are usually associated with perfectly periodic initial conditions, whereas natural settings involve perturbations that are spatially localized \cite{Kibler2015,Gelash2022}, evolving on slow transverse or temporal scales. Closing the gap calls for an analytical framework that can predict, for a given slowly varying initial perturbation, what breather structures will emerge and where they will be located.

For scalar nonlinear Schr\"{o}dinger equation, matched asymptotic expansions provide predictions for periodic recurrence \cite{Grinevich2018} and AB formation from localized perturbations \cite{Trillo2019}. By contrast, the vector system can admit coexisting unstable MI branches, offering richer physical scenarios absent in the scalar case \cite{Frisquet2015,Forest2000,Ling2017,Qin2023}. From the perspective of exact vector AB solutions, a selected branch can excite an eye-shaped, anti-eye-shaped, or four-petal pattern in different components \cite{Ling2019,Zhao2013,Ling2017,Liu2024,Qin2023}, and two excited branches admit interaction-dependent translations. However, although exact solutions describe these individual AB patterns, the usual periodic perturbations, such as a cosine modulation or a generic Fourier-mode perturbation, inevitably excite all available unstable branches simultaneously in a regime with multiple MI branches, making branch-selective excitation challenging. Furthermore, extending the scalar theory to vector systems faces severe mathematical difficulties, as solving the fourth-order Lax-pair spectral equation and inverting the $4\times4$ MI matrix in the two-component case yield algebraically cumbersome explicit forms \cite{Wright2000,Forest2000}, thereby further hindering the establishment of asymptotic vector AB solutions.

In this work, we develop a branch-resolved asymptotic theory for the Manakov system by directly matching each selected gMI eigenmode to the preformation expansion of the corresponding asymptotic vector AB. The resulting relations map the modulus and phase of the perturbation envelope to the location of the MI-induced breathers and their spatial translation. Numerical results for the single- and two-AB cases corroborate the asymptotic results. The theory provides analytical predictions for the deterministic branch-selective emergence of vector ABs from realistic initial perturbations.

\medskip \textit{The gMI analysis}---The Manakov system is the paradigmatic two-component integrable nonlinear-wave model, which is written as \cite{Manakov-74}
\begin{subequations}
\label{eq:cnls}
\begin{align}
\mathrm{i}q_{1,t}+\frac{1}{2}q_{1,xx}+\sigma (|q_{1}|^{2}+|q_{2}|^{2})q_{1}&
=0, \\
\mathrm{i}q_{2,t}+\frac{1}{2}q_{2,xx}+\sigma (|q_{1}|^{2}+|q_{2}|^{2})q_{2}&
=0,
\end{align}
\end{subequations}
where $\sigma =+1$ or $\sigma =-1$ is the focusing or defocusing sign of the nonlinearity. Here, $t$ is the evolution variable (the propagation distance in optics or time in Bose-Einstein condensates), $x$ is the spatial or temporal coordinate, and $q_{1,2}$ are the coupled field components. The Manakov system provides a fundamental framework for the study of MI and its nonlinear development leading to the formation of ABs \cite{Ling2019,Chen2021,Frisquet2015,Forest2000}, which also serve as building blocks creating recurrent nonlinear-wave patterns and rogue waves in coupled systems \cite{Baronio2012,Ling2014,Ling2017,Zhao2013,Qin2023,Liu2024,Mossman-MI,Baronio2014}. The Manakov system naturally arises in nonlinear optics, Bose-Einstein condensates, and hydrodynamics \cite{Agrawal-12,model3,Kang-96,model2}. Its two-component structure yields multi-branch MI spectra, supporting distinct vector AB states. The conventional linear-stability analysis typically addresses the continuous-wave (CW) background with superimposed perturbations \cite{Forest2000,Ling2017,Qin2023}. In experiments, the perturbation in general consists of a carrier wave modulated by a slowly varying envelope \cite{Kibler2015,Gelash2022}. Therefore, the analysis restricted to Fourier modes does not adequately represent such experimentally relevant excitations, motivating the development of a gMI framework for generalized Fourier modes.

The perturbed vector CW background is written as
\begin{equation*}
q_{i}(x,t)=a_{i}\mathrm{e}^{\theta _{i}}\left[1+p_{i}(x,t)\right], \quad i=1,2,
\end{equation*}
where $\theta _{i}=\mathrm{i}\{b_{i}x+[\sigma(a_{1}^{2}+a_{2}^{2})-b_{i}^{2}/2]t\}$, while $a_{i}$ and $b_{i}$ are the amplitude and wavenumber of the $i$-th component. The perturbation $p_{i}(x,t)$ is represented by the generalized Fourier-mode ansatz
\begin{equation}
p_{i}(x,t)=p_{i,+}(x,t)\mathrm{e}^{+\mathrm{i}kx}+p_{i,-}^{\ast }(x,t)
\mathrm{e}^{-\mathrm{i}kx},  \label{eq:pert}
\end{equation}
where $k$ is the modulation wavenumber and $p_{i,\pm}(x,t)=p_i^\pm f(x)\mathrm{e}^{\mathrm{i}k\Omega t}$ are spatially slowly varying sideband amplitudes. Here $p_i^\pm$ are constant mode coefficients, and $f(x)=\epsilon f_b(\mu x)$ denotes the prescribed, slowly varying complex envelope of the initial modulation imposed on the CW background, introduced so that the initial state can be readily prepared experimentally. The small parameter $\epsilon\ll1$ sets the initial modulation amplitude, and $\mu\ll1$ ensures a slowly varying envelope. Assuming that $f_b$ and its first two derivatives are bounded independently of $\epsilon$ and $\mu$, we have $f=\mathcal{O}(\epsilon)$, $f_x=\mathcal{O}(\epsilon\mu)$, and $f_{xx}=\mathcal{O}(\epsilon\mu^2)$. Substituting this ansatz into the linearized equations and neglecting the higher-order terms involving $f_x$ and $f_{xx}$ in the slow-envelope approximation yields the solvability condition $1+\sigma a_{1}^{2}/[(\Omega +b_{1})^{2}-k^{2}/4]+\sigma a_{2}^{2}/[(\Omega+b_{2})^{2}-k^{2}/4]=0$ \cite{Ling2019,Forest2000,Agrawal}. With a complex root $\Omega =\Omega _{R}+\mathrm{i}\Omega _{I}$, the factor $\mathrm{e}^{\mathrm{i}k\Omega t}$ grows as $\mathrm{e}^{Gt}$, where $G\equiv -k\Omega_{I}>0$ is the MI growth rate. The system with the defocusing nonlinearity has at most one unstable complex-conjugate eigenvalue pair \cite{Agrawal,Baronio2014,Qin2023}, whereas the focusing system may support two \cite{Forest2000,Ling2017,Qin2023}. The corresponding branch-resolved perturbation is written as $p_{i}(x,t)=\mathcal{P}_{i}[\Omega ,f](x,t)$, with
\begin{equation}
\mathcal{P}_{i}[\Omega ,f](x,t)=\frac{f(x)\mathrm{e}^{\mathrm{i}k(x+\Omega
t)}}{k+2b_{i}+2\Omega }+\frac{f^{\ast}(x)\mathrm{e}^{-\mathrm{i}k(x+\Omega
^{\ast }t)}}{(k-2b_{i}-2\Omega )^{\ast}}.  \label{eq:p-form}
\end{equation}

A common choice of $f(x)$ in previous studies was a constant envelope, for which the nonlinear development of MI leads to the establishment of an AB state \cite{Ling2019}. On the other hand, a slowly varying nonuniform envelope cannot be, in general, globally represented by an AB with fixed parameters. This motivates a locally matched asymptotic construction that determines the breather parameters of vector-AB states from $f(x)$ without requiring explicit solutions of the quartic spectral equation associated with the Lax spectral parameter.

\medskip \textit{Asymptotic AB solutions}---To connect the linear gMI stage with nonlinear breather formation, we first construct slowly varying approximate vector AB solutions from their exact counterparts~\cite{Ling2019}, as follows:
\begin{equation}
q_{i}^{[1]}(x,t;x_{1}(x),t_{1}(x))=a_{i}\mathrm{e}^{\theta _{i}}\frac{N_{i}}{M},  \label{eq:ab}
\end{equation}
where $N_{i}=\chi _{1,1}^{(i)}\mathrm{e}^{\omega _{R}}+\chi _{2,2}^{(i)}\mathrm{e}^{-\omega_{R}}+\chi_{1,2}^{(i)}\mathcal{B}\mathrm{e}^{\mathrm{i}\omega_{I}}+\chi _{2,1}^{(i)}\mathcal{B}^{\ast}\mathrm{e}^{-\mathrm{i}\omega _{I}}$, $\chi_{k,l}^{(i)}=(\chi _{k}^{\ast }+b_{i})/(\chi_{l}+b_{i})$, $M=2\cosh \omega_{R}+\mathcal{B}\mathrm{e}^{\mathrm{i}\omega_{I}}+\mathcal{B}^{\ast }\mathrm{e}^{-\mathrm{i}\omega _{I}}$, $\mathcal{B}=2\chi _{1I}/(2\chi _{1I}-\mathrm{i}\alpha )$, $\chi _{2}=\chi _{1}+\alpha$, $\omega _{I}=\alpha [(x-x_{1}(x))+(\chi _{1R}+\alpha/2)(t-t_{1}(x))]$, and $\omega_{R}=\alpha \chi _{1I}(t-t_{1}(x))$, with parameter $\chi _{1}=\chi _{1R}+\mathrm{i}\chi _{1I}$ satisfying the constraint $1+\sum_{i=1}^{2}\sigma a_{i}^{2}/[(\chi _{1}+b_{i})(\chi_{1}+\alpha +b_{i})]=0$. The occurrence frequency of each breather unit is $\alpha$. Its phase velocity is $V_{p}=-(\chi _{1R}+\alpha /2)$. The slowly varying functions $x_{1}(x)$ and $t_{1}(x)$ are introduced as the spatial and evolution-coordinate translation parameters of the local asymptotic AB solution, respectively.

To match the AB state to the gMI perturbation~\eqref{eq:p-form}, we define the phase-normalized field $q_{AB,i}^{[1]}=[(\chi _{2}+b_{i})/(\chi _{2}^{\ast }+b_{i})]q_{i}^{[1]}$
and expand it at fixed $x$ as $t-t_{1}(x)\rightarrow -\infty $. At leading order, we thus obtain
\begin{equation}
q_{AB,i}^{[1]}\approx a_{i}\mathrm{e}^{\theta _{i}}\left[ 1+\alpha \left(\chi_{2}^{\ast }-\chi _{2}\right) B_{i,-}\mathrm{e}^{\alpha \chi_{1I}(t-t_{1}(x))}\right] ,  \label{eq:asymp}
\end{equation}
with $B_{i,-}\equiv \mathrm{e}^{-A_{1}}/[(\chi _{1}+b_{i})(\chi _{2}^{\ast}-\chi _{1})]-\mathrm{e}^{A_{1}}/[(\chi _{2}^{\ast }+b_{i})(\chi _{1}^{\ast}-\chi _{2})]$, and $A_{1}\equiv \mathrm{i} \omega _{I}$. Matching the gMI perturbation from Eq.~\eqref{eq:p-form} to the AB's tail given by Eq.~\eqref{eq:asymp}, we find $\chi _{1}^{\ast }+\alpha /2=\Omega $, $\alpha =k$, and $f(x)=-4\Omega_{I}\alpha \mathrm{e}^{-\mathrm{i}\alpha \left( x_{1}(x)+\Omega t_{1}(x)\right) }/(\mathrm{i}\alpha +2\Omega _{I})$, which yields the following expressions for $x_{1}(x)$ and $t_{1}(x)$:
\begin{subequations}
\label{eq:dynamic}
\begin{align}
t_{1}(x)& =\frac{1}{G}\ln \left( \frac{\mathcal{K}}{|f(x)|}\right) ,
\label{eq:dynamic-a} \\
x_{1}(x)& =-\Omega _{R}t_{1}(x)-\frac{\arg f(x)+\phi }{k},
\label{eq:dynamic-b}
\end{align}%
\end{subequations}
where $\mathcal{K}=4|k\Omega _{I}|/\sqrt{k^{2}+4\Omega _{I}^{2}}$ and $\phi=\arg (\mathrm{i}k+2\Omega _{I})$. If $f(x)$ remains nonzero in the considered interval and has bounded logarithmic derivatives, Eqs.~\eqref{eq:dynamic} imply $\partial _{x}x_{1},\partial _{x}t_{1}=\mathcal{O}(\mu )$ and $\partial _{x}^{2}x_{1}, \partial_{x}^{2}t_{1}=\mathcal{O}(\mu^{2})$. Substituting $t_{1}(x)$ and $x_{1}(x)$ into the approximate solution~\eqref{eq:ab}  completes the construction of the asymptotic vector AB solution, establishing  its local compatibility  with the gMI mode. The accuracy of this asymptotic solution has been verified through numerical error analysis and local residual estimates (see Supplemental Material~\cite{supp}).  In particular, Eq.~\eqref{eq:dynamic-a} determines the AB emergence point $t_{1}(x)$ from the modulus $|f(x)|$. For the selected growing mode with gain $G$, decreasing $|f(x)|$ shifts the emergence point to larger $t$, with a delay that increases logarithmically with the inverse envelope modulus $1/|f(x)|$. Further, Eq.~\eqref{eq:dynamic-b} determines the spatial translation $x_{1}(x)$ of the AB formation position. It includes the drift term $-\Omega_{R}t_{1}$ accumulated before the formation, a fixed offset set by the selected mode, and a shift set by the phase $\arg f(x)$. To apply these conclusions, $f(x)$ must remain nonzero and satisfy the constraint $|\partial _{x}\ln f|\ll |k|$. These conditions are satisfied, for example, by the perturbation envelope $f(x)=\epsilon \sech(x/50)$, as well as by the oscillatory envelope, $f(x)=\epsilon \exp [0.4\cos (x/20)]$.

\begin{figure}[t]
\centering
\includegraphics[width=85mm]{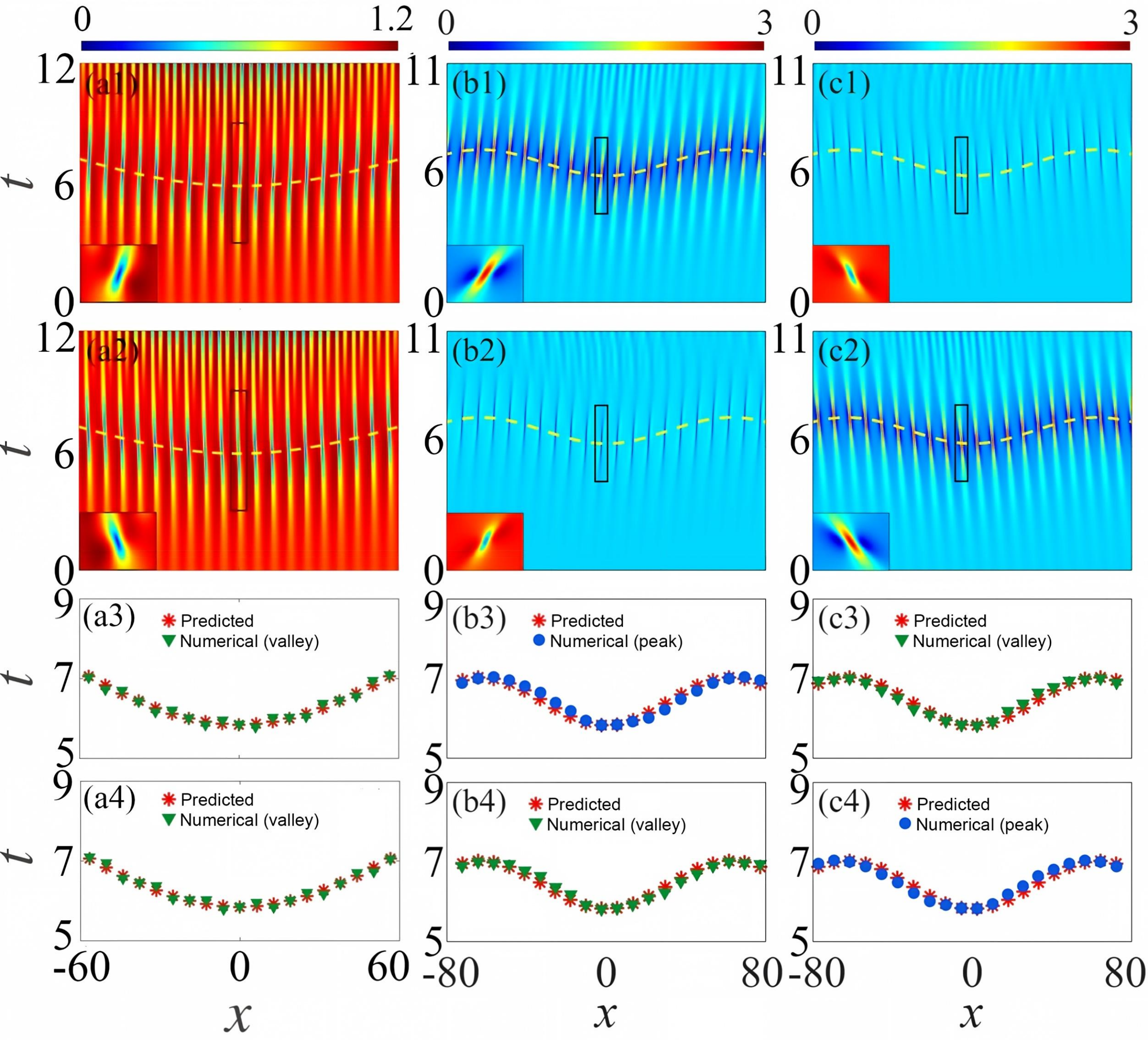}
\caption{The asymptotic AB excitation with $a_1=a_2=1$. Column (a) corresponds to the defocusing nonlinearity, with the input determined by $f(x)=\protect\epsilon
\sech(x/50)$ ($\protect\epsilon =10^{-1}$, $k=1$, and $b_{1}=-b_{2}=1.1$). Columns (b) and (c) show separate excitations of the two MI branches in the self-focusing regime, with input $f(x)=\protect\epsilon \exp [0.4\cos (x/20)]$ ($\protect\epsilon =2\times10^{-2}$, $k=0.8$, and $b_{1}=-b_{2}=1$). The upper two rows show $|q_{1}|$ and $|q_{2}|$, with the predicted trajectories overlaid as yellow dashed curves. The insets magnify individual breather units within the boxed regions. The lower two rows compare the analytical predictions (red asterisks) and numerical results for the locations of the centers of breather units (blue circles for peaks and green downward triangles for valleys).}
\label{fig1}
\end{figure}

To locate individual units in the emerging breather chain, we use the total density of the asymptotic vector AB evaluated at local translated positions. It is $|\mathbf{q}^{[1]}|^{2}=|a_{1}|^{2}+|a_{2}|^{2}+\sigma \partial_{x}^{2}\left(\ln M\right)$, where $\mathbf{q}^{[1]}=(q_{1}^{[1]},q_{2}^{[1]})^{\mathsf{T}}$, and $\partial_{x}$ acts only on the explicit $x$ dependence of $M$ [see Eq.~\eqref{eq:ab}]. At the local formation coordinate $t=t_{1}(x)$, the center condition in the translated coordinate is $x-x_{1}(x)=x_n$, where $x_{n}=[(2n+1)\pi-\arccos \delta]/\alpha $ and $\delta =2\chi _{1I}/\sqrt{\alpha ^{2}+4\chi _{1I}^{2}}$. The corresponding center coordinate $%
c_{n}$ of the slowly translated pattern is therefore determined by $c_{n}-x_{1}(c_{n})=x_{n}$. At the center $(c_{n},t_{1}(c_{n})) $, the total density is $|\mathbf{q}^{[1]}|^{2}=|a_{1}|^{2}+|a_{2}|^{2}+2\sigma \chi_{1I}(2\chi _{1I}+\sqrt{\alpha ^{2}+4\chi _{1I}^{2}})$, while the modulus of each component is $\left\vert q_{i}^{[1]}\left( c_{n},t_{1}(c_{n})\right)\right\vert =|a_{i}|B_{i}/|(\chi _{1}+b_{i})(\chi _{1}+\alpha +b_{i})|$, with $B_{i}=\big\vert\left( 2(1-\delta )\right) ^{-1}\alpha ^{2}\delta -\big[(\chi _{1R}+b_{i}+\alpha /2)^{2}-\alpha ^{2}/4+\chi _{1I}^{2}\big]\big\vert.$ At each center, the total density $\lvert \mathbf{q}^{[1]}\rvert ^{2}$ has a
local maximum for $\sigma =+1$ and a minimum for $\sigma =-1$.

We next test these predictions by performing direct simulations of the Manakov system with the initial condition $q_{i}(x,0)=a_{i}\mathrm{e}^{\theta _{i}}\left( 1+p_{i}(x,0)\right) $ with perturbation $p_{i}(x,0)$ given by Eq.~\eqref{eq:p-form}, and compare the positions of the numerically found centers with the predictions provided by Eq.~\eqref{eq:dynamic}. As shown in Fig.~\ref{fig1}, branch-selective asymptotic vector ABs are excited, whose analytically predicted trajectories closely follow the numerical breathers' paths in both the defocusing (column \ref{fig1}(a)) and focusing (columns \ref{fig1}(b) and \ref{fig1}(c)) cases. Specifically, the lower panels show close proximity between the predicted centers' positions and their counterparts extracted from the numerical simulations. These results demonstrate the asymptotic predictions and the ability to selectively excite particular vector-AB branches. The validity of the asymptotic theory is further supported by the extension to a nonintegrable version of the system, i.e., $\mathrm{i}q_{1,t}+\tfrac12q_{1,xx}+\sigma\big[(1+\epsilon_1)|q_1|^2+(1-\epsilon_1)|q_2|^2\big]q_1=0, \mathrm{i}q_{2,t}+\tfrac12q_{2,xx}+\sigma\big[(1-\epsilon_1)|q_1|^2+(1+\epsilon_1)|q_2|^2\big]q_2=0$. Here, $\epsilon_1$ is a small parameter representing the perturbation. Numerical simulations demonstrate that the breather patterns are preserved under weak nonintegrable perturbations (see Supplemental Material~\cite{supp}).

\medskip \textit{Asymptotic two-AB solutions}---A pair of ABs can be excited by introducing two perturbation modes of the form given by Eq.~\eqref{eq:p-form} into the initial condition. The two modes may be drawn either from the same MI branch or from two distinct branches, in the case of the focusing nonlinearity. To realize such selective two-AB excitations, it is necessary to match the gMI to the asymptotic two-AB solutions. We first construct the approximate two-AB field $q_i^{[2]}$ by replacing the four constant translations of the exact two-AB solution by slowly varying functions, and take its phase-normalized form $q_{AB,i}^{[2]}(x,t;x_{1}(x),t_{1}(x),x_{2}(x),t_{2}(x))$ as an asymptotic ansatz, which contains four eigenvalues $\chi _{1}, \cdots,\chi _{4}$, coupled by $\chi _{2j}=\chi _{2j-1}+\alpha $ ($j=1,2$) (see details in the Supplemental Material \cite{supp}). We here focus on the two-branch case, which offers a more generic and more transparent branch-resolved description, while the same-branch case can be treated analogously.

\begin{figure}[!t]
\centering
\includegraphics[width=85mm]{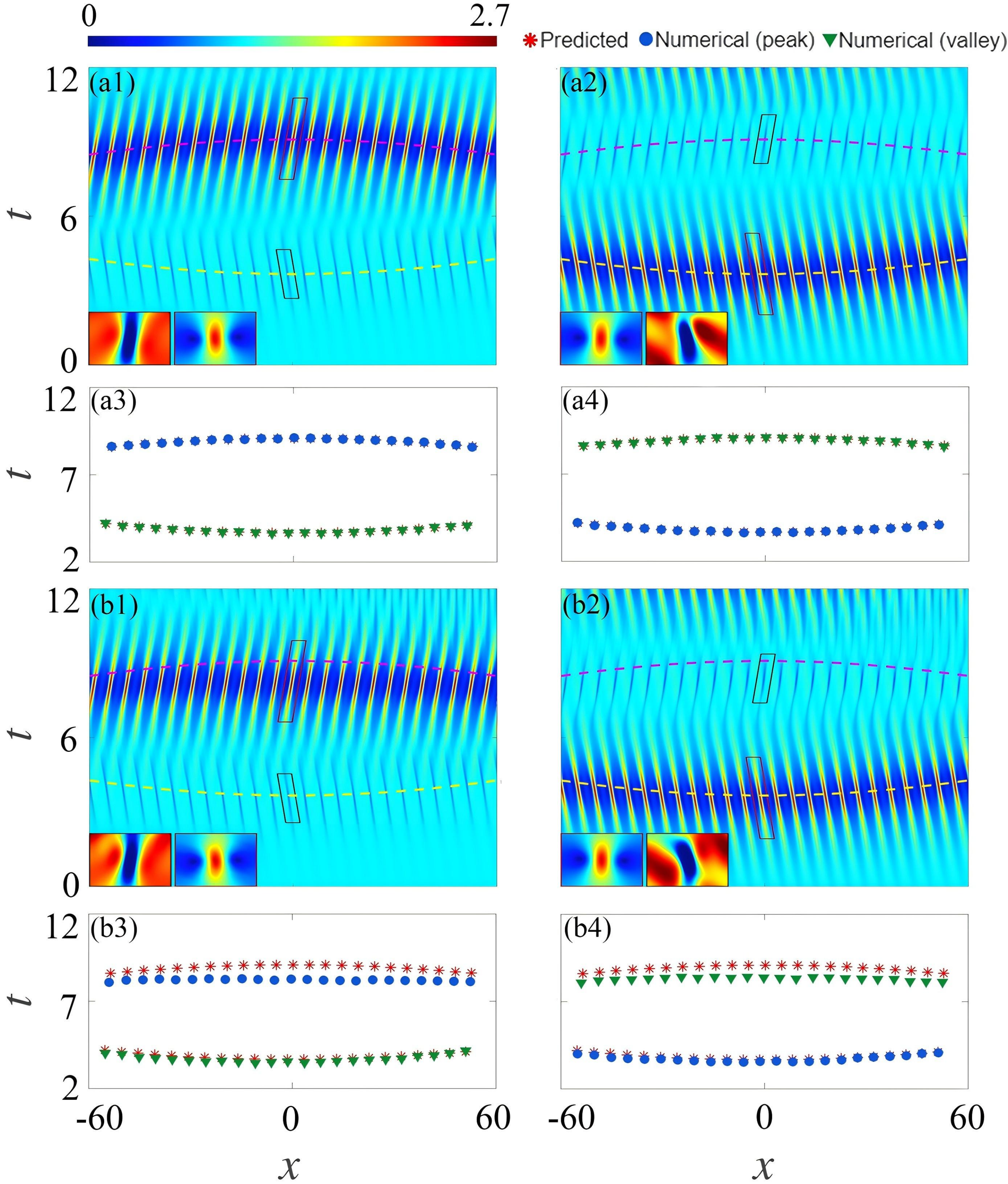}
\caption{Two-AB excitations with envelopes $f_{1}(x)=\protect\epsilon\exp[2\cos (x/80)]$ and $f_{2}(x)=\protect\epsilon \mathrm{e}^{-1}\exp [-2\cos(x/80)]$, with parameters $\protect\epsilon =10^{-2}$, $b_{1}=-b_{2} =1.3$, $k=1.3$, and $a_1=a_2=1$. Panel (a) illustrates the integrable Manakov case; 
panel (b) the weakly non-integrable case ($\protect\epsilon_1=0.02$). Yellow and magenta dashed curves represent the asymptotic trajectories $T_{j}^{(s)}(x)$. Panels (a3-a4) and (b3-b4) compare the analytical asymptotic predictions and numerical results.}\label{fig2}
\end{figure}

We write the two unstable conjugate pairs as $\Omega ^{(\ell)}=\Omega_{R}^{(\ell)}+\mathrm{i}\Omega _{I}^{(\ell)}$, $\ell=1,\cdots,4$, with $\Omega ^{(2)}=(\Omega ^{(1)})^{\ast }$, $\Omega ^{(4)}=(\Omega ^{(3)})^{\ast}$, and $\Omega _{I}^{(1)},\Omega _{I}^{(3)}<0$. We select $\Omega ^{(2j-1)}$ as the growing root satisfying the condition $-k\Omega _{I}^{(2j-1)}>0$, and $\Omega ^{(2j)}$ is its complex conjugate. Then, the perturbed vector CW background, containing the two-mode perturbation, is written as
\begin{equation}
q_{i}(x,t)=a_{i}\mathrm{e}^{\theta _{i}}\bigg[1+\sum_{j=1}^{2}\mathcal{P}_{i}[\Omega ^{(2j-1)},f_{j}](x,t)\bigg].  \nonumber
\end{equation}
Matching the two retained growing modes to the solution $q_{AB,i}^{[2]}$ at $t-t_{j}(x)\rightarrow-\infty $ yields $\alpha =k$ and $\chi _{2j-1}^{\ast }+\alpha /2=\Omega
^{(2j-1)}$. With $D_{m,l}\equiv \chi _{l}-\chi _{m}^{\ast }$, we define $C_{0}=1/(D_{2,2}D_{4,4})-1/(D_{2,4}D_{4,2})$, $\Delta_{1}=1/(D_{1,2}D_{4,4})-1/(D_{1,4}D_{4,2})$, and $\Delta_{2}=1/(D_{3,4}D_{2,2})-1/(D_{3,2}D_{2,4})$. Then, we obtain
\begin{subequations}
\label{2ABtx}
\begin{align}
t_{j}(x)& =\frac{1}{G_{j}}\ln \!\left( \frac{\mathcal{K}_{j}}{|f_{j}(x)|}\right), \\
x_{j}(x)& =-\Omega _{R}^{(2j-1)}t_{j}(x)-\frac{1}{k}\arg \!\left( -\frac{f_{j}(x)C_{0}}{2k\Delta _{j}}\right) ,
\end{align}
\end{subequations}
where $G_{j}\equiv -k\Omega_{I}^{(2j-1)}>0$ and $\mathcal{K}_{j}\equiv2|k\Delta _{j}|/|C_{0}|$. Substituting expressions~\eqref{2ABtx} in $q_{AB,i}^{[2]}$, we obtain the asymptotic two-AB solutions. To predict where the two breathers emerge, we must take their interaction into account. For well-separated breathers, the interaction produces center shifts $\delta_{j}^{(s)}$~\cite{supp} relative to the reference positions $t_{j}(x)$, obtained from Eq.~\eqref{2ABtx}. The shifts are obtained from the two-AB
solution, with $s$ labeling the corresponding asymptotic state. Thus, the predicted formation positions are $T_{j}^{(s)}(x)=t_{j}(x)+\delta _{j}^{(s)}$. For a fixed local ordering, let $s$ and $r$ label the asymptotic states of the second and first breathers, respectively. Their signed temporal separation is
\begin{small}
\begin{equation}
\Delta T^{(s,r)}\!(x)\!=\!\frac{1}{G_{2}}\ln\! \left( \!\frac{\mathcal{K}_{2}}{|f_{2}\!(x)|}\!\right)\!-\!\frac{1}{G_{1}}\ln \!\left(\! \frac{\mathcal{K}_{1}}{|f_{1}\!(x)|}\!\right)\!+\!\delta _{2}^{(s)}\!-\!\delta _{1}^{(r)}.
\label{eq:two-sector-gap}
\end{equation}
\end{small}

For a prescribed separation between the emerging breathers, Eq.~\eqref{eq:two-sector-gap} determines the required relation between the input envelopes. This is a guide for designing input modulations for the controlled excitation of well-separated vector breathers. Numerical results support the asymptotic prediction of two-AB dynamics in the
integrable Manakov system and its weakly non-integrable version, as shown in Fig.~\ref{fig2}. In the integrable system [Fig.~\ref{fig2}(a)], the prescribed envelopes excite two well-separated ABs from unstable modes with equal MI gain. Their peak and valley centers closely follow the predicted trajectories $T_{j}^{(s)}(x)$, confirming the envelope- and
interaction-induced separation given by Eq.~\eqref{eq:two-sector-gap}. Upon introducing the integrability-breaking terms with strength $\epsilon_{1}=0.02$ [Fig.~\ref{fig2}(b)], the two-AB pattern remains robust, although both breathers shift to earlier formation times.

\medskip \textit{Conclusion}---The above asymptotic single- and two-AB constructions go beyond the scalar asymptotic matching \cite{Trillo2019}, which links a single MI mode to the AB tail and yields the formation trajectory and minimum formation distance. Our framework resolves multiple unstable branches in coupled vector systems by directly matching each selected gMI eigenmode to the preformation tail of the corresponding vector AB, thus enabling branch-selective excitation. It determines the coordinates and densities of individual vector-breather units, as well as the separation between two ABs. It is worth noting that this framework requires comparable gains among the relevant gMI branches. If gains differ substantially, the higher-gain branch eventually overtakes the initially selected lower-gain branch, thereby limiting selective excitation. Within this regime, the framework generalizes directly to an arbitrary number of coupled components with unstable modes. 

These results demonstrate that our asymptotic theory provides a direct route to branch-selective excitation and positioning of vector ABs, extending the range of dynamics covered by exact solutions of integrable systems. This enables asymptotic predictions beyond the reach of existing theories for coexisting unstable MI branches, and offers a practical guide for multicomponent experiments. The robustness of the predicted breather patterns under weak nonintegrable perturbations and weak random noise further supports their experimental observability. For an optical realization, the prescribed gMI perturbations can be imposed on two orthogonally polarized CW pumps using programmable amplitude and phase shaping \cite{Frisquet2014,Kibler2015,Gelash2022}. The orthogonally polarized propagation in the Manakov regime has also been demonstrated \cite{OF2018,OF2016,OFAB,Frisquet2015}. For example, choosing a physical modulation frequency of $40\,\mathrm{GHz}$, the dimensionless parameters in Fig.~\ref{fig1}(b,c) correspond to a pump frequency separation of \(100\,\mathrm{GHz}\) and approximately \(2.2\,\mathrm{W}\) per pump for standard-fiber parameters. These frequency and power scales are comparable with those used in existing two-component AB experiments \cite{OFAB}. These findings provide a reliable theoretical foundation for the controllable engineering of vector breathers.

\setlength{\bibsep}{0pt}

\onecolumngrid
\clearpage
\setcounter{equation}{0}
\setcounter{figure}{0}
\renewcommand{\theequation}{S\arabic{equation}}
\renewcommand{\thefigure}{S\arabic{figure}}
\renewcommand{\theHequation}{supp.\arabic{equation}}
\renewcommand{\theHfigure}{supp.\arabic{figure}}

\begin{center}
{\large\bfseries Supplemental Material: Asymptotic breathers in the Manakov system\par}

\end{center}

\section{The Darboux transformation and vector Akhmediev breather solutions}

The Manakov system for the two-component complex field
$\mathbf{q}=(q_{1},q_{2})^{\mathsf T}$, with
$q_i(x,t):\mathbb{R}^{2}\rightarrow\mathbb{C}$ for $i=1,2$, is
\begin{equation}
\mathrm{i}\mathbf{q}_{t}+\frac{1}{2}\mathbf{q}_{xx}+\sigma \mathbf{q}\mathbf{%
q}^{\dag }\mathbf{q}=0,  \label{eq:cnls}
\end{equation}%
where $\sigma =+1$ and $\sigma =-1$ denote the focusing and defocusing
regimes, respectively. The superscripts $\mathsf T$, $*$, and $\dagger$ denote
transpose, complex conjugation, and Hermitian conjugation, respectively;
$|\mathbf q|^2=\mathbf q^\dagger\mathbf q=|q_1|^2+|q_2|^2$. Here $t$ denotes the evolution variable, $x$ the
spatial coordinate, and the two entries of $\mathbf q$ represent polarization,
spatial, or internal-state components, depending on the physical realization.
We denote the real plane-wave amplitudes by $a_1,a_2$, the real background
wavenumbers by $b_1,b_2$, and set $|a|^2=a_1^2+a_2^2$.
Let $\Phi$ be a three-component auxiliary eigenfunction and
$\lambda\in\mathbb C\cup\{\infty\}$ its spectral parameter. The Lax pair for
Eq.~\eqref{eq:cnls} is
\begin{equation}
\Phi _{x}=U(\lambda ;Q)\Phi ,\quad \Phi _{t}=V(\lambda ;Q)\Phi .
\label{eq:lax-pair}
\end{equation}%
With
\[
Q=\begin{pmatrix}0&\sigma\mathbf q^\dagger\\\mathbf q&0\end{pmatrix},
\qquad \sigma_3=\operatorname{diag}(1,-\mathbb I_2),
\]
where $\mathbb I_n$ is the $n\times n$ identity matrix, the matrices are
\begin{equation}
\begin{aligned} U(\lambda; Q)&={\rm
i}\left[\frac{\lambda}{2}\left(\sigma_3+\mathbb{I}_3\right)+Q\right],\\
V(\lambda;Q)&={\rm i}\left[ \frac
{\lambda^2}{4}(\sigma_3+\mathbb{I}_3)+\frac {\lambda}{2}Q-\frac
{1}{2}\sigma_3\left(\mathrm{i}Q_x+Q^2\right)+\sigma|a|^2\mathbb{I}_3\right].
\end{aligned}
\end{equation}%
A new Lax eigenfunction $\Phi^{[1]}$ and field $\mathbf q^{[1]}$
are generated by means of the following Darboux transform~\cite{Wright2000,Ling2019}:
\begin{equation}
\Phi ^{[1]}=T_{1}\Phi ,\quad T_{1}=\mathbb{I}_{3}+\frac{\lambda
_{1}^{\ast }-\lambda _{1}}{\lambda -\lambda _{1}^{\ast }}P_{1},\quad P_{1}=%
\frac{\Phi _{1}\Phi _{1}^{\dagger }\Sigma }{\Phi _{1}^{\dagger }\Sigma \Phi
_{1}},  \label{eq:dt}
\end{equation}%
where $\Phi _{1}=(\psi _{1},\psi _{2},\psi _{3})^{\mathsf T}$ is a particular
solution of the Lax pair at $\lambda =\lambda _{1}$, and $\Sigma =%
\mathrm{diag}(1,\sigma ,\sigma )$. The corresponding B\"{a}cklund
transformation is
\begin{equation}
\mathbf{q}^{[1]}=\mathbf{q}+\frac{(\lambda _{1}^{\ast }-\lambda _{1})\psi
_{1}^{\ast }}{|\psi _{1}|^{2}+\sigma (|\psi _{2}|^{2}+|\psi _{3}|^{2})}%
\begin{bmatrix}
\psi _{2} \\
\psi _{3}%
\end{bmatrix}%
.  \label{eq:bt}
\end{equation}

To derive vector Akhmediev breather (AB) solutions of the Manakov system \eqref{eq:cnls}, we
start with the plane-wave background solution $q_{0,i}=a_{i}\mathrm{e}^{\theta _{i}}$,
where $\theta _{i}=\mathrm{i}\left( b_{i}x+\left( \sigma
|a|^2-\frac{1}{2}b_{i}^{2}\right) t\right) $. Substituting
it into the Lax pair \eqref{eq:lax-pair}
yields the vector solution $\Phi _{1,m}=D_{3}[\mathrm{e}^{\omega _{m}},\frac{%
\mathrm{e}^{\omega _{m}}}{\chi _{m}+b_{1}},\frac{\mathrm{e}^{\omega _{m}}}{\chi _{m}+b_{2}}%
]^{\mathsf T}$, where $D_{3}=\operatorname{diag}\left( 1,a_{1}\mathrm{e}^{\theta _{1}},a_{2}\mathrm{e}^{\theta
_{2}}\right)$, $\omega _{m}=\mathrm{i}\chi _{m}\left( x+\frac{1}{2}\chi
_{m}t+\vartheta _{m}\right)$, and $\vartheta _{m}$ is a complex parameter.
For fixed $\lambda_1$, the $\chi_m$ are roots of the following cubic equation in $\chi$.
\begin{equation}\label{eq:lax-cubic}
(\chi +b_{1})(\chi +b_{2})\left[ \chi -\lambda _{1}-\sum_{i=1}^{2}\frac{%
\sigma a_{i}^{2}}{\chi +b_{i}}\right] =0.
\end{equation}

Using the identity \cite{Ling2019}
\begin{equation}
\frac{\lambda _{1}^{\ast }-\lambda _{1}}{\chi _{l}^{\ast }-\chi _{m}}%
=1+\sum_{i=1}^{2}\frac{\sigma a_{i}^{2}}{(\chi _{m}+b_{i})(\chi _{l}^{\ast
}+b_{i})},
\end{equation}%
we combine the eigenfunctions associated with two selected roots as $\Phi_1=\Phi_{1,1}+\Phi_{1,2}$
(the relative weights are absorbed into $\vartheta_m$) and obtain
the following one-fold Darboux solution:
\begin{equation}
q_{i}^{[1]}=a_{i}\frac{N_i}{M}\mathrm{e}^{\theta _{i}},  \label{eq:AB}
\end{equation}%
where
\begin{equation}
\begin{aligned}
M=\sum_{l=1}^{2}\sum_{m=1}^{2}\frac{\exp\left(\omega_{l}^*+\omega_{m}\right)}{%
\chi_{m}-\chi_{l}^*},\qquad
N_i=\sum_{l=1}^{2}\sum_{m=1}^{2}\frac{\chi_{l}^*+b_i}{\chi_{m}+b_i}\frac{\exp%
\left(\omega_{l}^*+\omega_{m}\right)}{\chi_{m}-\chi_{l}^*}. \end{aligned}
\end{equation}%
Different choices of the two roots produce different nonlinear waves on
the plane-wave background. An AB is localized along the evolution coordinate,
and its modulus is periodic in space. To construct it, we choose
$\chi_2=\chi_1+\alpha$ with $\alpha>0$ and
$\chi_{1I}=\operatorname{Im}\chi_1>0$, so the roots have the same imaginary
part. We also write $\chi_{1R}=\operatorname{Re}\chi_1$. For brevity, we leave their dependence on \(\lambda_1\) implicit below. One root is determined by $\chi _{1}-\lambda
_{1}-\sum_{i=1}^{2}\frac{\sigma a_{i}^{2}}{\chi _{1}+b_{i}}=0$, while the
other one is $\chi _{2}=\chi _{1}+\alpha $, satisfying the
relation $\chi _{1}+\alpha -\lambda _{1}-\sum_{i=1}^{2}\frac{\sigma a_{i}^{2}%
}{\chi _{1}+\alpha +b_{i}}=0$. Subtracting the former equation from the
latter one gives:
\begin{equation}
1+\sum_{i=1}^{2}\frac{\sigma a_{i}^{2}}{\left( \chi _{1}+b_{i}\right) \left(
\chi _{1}+\alpha +b_{i}\right) }=0,  \label{eq:ab-deter}
\end{equation}%
which is exactly the parameter equation determining the vector AB solutions.
Let $x_1$ and $t_1$ be the constant spatial and evolution-coordinate
translations of the exact AB, and define
\[
\omega_R=\alpha\chi_{1I}(t-t_1),\qquad
\omega_I=\alpha\left[(x-x_1)+(\chi_{1R}+\alpha/2)(t-t_1)\right],
\qquad \theta=\arccos\frac{2\chi_{1I}}{\sqrt{\alpha^2+4\chi_{1I}^2}}.
\]
After canceling a common factor in $N_i/M$, the numerator and
denominator in Eq.~\eqref{eq:AB} can be written as
\begin{equation}
\begin{split}
& M=2\cosh \left( \omega _{R}\right) +2\frac{2\chi _{1I}}{\sqrt{\alpha
^{2}+4\chi _{1I}^{2}}}\cos (\theta +\omega _{I}), \\
& N_i=\frac{\chi _{1}^{\ast }+b_{i}}{\chi _{1}+b_{i}}\mathrm{e}^{\omega _{R}}+%
\frac{\chi _{2}^{\ast }+b_{i}}{\chi _{2}+b_{i}}\mathrm{e}^{-\omega _{R}}+\frac{\chi
_{1}^{\ast }+b_{i}}{\chi _{2}+b_{i}}\frac{2\chi _{1I}\mathrm{e}^{\mathrm{i}\omega _{I}}}{2\chi
_{1I}-\mathrm{i}\alpha }+\frac{\chi _{2}^{\ast }+b_{i}}{\chi _{1}+b_{i}}\frac{2\chi
_{1I}\mathrm{e}^{-\mathrm{i}\omega _{I}}}{2\chi _{1I}+\mathrm{i}\alpha }.
\end{split}%
\end{equation}%
The spatial modulation period of the exact AB is $2\pi/\alpha$, and
the phase velocity of the periodic breather pattern is
$-(\chi _{1R}+\alpha/2)$. The root $\chi_1$ is determined by Eq.~\eqref{eq:ab-deter}.
Each component can assume one of three distinct shapes: the eye-shaped (ES),
anti-eye-shaped (AES), and four-petal (FP) patterns \cite{Ling2017,Chen2021}.

We introduce slow spatial modulation by replacing the constants $x_{1}$ and
$t_{1}$ with slowly varying functions $x_{1}(x)$ and $t_{1}(x)$. Let
$\epsilon\ll1$ denote the amplitude of the generalized modulational-instability
(gMI) perturbation to be
matched below, and let $\mu\ll1$ denote the inverse
envelope scale. We assume
$\partial_x x_{1},\partial_x t_{1}=\mathcal O(\mu)$ and
$\partial_x^2x_{1},\partial_x^2t_{1}=\mathcal O(\mu^2)$.
Thus $\epsilon$ controls the perturbation amplitude, whereas $\mu$ controls the
spatial variation. The resulting
slowly translated AB is an asymptotic ansatz rather than an exact solution of
the Manakov system \eqref{eq:cnls}. The translations $x_1(x)$ and $t_1(x)$
are determined by the tail matching described in the main text [Eq.~(6)].
The modulated profile is generally not globally periodic;
$2\pi/\alpha$ remains the reference modulation length inherited from the
underlying exact AB. 

\section{Modulational instability (MI) spectra}

To facilitate understanding of the complex MI dynamics in the Manakov system %
\eqref{eq:cnls}, we present explicit MI spectra here. We consider the basic
case of the equal-amplitude background, $a_{1}=a_{2}=1$, and choose the
background-wavenumber condition $b_{1}=-b_{2}=\beta >0$.
Let $k>0$ be the perturbation wavenumber of a linear mode proportional to
$\exp[\mathrm i k(x+\Omega t)]$. Where both denominators are nonzero, the MI dispersion relation is
\begin{equation}
1+\frac{\sigma a_{1}^{2}}{\left( \Omega +b_{1}\right) ^{2}-\frac{1}{4}k^{2}}+%
\frac{\sigma a_{2}^{2}}{\left( \Omega +b_{2}\right) ^{2}-\frac{1}{4}k^{2}}=0.
\label{MI1}
\end{equation}
With $d=\beta^2+k^2/4-\sigma$ and $\nu=\sqrt{(k^2-4\sigma)\beta^2+1}$
(the principal square root), the four roots of the quartic characteristic polynomial of the linearized
system, counted with multiplicity, are
\begin{figure}[!b]
	\centering
	\includegraphics[width=0.98\textwidth]{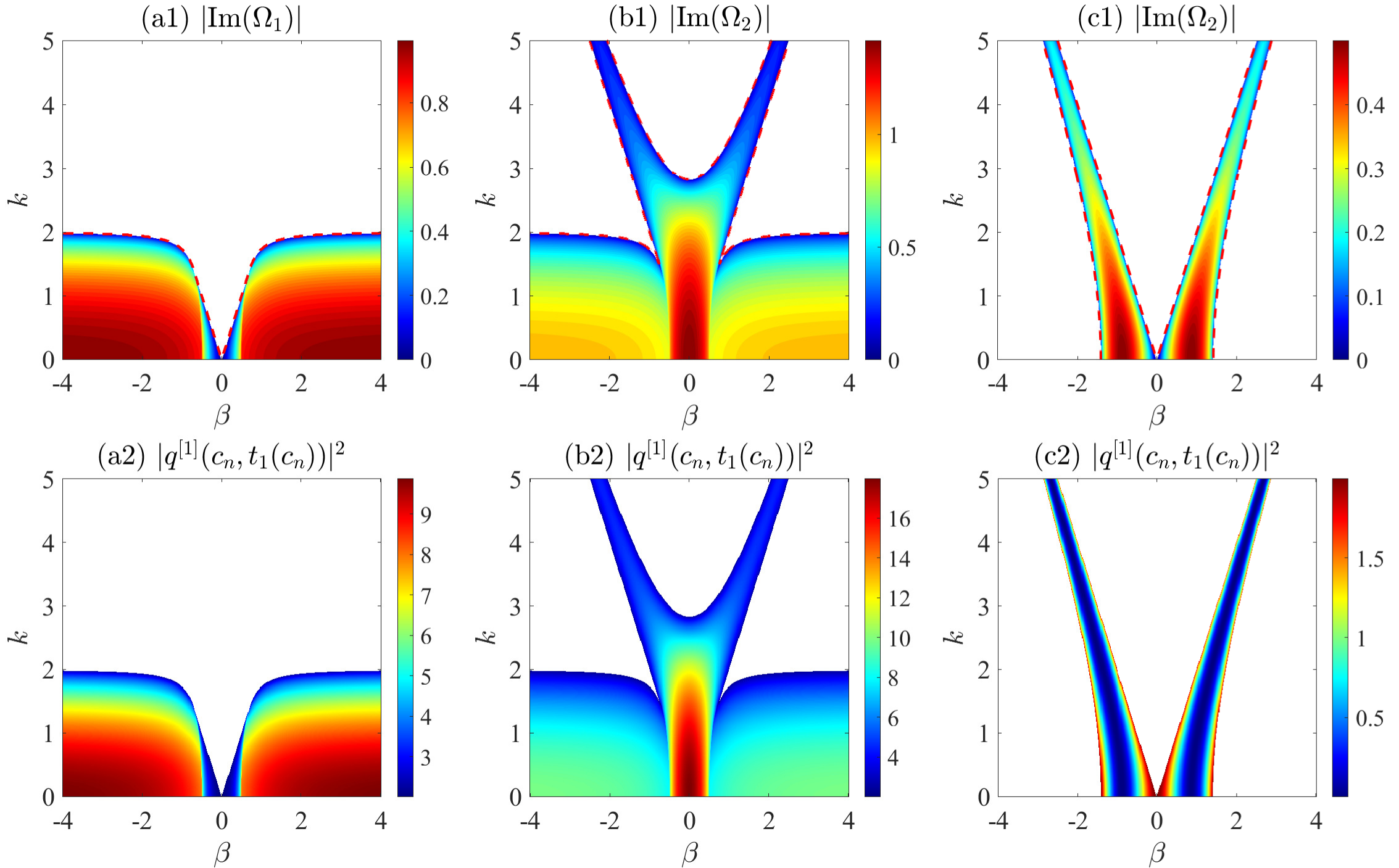}
\caption{MI regions and normalized gains for background wavenumber $\beta$ and
perturbation wavenumber $k$ in the focusing (a,b) and defocusing (c) regimes.
The upper panels show $|\operatorname{Im}\Omega|=G/|k|$ and its limit at $k=0$.
Red dashed curves in these panels mark the MI boundaries. The lower panels show the
maximum total density $|\mathbf q|^2$ for focusing and its minimum for defocusing.}
	\label{figS1}
\end{figure}
\begin{equation}
{\Omega_{1,\pm}=\pm \sqrt{d+\nu },\,\,\,\,
\Omega_{2,\pm}=\pm \sqrt{d-\nu }}.
\end{equation}%
The MI gain (growth-rate magnitude) is $G=|\operatorname{Im}(k\Omega)|$.
For each root family $\Omega_{j,\pm}$ containing nonreal roots, let $\Omega_j$
denote the root with $\operatorname{Im}\Omega_j<0$. For $k>0$, the corresponding
MI mode grows with gain $G_j=-k\operatorname{Im}\Omega_j>0$. When considering
a single mode, we omit the index $j$ and write $\Omega$ and $G$.
When $d+\nu$ and $d-\nu$ are real, each negative value gives a purely
imaginary conjugate pair with nonzero MI gain, and each nonnegative value gives
real roots with zero gain. When $\nu$ is nonreal, complex conjugation pairs
roots from different families. In the defocusing case ($\sigma=-1$), there
is at most one nonreal conjugate pair and hence at most one unstable MI branch.
Figure~\ref{figS1}(c1) shows the corresponding normalized-gain spectrum in
the $(\beta,k)$ plane. In the focusing case ($\sigma=+1$), two nonreal
conjugate pairs and hence two unstable MI branches can coexist.
Figures~\ref{figS1}(a1) and (b1) show the normalized gains associated with
$\Omega_{1,\pm}$ and $\Omega_{2,\pm}$,
respectively. The two gains are equal when
$0<k<2$ and $(4-k^{2})\beta ^2>1$, which is precisely the region where the
inner radical in the explicit dispersion relation is purely imaginary.
Outside this region the two gains are generally unequal, and in part of the
parameter plane only one MI branch is unstable. At $\beta=0$, the modes
associated with $\Omega_{1,\pm}$ have zero gain, so only the $\Omega_{2,\pm}$
root family can give a growing mode. Modes on different MI branches give rise to different breathers in the nonlinear
stage, and the wavenumber limit $k\to0$ is related to the generation
mechanism of rogue waves (RWs).

\begin{figure}[!t]
	\centering
	\includegraphics[width=0.8\textwidth]{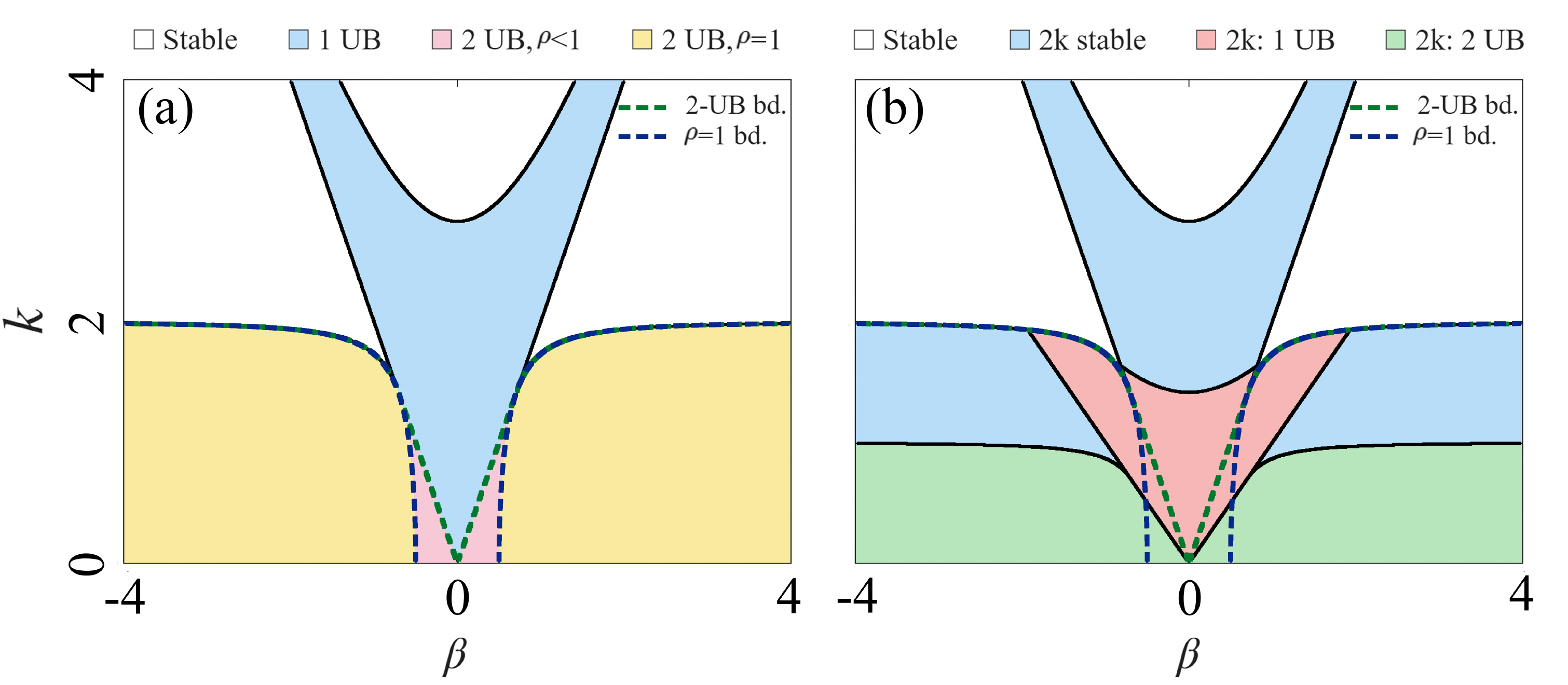}
\caption{Focusing-instability diagram used for parameter selection
		at unit background amplitude. Panel (a) shows the number of unstable MI branches
		at the fundamental wavenumber $k$ and their relative gains. Here $G_{+}$ and $G_{-}$ are the
		larger and smaller gains, respectively, and $\rho=G_{-}/G_{+}$.
		The abbreviation ``UB'' denotes an unstable branch, while the dark-blue
		dashed curve marks the equal-growth boundary $\rho=1$. Panel (b) shows the stability classification of the second
		harmonic $2k$ on the same $(\beta,k)$ plane. The green dashed curve marks
		the boundary of the region with two unstable MI branches at $k$.}
	\label{figS2}
\end{figure}

{
Notably, in the MI regions, the total density $|\mathbf{q}^{[1]}|^{2}$ at
the center $(x,t)=\left(c_n,t_1(c_n)\right)$ of a breather unit
exhibits distinct properties for the defocusing and focusing signs of the
nonlinearity. A breather unit is one local oscillation cell of the modulated vector AB,
identified relative to the underlying exact-AB modulation length. Here, $c_n-x_1(c_n)=x_n$, where
\[
x_n=\frac{(2n+1)\pi-\arccos\delta}{\alpha},\qquad
\delta=\frac{2\chi_{1I}}{\sqrt{\alpha^2+4\chi_{1I}^2}},
\qquad n\in\mathbb Z.
\]
The corresponding total density is
\[
\left|\mathbf q^{[1]}(c_n,t_1(c_n))\right|^2
=|a|^2+2\sigma\chi_{1I}
\left(2\chi_{1I}+\sqrt{\alpha^2+4\chi_{1I}^2}\right).
\]
In the focusing case ($\sigma=+1$), these total-density centers are
local maxima, as shown in Fig.~\ref{figS1}(a2, b2). In the defocusing case
($\sigma=-1$), they are local minima, as shown in Fig.~\ref{figS1}(c2).
For an ES or AES component the same coordinates give a componentwise peak or
valley, whereas for an FP component they specify the symmetry center rather
than a simple extremum.
}
For fixed background parameters, the total-density value at each center
depends on $\alpha$ and $\chi_{1I}$. The reference modulation length is
$2\pi/\alpha$; it need not equal the spacing between successive modulated
centers $c_n$.
Together, these relations connect the morphology near the center of each AB unit to
the same spectral parameters that enter the gMI analysis.

For selecting parameters in the numerical tests below, we further resolve
the focusing spectrum. Let $G_{+}$ and $G_{-}$ be the larger and smaller
gains of the two unstable MI branches at wavenumber $k$, and define
$\rho=G_{-}/G_{+}$. Panel (a) of Fig.~\ref{figS2} separates the stable,
one-branch, unequal-growth two-branch, and equal-growth two-branch regions.
For input envelopes $f_1(x)$ and $f_2(x)$ of the same order in $\epsilon$,
unequal gains cause the two breather formation coordinates to separate
logarithmically as $\epsilon\to0$. When the gains are equal, the logarithmic
term cancels, allowing the two first-order predictions to be compared on
the same formation scale. The remaining separation along $t$ can be
adjusted by varying the ratio $|f_1(x)|/|f_2(x)|$.
Panel (b) of Fig.~\ref{figS2} shows where the $2k$ harmonic generated at
second order in the perturbation can grow through MI. This classification
provides a simple guide for selecting the parameters used in the numerical tests.

\FloatBarrier
\section{The classification of single AB solutions}

\begin{figure}[!b]
\centering
\includegraphics[width=0.65\textwidth]{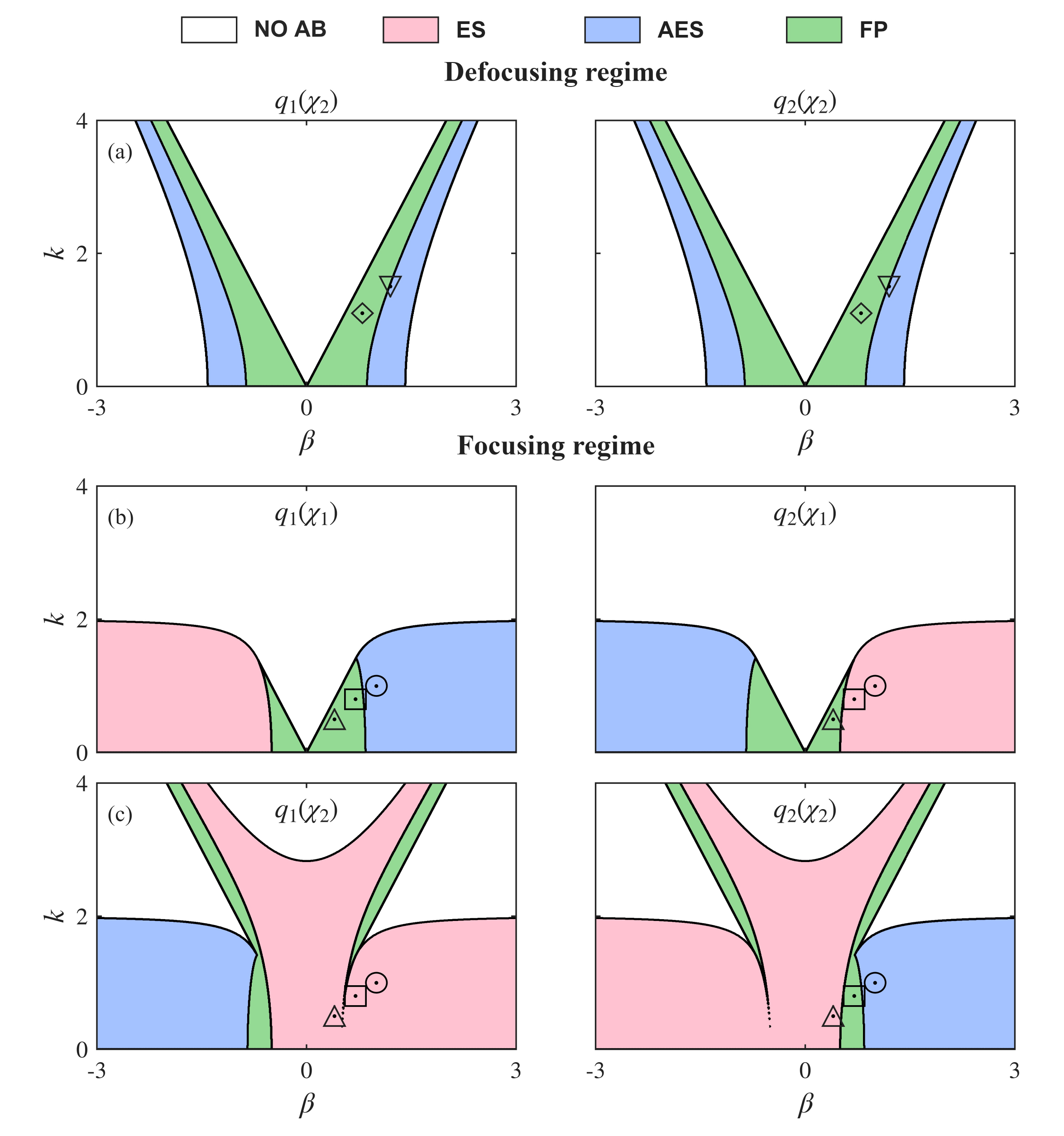}
\caption{The phase diagram of vector AB unit patterns in the plane of the
continuous-wave background wavenumber $\protect\beta $ and perturbation wavenumber
$k$. The root-family labels are defined in the text. The pink,
green, and blue regions represent AB solutions of the ES, FP, and AES types,
respectively. Panel (a) shows vector AB unit patterns in the defocusing
regime, using the root with positive imaginary part in the second
family. Panels (b) and (c) show the first and second exact-solution root
families, respectively, in the focusing regime.}
\label{figS3}
\end{figure}

\begin{figure}[!t]
\centering
\includegraphics[width=7cm,height=7cm]{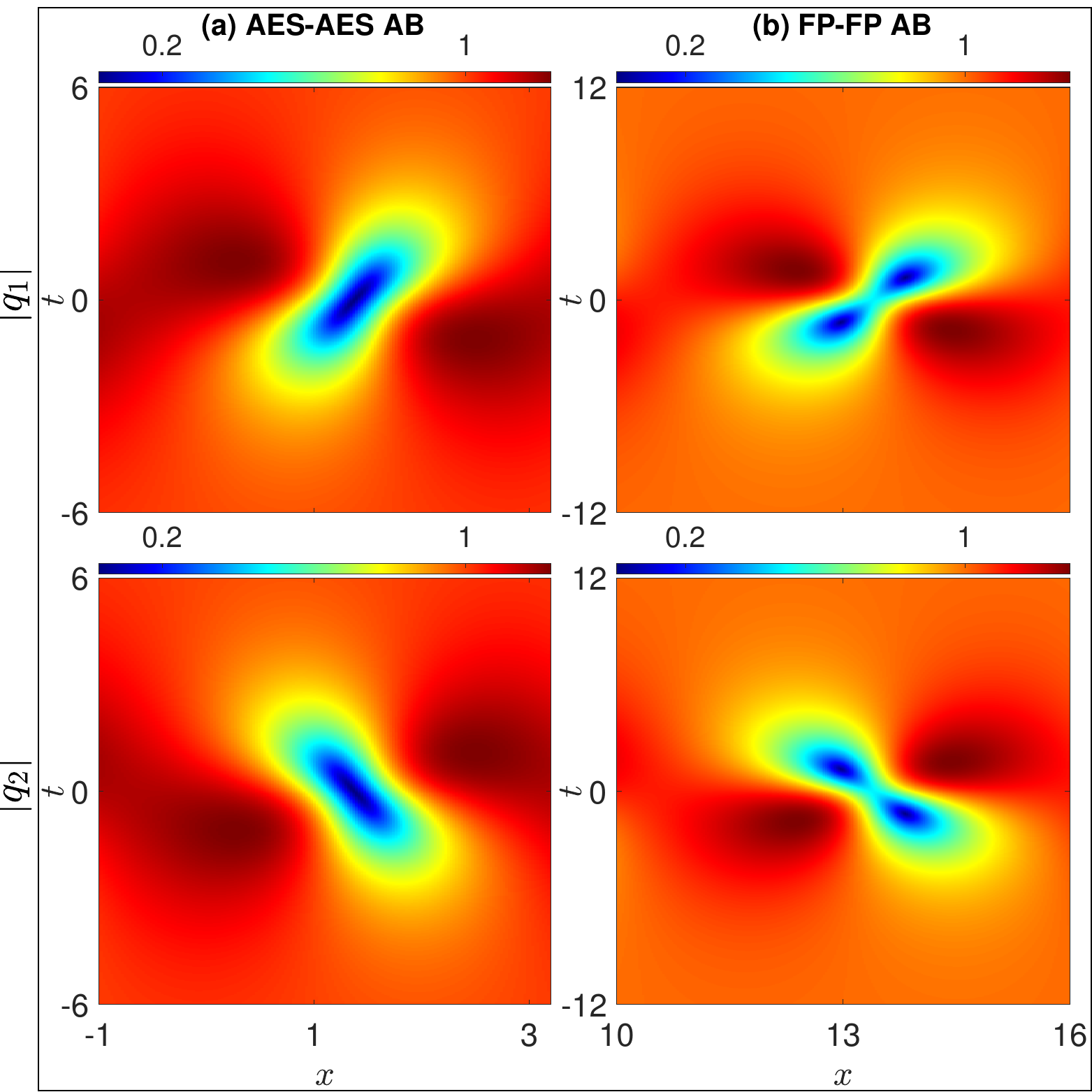}
\caption{Examples of two vector breather units in the defocusing regime with
(a) $k=1.5,\protect\beta =1.2$ and (b) $k=1.1,\protect\beta =0.8$. These
parameter sets correspond, respectively, to the inverted-triangle and rhombus
symbols in the phase diagram in Fig.~\protect\ref{figS3}(a).}
\label{figS4}
\end{figure}

\begin{figure}[!t]
	\centering
	\includegraphics[width=\textwidth]{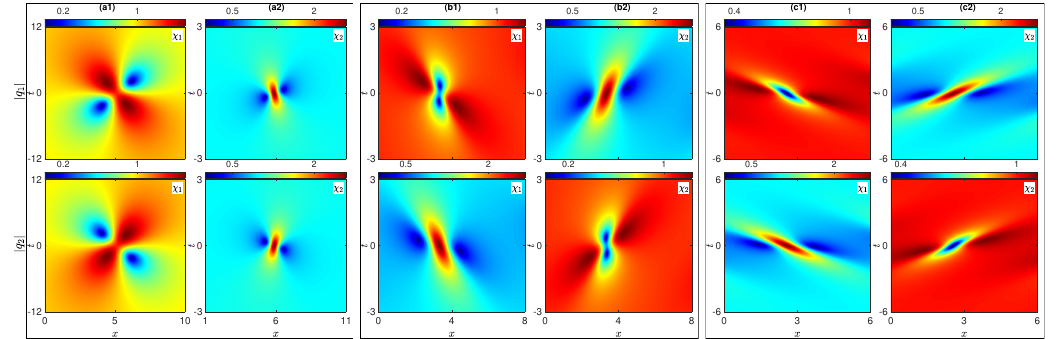}
\caption{Examples of three sets of vector breather units in the focusing
		regime. The parameter sets for panels (a1) and (a2); (b1) and (b2); and (c1) and (c2)
		are $k=0.5,\protect\beta =0.4$; $k=0.8,\protect\beta =0.7$; and $k=1,\protect%
		\beta =1$, respectively. They correspond to the triangle, square, and circle
		symbols depicted in Fig.~\protect\ref{figS3}(b) and (c).}
	\label{figS5}
\end{figure}

The two families of roots for the exact AB are given by
\[
\chi=-k/2\pm\sqrt{d+(-1)^{j-1}\nu},\qquad j=1,2.
\]
When family $j$ contains a root with positive imaginary part, we denote it by
$\chi_1^{(j)}$. In Figs.~\ref{figS3}--\ref{figS5}, the labels $\chi_1$ and
$\chi_2$ denote the first and second families, respectively. In the single-AB
formula, the same symbols denote two roots related by $\chi_2=\chi_1+\alpha$.
The root corresponding to the growing MI mode is
$\Omega=(\chi_1^{(j)})^*+k/2$, with $\alpha=k$.
For real $\nu$, $\Omega$ belongs to the root family $\Omega_{j,\pm}$.
For nonzero purely imaginary $\nu$, complex conjugation interchanges the two family labels.


Building on the AB solution given by Eq.~\eqref{eq:AB}, we classify the
fundamental vector AB unit patterns in both focusing and defocusing regimes.
Figure~\ref{figS3} summarizes the resulting phase diagram in the
$(\beta,k)$ plane, where $\beta$ is the continuous-wave background
wavenumber and $k$ is the perturbation wavenumber. In the focusing regime,
three distinct
breather unit morphologies are identified: ES, FP, and AES. In the
defocusing regime, only FP and AES patterns occur.

Examples of vector AB units in the defocusing regime are
shown in Fig.~\ref{figS4}. The symbols in Fig.~\ref{figS3}%
(a), namely inverted triangles and rhombuses, correspond to the parameters
associated with each type of vector AB in Figs.~\ref{figS4}%
(a) and (b). Figure~\ref{figS3}(a) indicates that, in the defocusing
regime, both components have the same classification in the $(\beta,k)$
plane, yielding only AES-AES and FP-FP patterns. For the focusing system,
panels (b) and (c) of Fig.~\ref{figS3} show the phase diagrams for the first
and second exact-solution root families, respectively. The focusing regime includes four types of two-component AB
units: ES-ES, ES-FP, ES-AES, and FP-FP. Consequently, the FP-FP breather
unit may exist in both focusing and defocusing regimes. Interestingly, for
the same $k$ and $\beta$, the two exact-solution root families can
produce AB types that differ beyond a simple exchange of components---a phenomenon impossible in the
defocusing regime. Figure~\ref{figS5} shows three sets of
vector ABs on three different backgrounds. Panels (a1) and (a2) show ABs of
the FP-FP and ES-ES types, from the first and second root families,
respectively, matching the triangle symbols in Figs.~\ref{figS3}(b) and
(c). Panels (b1) and (b2) show ABs of the FP-ES and ES-FP types,
respectively, matching the square symbols. Panels (c1) and (c2) show ABs of
the AES-ES and ES-AES types, from the first and second root families,
respectively, matching the circle symbols. Thus Eqs.~\eqref{eq:AB} and \eqref{eq:ab-deter} connect the spectral
parameters to the local two-component AB morphology.

\section{Numerical simulations of asymptotic AB solutions}

To test the asymptotic AB approximation, we perform numerical simulations
with four slowly varying envelopes. For a growing mode with root $\Omega$,
we treat the envelope $f(x)$ as locally constant and write the leading-order
gMI perturbation as
\begin{equation*}
\mathcal P_i[\Omega,f](x,t)=\frac{f(x)\mathrm{e}^{\mathrm{i}k(x+\Omega t)}}
{k+2b_i+2\Omega}
+\frac{f^*(x)\mathrm{e}^{-\mathrm{i}k(x+\Omega^*t)}}
{(k-2b_i-2\Omega)^*}.
\end{equation*}
The initial condition is
$q_i(x,0)=a_i\mathrm e^{\mathrm i b_i x}[1+\mathcal P_i[\Omega,f](x,0)]$.
Matching the fundamental harmonics to the pre-formation tail of the exact
single AB yields
\begin{equation*}
\alpha=k,\qquad \chi_1^*+\frac{k}{2}=\Omega.
\end{equation*}
Accordingly, the matched single-AB formulas below use $k$ as the spatial
modulation wavenumber. The trajectories follow from Eq.~(6) of the main text,
with spectral matching factor $\mathcal K=4|k|G/\sqrt{k^4+4G^2}$.
The asymptotic field used in the numerical
comparisons is the phase-normalized $q_{AB,i}^{[1]}$ defined in the main text,
evaluated with these translations.
We examine the following four envelope choices:

\begin{itemize}
\item Defocusing regime with $f(x)=\epsilon\operatorname{sech}(\mu x)$.
\item Focusing regime with $f(x)=\epsilon \mathrm{e}^{-(\mu x)^2}$.
\item Focusing regime with $f(x)=\epsilon \mathrm{e}^{\cos(\mu x)}$.
\item Focusing regime with $f(x)=\epsilon\operatorname{sech}(\mu x)$.
\end{itemize}
These choices have the common form
$f(x)=\epsilon f_b(\mu x)$. The bounded function
$f_b$ and its first two derivatives with respect
to its argument are $\mathcal O(1)$. Consequently, the perturbation amplitude
and its derivatives with respect to the physical coordinate $x$ have the
distinct orders
$f=\mathcal O(\epsilon)$,
$\partial_xf=\mathcal O(\epsilon\mu)$, and
$\partial_x^2f=\mathcal O(\epsilon\mu^2)$.

\begin{figure}[!b]
	\centering
	\includegraphics[width=\textwidth]{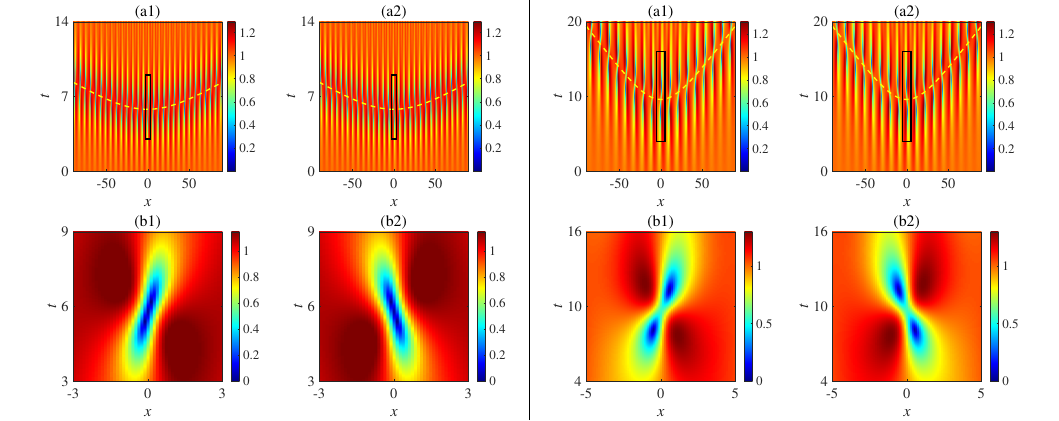}
	\caption{Numerical solutions of the Manakov system \eqref{eq:cnls} in the
		defocusing regime with input $q_{i}(x,0)=a_{i}\mathrm{e}^{\mathrm{i}%
			b_{i}x}[1+\mathcal P_i[\Omega,f](x,0)]$, where
		$f(x)=\protect\epsilon \operatorname{sech}(\mu x)$. The parameters are
		$\protect\epsilon=10^{-1}$ and
		$a_{1}=a_{2}=1$. The yellow dashed line represents $t_{1}(x)$. In the left panel, the
		modulation wavenumber is $k=1$, the background wavenumbers satisfy
		$b_{1}=-b_{2}=1$, and $\mu=1/50$,
		corresponding to the breather unit of the AES-AES type; in the right panel,
		the parameters are $k=0.6$, $b_{1}=-b_{2}=0.6,\mu=1/30$, corresponding to
		the breather unit of the FP-FP type.}
	\label{figS6}
\end{figure}

\begin{figure}[!t]
	\centering
	\includegraphics[width=\textwidth]{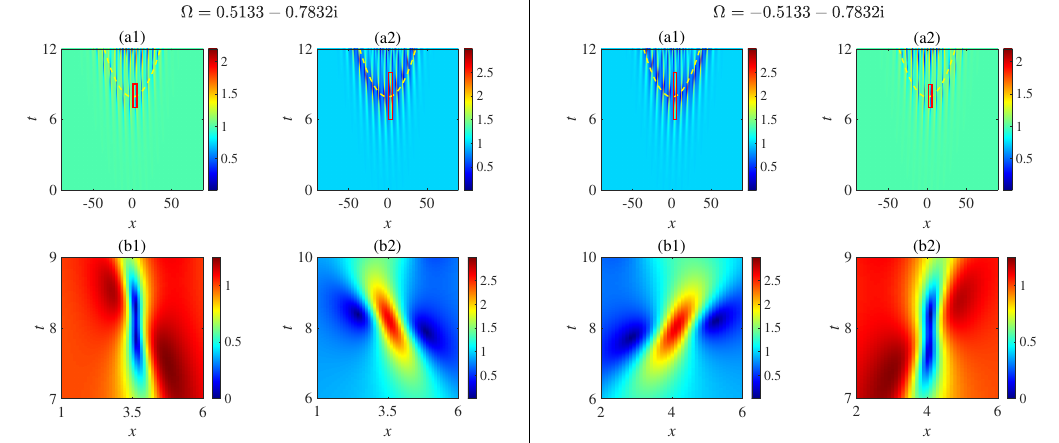}
	\caption{Numerical solutions of the Manakov system \eqref{eq:cnls} in the
		focusing regime with input $q_{i}(x,0)=a_{i}\mathrm{e}^{\mathrm{i}%
			b_{i}x}[1+\mathcal P_i[\Omega,f](x,0)]$, with the Gaussian envelope
		$f(x)=\protect\epsilon \mathrm{e}^{-(\mu x)^2}$. The parameters are $\protect%
		\epsilon =10^{-2},\mu=1/\sqrt{500},k=0.8,a_{1}=a_{2}=1,b_{1}=-b_{2}=0.7$. The
		yellow dashed curve represents $t_{1}(x)$. The left and right panels
		show the evolution of growing MI modes with $\operatorname{Re}\Omega>0$ and
		$\operatorname{Re}\Omega<0$, respectively. The resulting units are FP-ES
		on the left and ES-FP on the right.}
	\label{figS7}
\end{figure}

\begin{figure}[!t]
	\centering
	\includegraphics[width=\textwidth]{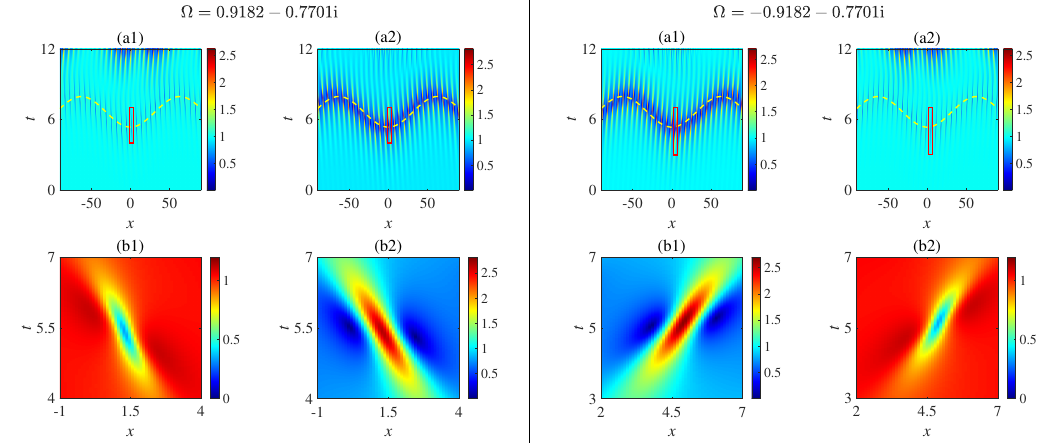}
	\caption{Numerical solutions of the Manakov system \eqref{eq:cnls} in the
		focusing regime with input $q_{i}(x,0)=a_{i}\mathrm{e}^{\mathrm{i}%
			b_{i}x}[1+\mathcal P_i[\Omega,f](x,0)]$, where
		$f(x)=\protect\epsilon \mathrm{e}^{\cos (\mu x)}$ is a bounded periodic envelope.
		The parameters are $\protect\epsilon %
		=10^{-2},\mu=1/20,k=1,a_{1}=a_{2}=1,b_{1}=-b_{2}=1$. The yellow dashed line
		represents $t_{1}(x)$. The left and right panels show the evolution of
		growing MI modes with $\operatorname{Re}\Omega>0$ and $\operatorname{Re}\Omega<0$,
		respectively. The resulting units are AES-ES on the left and ES-AES on the right.}
	\label{figS8}
\end{figure}

\begin{figure}[!t]
	\centering
	\includegraphics[width=\textwidth]{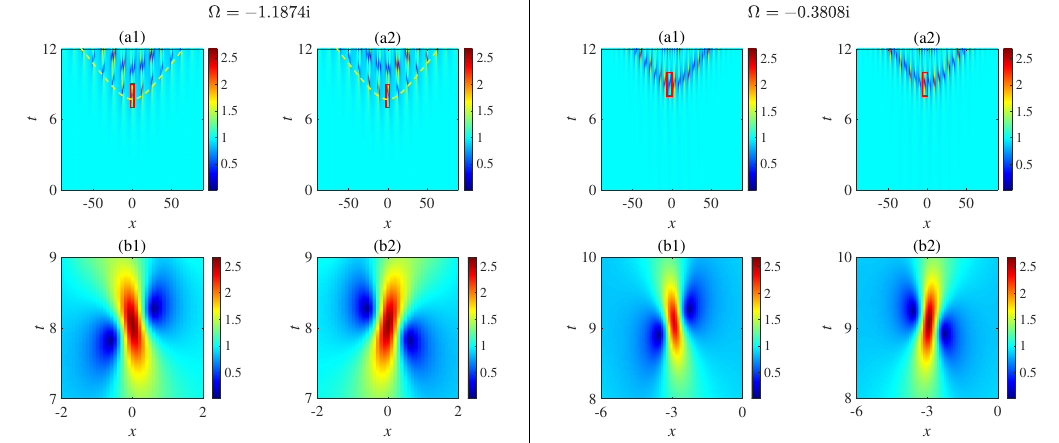}
	\caption{Numerical solutions of the Manakov system \eqref{eq:cnls} in the
		focusing regime, with input $q_{i}(x,0)=a_{i}\mathrm{e}^{\mathrm{i}%
			b_{i}x}[1+\mathcal P_i[\Omega,f](x,0)]$, where the sech envelope is
		$f(x)=\protect\epsilon\operatorname{sech}(\mu x)$. The parameters are $\protect\epsilon %
		=10^{-2},\mu=1/20,k=0.5,a_{1}=a_{2}=1,b_{1}=-b_{2}=0.4$.
		The growing MI modes used in the left and right simulations correspond to the roots
		$\Omega\simeq-1.1874\mathrm i$ and $\Omega\simeq-0.3808\mathrm i$,
		respectively. These are separate single-mode initial conditions.
		The yellow dashed curves in the left panels show the matching prediction
		\eqref{eq:single-sech-formation}, for which $t_1(0)\simeq7.72$.
		For the right input mode, the same formula gives
		$t_1(x)\ge t_1(0)\simeq23.24$, outside the displayed interval
		$0\le t\le12$, so no predicted trajectory is shown.
		The displayed later-time structures should not be interpreted as a
		quantitative validation of both branch-specific AB formation coordinates.}
	\label{figS9}
\end{figure}

Figure~\ref{figS6} displays the dynamics governed by the Manakov
system with the defocusing nonlinearity. For the parameter sets shown,
a sech envelope excites vector breathers whose center trajectories
are close to the asymptotic predictions. For this envelope, in either the
focusing or defocusing regime, the matching formula gives
\begin{equation}
t_1(x)=\frac{1}{G}\left[\ln\!\left(\frac{\mathcal K}{\epsilon}\right)
+\ln\cosh\!\left(\mu x\right)\right].
\label{eq:single-sech-formation}
\end{equation}%
The dashed curves in Fig.~\ref{figS6} follow this trajectory, which is
approximately parabolic near $x=0$.

Figure~\ref{figS7} presents a typical density plot of the evolution
governed by the focusing Manakov system, starting from the initial condition $%
q_{i}(x,0)=a_{i}\mathrm{e}^{\mathrm{i}b_{i}x}
[1+\mathcal P_i[\Omega,f](x,0)]$, with the Gaussian envelope
$f(x)=\epsilon \mathrm{e}^{-(\mu x)^2}$. The corresponding
trajectory is
\begin{equation}
t_1(x)=\frac{1}{G}\left[\ln\!\left(\frac{\mathcal K}{\epsilon}\right)
+\left(\mu x\right)^2\right].
\end{equation}%
The breather units form along a parabolic center trajectory, with additional
modulation developing at later times. The two components have different
local patterns. At the tested parameters, the yellow dashed trajectories
follow the observed breather-unit centers; an FP center is a symmetry
center rather than a simple local extremum. The two coexisting MI branches
are tested separately by selecting one growing mode in each simulation.

Figure~\ref{figS8} shows another typical example of the evolution,
starting from the initial condition $q_{i}(x,0)=a_{i}\mathrm{e}^{\mathrm{i}b_{i}x}
[1+\mathcal P_i[\Omega,f](x,0)]$ with a bounded periodic envelope
$f(x)=\epsilon \mathrm{e}^{\cos (\mu x)}$. The
corresponding trajectory is
\begin{equation}
t_1(x)=\frac{1}{G}\left[\ln\!\left(\frac{\mathcal K}{\epsilon}\right)
-\cos\!\left(\mu x\right)\right].
\end{equation}

Figure~\ref{figS9} shows the evolution starting from the initial condition $%
q_{i}(x,0)=a_{i}\mathrm{e}^{\mathrm{i}b_{i}x}
[1+\mathcal P_i[\Omega,f](x,0)]$ with the sech envelope
$f(x)=\epsilon \operatorname{sech}(\mu x)$, in the case of the
focusing nonlinearity. For each selected input mode, its gain $G$ and
matching factor $\mathcal K$ determine the formation trajectory through
the same sech-envelope formula~\eqref{eq:single-sech-formation}.

The periodic envelope produces a periodic center trajectory, whereas the
sech envelope produces a smoothly curved one. These examples illustrate
how the envelope enters the predicted formation coordinate. The limitations
of the branch-specific interpretation of Fig.~\ref{figS9} are
noted in its caption.

The tail-matching construction retains $\epsilon$ and $\mu$ as separate
small parameters.
To examine the numerical accuracy as both parameters decrease, we choose
three fixed-ratio paths $c=\mu/\epsilon=1/4$, $1$, and $4$, setting
$\epsilon=\mu/c$ while $\mu\to0$.
These paths specify the numerical tests; they do not establish a general
condition for the accuracy of the predicted formation trajectories.
At fixed evolution coordinate the joint limit returns the plane-wave background;
the nontrivial comparison must therefore follow the breather formation
coordinate. Let $C>0$ collect the local envelope and matching factors,
with $C$ bounded above and below independently of $\epsilon$.
Balancing the growing linear perturbation
$\epsilon\mathrm{e}^{Gt}$ with an $\mathcal O(1)$ breather amplitude gives
the matched formation-coordinate scaling $t_1(x)\sim G^{-1}\ln(C/\epsilon)$.
We consequently compare the fields in the moving window
$t=t_1(x)+\tau$, where $\tau=\mathcal O(1)$ is the
evolution-coordinate offset from the local formation coordinate.

Along each fixed-$c$ path, the computations in
Table~\ref{tab:single-ab-joint-convergence} use $x\in[-L/2,L/2)$ with $L=2\pi/\mu$ and periodic
boundary conditions for the fields after removing their background carriers.
Writing $u=\mu x$ and $f(x)=\epsilon f_b(u)$ as above, we use
$f_b(u)=\exp(\cos u)$ for focusing and the smooth periodicized sech profile
$f_b(u)=\operatorname{sech}[2\sin(u/2)]$ for defocusing
(equivalently, $f(x)=\epsilon\operatorname{sech}[2\sin(\mu x/2)]$). The perturbation carrier has $k=1$ and is periodic on
this domain for all listed values of $\mu$.
The spatial grids have $N_x=4096$ for $\mu^{-1}=30,45$ and $N_x=8192$
for $\mu^{-1}=70,100$, giving $\Delta x=L/N_x\leq0.077$.
The dealiased split-step Fourier evolution uses an integration step
$\Delta t_{\rm int}=0.00625$ and saves fields at intervals
$\Delta t_{\rm save}=0.025$. The integration starts at $t=0$ and ends
at the first saved time at or beyond $t_{\rm c}+2.05/G$, where
$t_{\rm c}:=t_1(0)$ is the matched formation coordinate at $x=0$.
Let $L_{\mathrm{AB}}=2\pi/k$ denote the exact-AB reference modulation
length. For both focusing and defocusing solutions, define the two-component modulus field
$\mathbf A=(|q_1|,|q_2|)$ and the background
$\mathbf A_0=(a_1,a_2)$ for the positive amplitudes used here.
The subscripts $\mathrm{num}$ and $\mathrm{asy}$ denote numerical and
asymptotic fields, respectively. For center alignment, define the translation operator by
$(\mathcal T_{\xi,\tau}\mathbf A)(x,t)=\mathbf A(x-\xi,t-\tau)$.
On the common natural-scale integration window
$W=\{|x|\leq2L_{\rm AB},\ |t-t_{\rm c}|\leq1.35/G\}$, the normalized modulus-field
discrepancy and the discrepancy after a fitted translation
$(\delta x,\delta t)$ are
\begin{equation*}
\begin{aligned}
E_{\rm raw}&=
\frac{\|\mathbf A_{\rm num}-\mathbf A_{\rm asy}\|_{L^2(W)}}
 {\|\mathbf A_{\rm asy}-\mathbf A_0\|_{L^2(W)}}, \qquad
E_{\rm shape}=
\frac{\|\mathbf A_{\rm num}-\mathcal T_{\delta x,\delta t}
\mathbf A_{\rm asy}\|_{L^2(W)}}
{\|\mathcal T_{\delta x,\delta t}\mathbf A_{\rm asy}-
\mathbf A_0\|_{L^2(W)}}.
\end{aligned}
\end{equation*}
Let $\mathcal S$ be the sampled set of trial translations described below.
The fitted translation minimizes the same normalized discrepancy
used in $E_{\rm shape}$:
\begin{equation*}
\mathcal E(\xi,\tau)=
\frac{\|\mathbf A_{\rm num}-\mathcal T_{\xi,\tau}\mathbf A_{\rm asy}\|_{L^2(W)}}
{\|\mathcal T_{\xi,\tau}\mathbf A_{\rm asy}-\mathbf A_0\|_{L^2(W)}},
\qquad
(\delta x,\delta t)=\operatorname*{arg\,min}_{(\xi,\tau)\in\mathcal S}
\mathcal E(\xi,\tau).
\end{equation*}
The sampled translation set $\mathcal S$ is constructed by first
searching a $13\times13$ grid over $|\xi|\leq0.28L_{\rm AB}$ and
$|\tau|\leq0.55/G$, then an $11\times11$ grid spanning one coarse-grid
step on either side of the best pair in each coordinate. Translated
templates are evaluated by bilinear interpolation in $x$ and $t$.
Thus, the fitted translation $(\delta x,\delta t)$ minimizes
$\mathcal E(\xi,\tau)$ over the sampled set $\mathcal S$;
the normalization denominator is evaluated separately
for each trial translation. Positive
$\delta x$ and $\delta t$ translate the template toward increasing $x$
and later $t$,
respectively.  These fitted translations are numerical
alignment diagnostics and are distinct from the analytical interaction-induced
shifts derived in the two-AB analysis below.  Thus $E_{\rm raw}$
measures the modulus-field discrepancy before alignment, while
$E_{\rm shape}$ measures the remaining discrepancy after the fitted
translation. The norms are evaluated with equal-weight sums over
both components and the same uniform saved space--time grid in $W$;
the common quadrature factors cancel in each ratio. Table~\ref{tab:single-ab-joint-convergence}
therefore reports dimensionless modulation-normalized errors.
Both $E_{\rm raw}$ and $E_{\rm shape}$ apply to components with either
peaks or valleys.  Subtracting $\mathbf A_0$ normalizes the error
by the coherent modulation about the plane-wave
background.

\begin{table}[!tb]
	
	\centering
	\small
	\begin{tabular}{ccccccc}
		\toprule
		$c$ & $\mu^{-1}$ & $\epsilon$ & $E_{\rm raw}^{\rm(def)}$ &
		$E_{\rm shape}^{\rm(def)}$ & $E_{\rm raw}^{\rm(foc)}$ &
		$E_{\rm shape}^{\rm(foc)}$\\
		\midrule
		$1/4$ & $30$  & $0.1333$ & $0.3196$ & $0.3123$ & $0.3103$ & $0.2932$\\
		      & $45$  & $0.0889$ & $0.2265$ & $0.2212$ & $0.1527$ & $0.1376$\\
		      & $70$  & $0.0571$ & $0.1545$ & $0.1510$ & $0.0701$ & $0.0648$\\
		      & $100$ & $0.0400$ & $0.1145$ & $0.1115$ & $0.0371$ & $0.0371$\\
		\midrule
		$1$   & $30$  & $0.0333$ & $0.0960$ & $0.0935$ & $0.0416$ & $0.0416$\\
		      & $45$  & $0.0222$ & $0.0669$ & $0.0658$ & $0.0211$ & $0.0211$\\
		      & $70$  & $0.0143$ & $0.0448$ & $0.0448$ & $0.0099$ & $0.0099$\\
		      & $100$ & $0.0100$ & $0.0327$ & $0.0327$ & $0.0054$ & $0.0054$\\
		\midrule
		$4$   & $30$  & $0.0083$ & $0.0510$ & $0.0510$ & $0.0557$ & $0.0557$\\
		      & $45$  & $0.0056$ & $0.0289$ & $0.0289$ & $0.0293$ & $0.0293$\\
		      & $70$  & $0.0036$ & $0.0162$ & $0.0162$ & $0.0144$ & $0.0144$\\
		      & $100$ & $0.0025$ & $0.0106$ & $0.0106$ & $0.0078$ & $0.0078$\\
		\bottomrule
	\end{tabular}
	\caption{Fixed-ratio joint-limit normalized modulus-field errors (superscripts
            $\mathrm{def}$ and $\mathrm{foc}$ denote the defocusing and focusing cases). Each four-row block follows
			$\epsilon=\mu/c$ at fixed $c$ and uses the same
			two-component modulus norm. Both cases use $a_1=a_2=1$ and $k=1$,
			with $b_1=-b_2=1.1$ for defocusing and $b_1=-b_2=1$ for focusing.
			The focusing entries use the growing mode on one MI branch. For the real even
			envelope and symmetric observation window used here, the modulus fields on the
			two branches are related by component exchange and $x\mapsto-x$, so their
			two-component modulus-norm diagnostics agree.}
	\label{tab:single-ab-joint-convergence}
\end{table}

Table~\ref{tab:single-ab-joint-convergence} shows a monotone decrease
of both field errors along every fixed-$c$ path.  For $c=1/4$, $1$, and
$4$, the defocusing $E_{\rm shape}$ decreases respectively from
$0.3123$ to $0.1115$, from $0.0935$ to $0.0327$, and from $0.0510$ to
$0.0106$ as $\mu$ decreases from $1/30$ to $1/100$.  The corresponding
focusing reductions are $0.2932$ to $0.0371$, $0.0416$ to $0.0054$, and
$0.0557$ to $0.0078$.  Log--log least-squares fits of
$E_{\rm shape}$ against $\mu$ give finite-range effective slopes
$0.86$, $0.87$, and $1.30$ for defocusing and $1.72$, $1.69$, and
$1.62$ for focusing.  These slopes summarize the tested range and are
not asserted as universal convergence exponents.  Values at different
$c$ but fixed $\mu$ should likewise not be read as a monotone dependence
on $c$, because changing $c$ also changes $\epsilon$ and the formation
coordinate.

A separate nested-grid calculation assesses discretization sensitivity
at two representative parameter sets. Let
$h=(\Delta x,\Delta t_{\rm int})$ denote a spatial/evolution-coordinate discretization and
$h/2=(\Delta x/2,\Delta t_{\rm int}/2)$ its refinement.  After interpolating
the fine-grid field onto the coarse grid, we use the diagnostic
$E_{\rm grid}^{(h)}=\|\mathbf A_{\rm num}^{(h)}-\mathbf A_{\rm num}^{(h/2)}\|_{L^2(W)}/
\|\mathbf A_{\rm num}^{(h/2)}-\mathbf A_0\|_{L^2(W)}$.
We evaluate it at two representative benchmark parameter sets: the
defocusing case uses $\epsilon=10^{-1}$, $\mu=1/50$, $\beta=1.1$, and $k=1$,
whereas the focusing case uses $\epsilon=10^{-2}$, $\mu=1/30$, $\beta=1$,
and $k=1$. Both use $a_1=a_2=1$ and the domain and envelope conventions
specified above. The three levels use $N_x=1024,2048,4096$ with
$\Delta t_{\rm int}=0.0125,0.00625,0.003125$, respectively;
$\Delta x=2\pi/(\mu N_x)$ and $\Delta t_{\rm save}=0.025$.
The successive-grid errors between the two finest comparable
levels are $3.047\times10^{-5}$ for defocusing and
$3.468\times10^{-4}$ for focusing.  The largest corresponding change in
$E_{\rm raw}$ or $E_{\rm shape}$ is $2.4\times10^{-3}$ between these
levels.  This grid
check concerns solver resolution at these two parameter sets; it is not
a separate refinement check for every entry of
Table~\ref{tab:single-ab-joint-convergence}. The fixed-$c$ rows of that table
measure the modulus-field discrepancies along the tested joint-limit paths.
Decreasing $\epsilon$ at fixed $\mu$ explores a different parameter path,
for which the accuracy requires a separate numerical assessment.
In either case, the terminal evolution coordinate must follow the predicted
formation coordinate $t_{\mathrm c}$. For the domain convention used here,
$L$ grows as $\mu$ decreases, and the grid must be refined as needed to keep the physical
$\Delta x$ and $\Delta t_{\rm int}$ controlled.

Having observed decreasing field-norm discrepancies along the tested
joint-limit paths, we next use local extrema only to quantify how accurately the asymptotic solution
reproduces the observable peak or valley in each component.  Let
$(x_n^{\mathrm{asy}},t_n^{\mathrm{asy}})=(c_n,t_1(c_n))$ be the predicted center of the
$n$th breather unit and
$(x_{i,n}^{\mathrm{num}},t_{i,n}^{\mathrm{num}})$ the peak or valley
extracted from component $q_i$.  With
$A_{i,n}^{\mathrm{num}}=
|q_i^{\mathrm{num}}(x_{i,n}^{\mathrm{num}},t_{i,n}^{\mathrm{num}})|$
and
$A_{i,n}^{\mathrm{asy}}=
|q_i^{\mathrm{asy}}(x_n^{\mathrm{asy}},t_n^{\mathrm{asy}})|$, define
\begin{equation*}
\begin{gathered}
\mathrm{AE}_{A,i,n}=
|A_{i,n}^{\mathrm{num}}-A_{i,n}^{\mathrm{asy}}|,
\quad
\mathrm{AE}_{x,i,n}=|x_{i,n}^{\mathrm{num}}-x_n^{\mathrm{asy}}|,
\quad
\mathrm{AE}_{t,i,n}=|t_{i,n}^{\mathrm{num}}-t_n^{\mathrm{asy}}|.
\end{gathered}
\end{equation*}
These absolute errors are local feature diagnostics, not a
second convergence norm.

\begin{figure}[!t]
	\centering
	\includegraphics[width=0.96\textwidth]{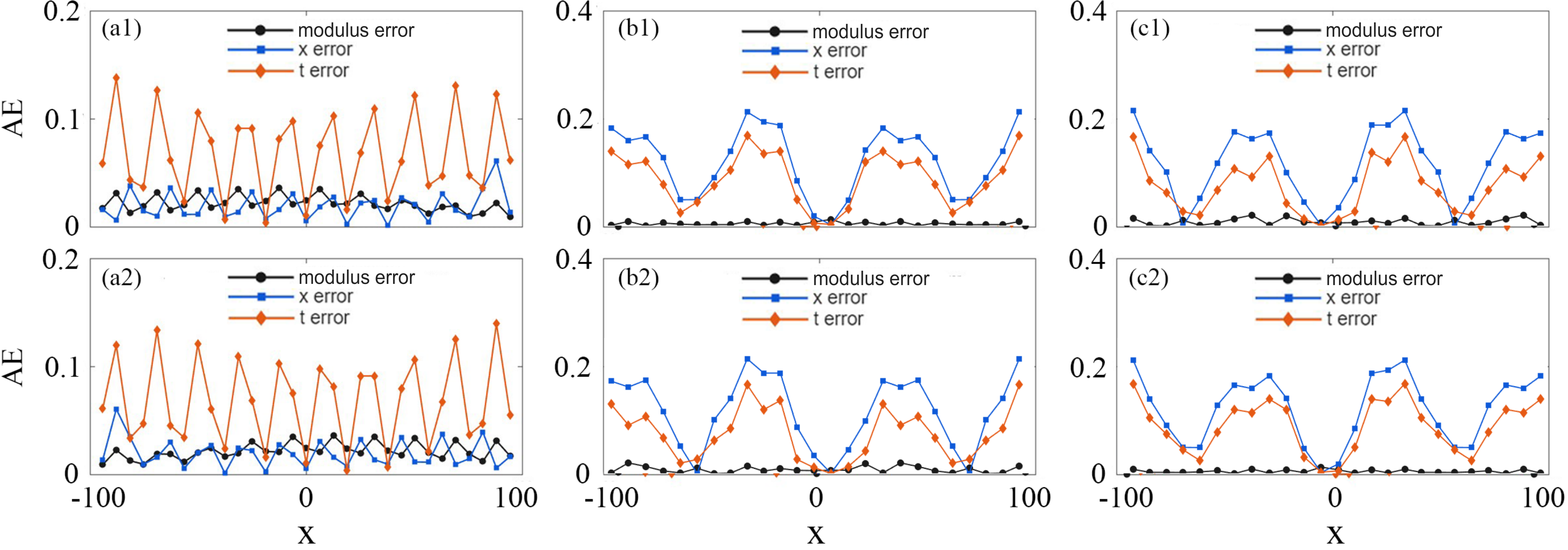}
\caption{Component-resolved local errors for three single-AB benchmark simulations.
            Column (a) corresponds to the unstable MI branch in the defocusing case,
            and columns (b) and (c) to the two unstable MI branches in the focusing case.
            Rows 1 and 2 correspond to $q_1$ and $q_2$. All cases use $a_1=a_2=1$.
            The defocusing case uses $f(x)=0.1\operatorname{sech}(x/50)$
            and $k=1$, $\beta=1.1$; the focusing cases use
            $f(x)=0.02\exp[0.4\cos(x/20)]$, $k=0.8$, and $\beta=1$.
            The focusing data are from the same numerical evolution as main-text Fig.~1.
            The errors are evaluated over $|x|\le100$, whereas Fig.~1
            displays $|x|\le80$ for these cases.
            Black circles show $\mathrm{AE}_{A,i,n}$, blue squares show
			$\mathrm{AE}_{x,i,n}$, and orange diamonds show
			$\mathrm{AE}_{t,i,n}$. The horizontal coordinate is the predicted
			center position $x_n^{\mathrm{asy}}$.}
	\label{figS10}
\end{figure}

Figure~\ref{figS10} shows that, in the defocusing
example, the modulus and spatial errors are
typically only a few hundredths, while the evolution-coordinate error remains below
approximately $0.15$.  For the focusing branches, the spatial center
error is generally the largest contribution and reaches about $0.22$ in
the most sensitive component; the evolution-coordinate and modulus errors remain
below approximately $0.17$ and $0.021$, respectively.  The increase toward the edges of the observation interval is associated
with the larger trajectory slope and the
component-dependent peak/valley profile.  These data complement the
field-norm test by locating where the remaining discrepancy is
concentrated; they are not used to infer a convergence rate.

\FloatBarrier
\section{Persistence under a small nonintegrable perturbation}

\begin{figure}[!b]
	\centering
	\includegraphics[width=\textwidth]{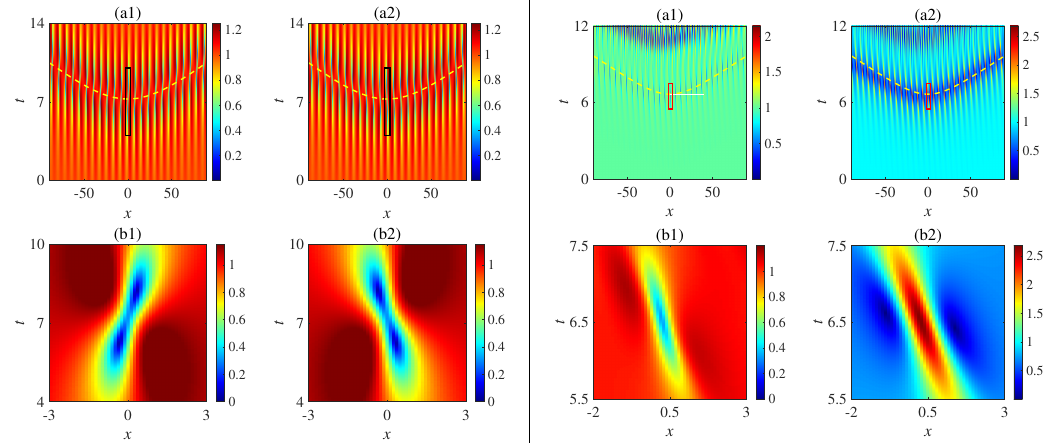}
\caption{Numerical solutions of the Manakov system with a small nonintegrable
		perturbation (Eq.~\eqref{eq:cnls1}), with $\epsilon_1=0.02$. The input is $q_{i}(x,0)=a_{i}\mathrm{e}^{%
			\mathrm{i}b_{i}x}[1+\mathcal P_i[\Omega,f](x,0)]$ with
		$f(x)=\protect\epsilon \operatorname{sech}(\mu x)$. For the left panel, the parameters are $%
		\protect\epsilon =10^{-1},k=0.8,a_{1}=a_{2}=1,b_{1}=-b_{2}=0.8,\mu=1/50$,
		corresponding to the AB unit of the AES-AES type in the defocusing regime.
		In the right panel, the parameters are $\protect\epsilon %
		=10^{-2},k=1,a_{1}=a_{2}=1,b_{1}=-b_{2}=1,\mu=1/30$, corresponding to the
		AB unit of the AES-ES type in the focusing regime.}
	\label{figS11}
\end{figure}

We further consider the weakly nonintegrable system described in the main text.
Its coefficient $\epsilon_1$ controls the mismatch between self- and
cross-phase modulation. This perturbation preserves the discrete
component-exchange symmetry but breaks the continuous Manakov
component-rotation symmetry. In component form, the resulting system is
\begin{equation}\label{eq:cnls1}
\left\{ \begin{aligned} &\mathrm{i} q_{1,t}+
\frac{1}{2}q_{1,xx}+\sigma\left(\left|q_1\right|^2+\left|q_2\right|^2+%
\epsilon_1\left(\left|q_1\right|^2-\left|q_2\right|^2\right)\right) q_1=0,
\\ & \mathrm{i} q_{2,t}+
\frac{1}{2}q_{2,xx}+\sigma\left(\left|q_1\right|^2+\left|q_2\right|^2-%
\epsilon_1\left(\left|q_1\right|^2-\left|q_2\right|^2\right)\right) q_2=0.\\
\end{aligned}\right.
\end{equation}%
We performed numerical simulations to investigate the qualitative persistence
of the asymptotic AB solutions under the action of this nonintegrable
perturbation. The initial condition is chosen as $q_{i}(x,0)=a_{i}\mathrm{e}^{\mathrm{%
i}b_{i}x}[1+\mathcal P_i[\Omega,f](x,0)]$. We consider sech, Gaussian, and periodic exponential envelopes to examine the persistence of the breather
dynamics in the weakly nonintegrable system.

\begin{figure}[!htbp]
	\centering
	\includegraphics[width=\textwidth]{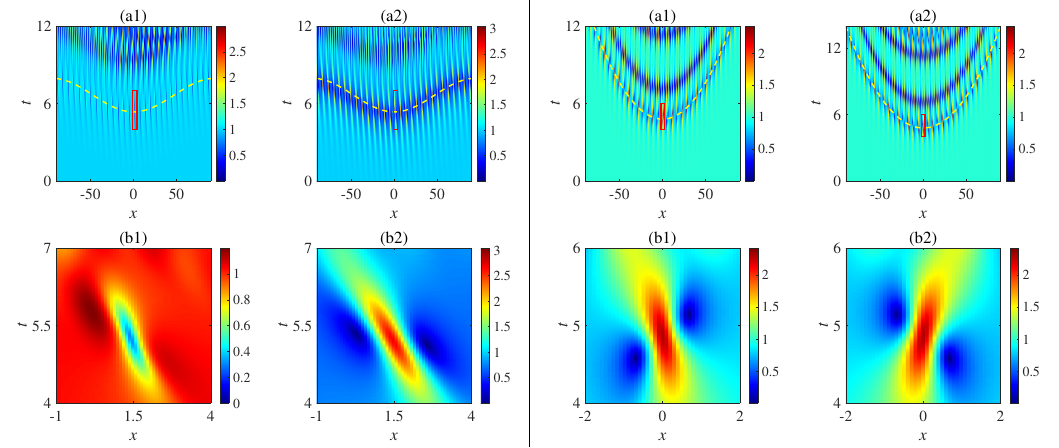}
\caption{Numerical solutions of the perturbed system \eqref{eq:cnls1} in
		the focusing regime, with $\epsilon_1=0.02$. In the left panel, $f(x)=\protect%
		\epsilon \mathrm{e}^{\cos (\mu x)}$ with parameters $\protect\epsilon %
		=10^{-2},k=1,a_{1}=a_{2}=1,b_{1}=-b_{2}=1,\mu=1/30$, corresponding to the AB
		unit of the AES-ES type. In the right panel, $f(x)=\protect\epsilon %
		\mathrm{e}^{-(\mu x)^2}$ with parameters $\protect\epsilon %
		=10^{-2},k=1,a_{1}=a_{2}=1,b_{1}=-b_{2}=0.5,\mu=1/(20\sqrt{2})$, corresponding to the
		AB unit of the ES-ES type.}
	\label{figS12}
\end{figure}

\begin{figure}[!htbp]
	\centering
	\includegraphics[width=\textwidth]{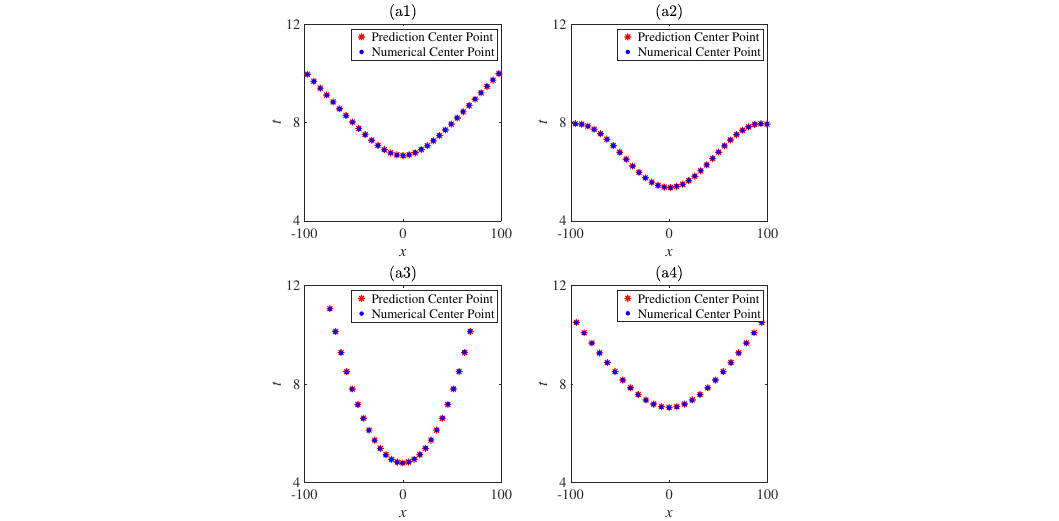}
\caption{Comparison of the predicted and numerically extracted centers
        for the perturbed system \eqref{eq:cnls1}.
		(a1): The focusing regime, with the envelope $f(x)=\protect\epsilon
		\operatorname{sech}(\mu x)$ and parameters $\protect\epsilon =10^{-2},\protect\epsilon %
		_{1}=0.02,k=1,\protect\beta =1,\mu=1/30$; (a2): The focusing regime, with
		the envelope $f(x)=\protect\epsilon \mathrm{e}^{\cos (\mu x)}$ and parameters $%
		\protect\epsilon =10^{-2},k=1,\protect\epsilon _{1}=0.02,\protect\beta %
		=1,\mu=1/30$; (a3): The focusing regime, with the envelope $f(x)=%
		\protect\epsilon \mathrm{e}^{-(\mu x)^2}$ and parameters $\protect\epsilon =10^{-2},%
		\protect\epsilon _{1}=0.02,k=1,\protect\beta =0.5,\mu=1/(20\sqrt{2})$; (a4): The
		defocusing regime, with the envelope $f(x)=\protect\epsilon \operatorname{%
			sech}(\mu x)$ and parameters $\protect\epsilon =10^{-1},\protect\epsilon %
		_{1}=0.02,k=0.8,\protect\beta =0.8,\mu=1/50$.}
	\label{figS13}
\end{figure}

We first use the envelope $f(x)=\epsilon\operatorname{sech}(\mu x)$
in the initial condition specified above, in both the defocusing and
focusing regimes. The perturbation coefficient is fixed at $\epsilon
_{1}=0.02$ in Eq.~\eqref{eq:cnls1}.

Figure~\ref{figS11} shows simulations of the nonintegrable model %
\eqref{eq:cnls1} with $\epsilon _{1}=0.02$ and the sech-envelope initial
condition specified above. The observed breather-unit
dynamics is consistent with the persistence of the corresponding asymptotic
AB structures under the weak nonintegrable perturbation, in both defocusing and focusing
settings. These results support structural persistence of the main dynamical
features beyond exact integrability; they do not establish quantitative
validity of the uncorrected integrable matching formula.

We next consider two further envelopes in the focusing regime:
$f(x)=\epsilon \mathrm{e}^{\cos
(\mu x)}$ and $f(x)=\epsilon \mathrm{e}^{-(\mu x)^2}$, with the perturbation
coefficient fixed at $\epsilon_1 =0.02$.

Figure~\ref{figS12} presents additional focusing-regime simulations
with the same perturbation strength, for distinct envelope choices. The
resulting breather-unit patterns retain the principal vector-breather
geometry for the tested parameter sets. These examples provide evidence of
qualitative persistence at $\epsilon _1=0.02$; they do not establish a
finite perturbation interval of structural stability or quantitative
accuracy of the uncorrected integrable matching formula.

Figure~\ref{figS13} compares the analytically predicted centers with
their counterparts extracted from the
direct numerical simulations across multiple envelopes for both the
focusing and defocusing regimes. The uncorrected integrable prediction
captures the overall center geometry, but systematic evolution-coordinate offsets remain
under the equation perturbation. Thus these simulations test structural
persistence rather than quantitative validity of the integrable matching
formula in the nonintegrable model: the visible breather geometry is more
robust here than its absolute formation coordinate.

\section{The asymptotic two-AB solution}\label{sec:two-ab}

For the integrable Manakov system, exact multi-breather solutions can be
constructed by iterated Darboux transformations. For the backgrounds considered
in Fig.~\ref{figS1}, at most one MI branch is unstable in the defocusing regime
at a given wavenumber. For the two-AB construction below, we therefore use the
focusing Manakov system and select growing modes from two distinct MI branches
at the same wavenumber.

The exact two-AB solution on the plane-wave background has the determinant
form below. Here $M$ and $M^{(i)}$ denote $2\times2$ matrices, distinct
from the scalar numerator $N_i$ and denominator $M$ in the single-AB formula:
\begin{equation}
q_{i}^{[2]}(x,t)=a_{i}\frac{\det (M^{(i)})}{\det (M)}\exp (\theta _{i}),
\label{eq:two-AB-exact}
\end{equation}%
where the elements of the matrices $M=(m_{rs})_{1\leq r,s\leq 2}$ and $%
M^{(i)}=(m_{rs}^{(i)})_{1\leq r,s\leq 2}$ are
\begin{align}
m_{r,s}& =\sum_{\ell=1}^{2}\sum_{m=1}^{2}\frac{\exp \left( \omega _{r,\ell}^{\ast
}+\omega _{s,m}\right) }{\chi _{s,m}-\chi _{r,\ell}^{\ast }}, \\
m_{r,s}^{(i)}& =\sum_{\ell=1}^{2}\sum_{m=1}^{2}\frac{\chi _{r,\ell}^{\ast }+b_{i}}{%
\chi _{s,m}+b_{i}}\frac{\exp \left( \omega _{r,\ell}^{\ast }+\omega
_{s,m}\right) }{\chi _{s,m}-\chi _{r,\ell}^{\ast }}.
\end{align}%
Here, $\chi _{j,m}$ ($m=1,2$) are two roots of the cubic equation~\eqref{eq:lax-cubic}
evaluated at the spectral parameter $\lambda _{j}$, and $\omega _{j,m}=%
\mathrm{i}\chi _{j,m}\Bigl(x+\tfrac{1}{2}\chi _{j,m}t+\vartheta _{j,m}\Bigr)$
are the corresponding exponential phases, $\vartheta _{j,m}$ being complex
parameters. For the two-AB construction, write
$\Omega^{(\ell)}=\Omega_R^{(\ell)}+\mathrm{i}\Omega_I^{(\ell)}$
for the four roots of the MI dispersion relation,
$\ell=1,\ldots,4$. For $j=1,2$,
we select the root $\Omega^{(2j-1)}$ for which
$-k\Omega_I^{(2j-1)}>0$ and denote its conjugate by $\Omega^{(2j)}$. This four-root notation distinguishes the two
conjugate pairs without using $j$ as a field-component index. The parameters $\chi_{j,m}$ in the exact
solution are labeled as
$(\chi_{1,1},\chi_{1,2},\chi_{2,1},\chi_{2,2})
=(\chi_1,\chi_2,\chi_3,\chi_4)$,
where $\operatorname{Im}\chi_1=\operatorname{Im}\chi_2$ and
$\operatorname{Im}\chi_3=\operatorname{Im}\chi_4$. The common-period
construction imposes the single real separation $
\alpha\equiv\chi_2-\chi_1=\chi_4-\chi_3
>0$,
which fixes the common spatial period $2\pi/|\alpha|$. The relations between
these roots and $\Omega^{(2j-1)}$ are determined by the matching calculation below.

For each constituent $j$, the phase constants $\vartheta_{j,1}$ and
$\vartheta_{j,2}$ affect the solution only through the relative exponential
phase: a common nonzero rescaling of both exponentials in the same Darboux
vector cancels from the determinant ratio. We express this relative phase
in terms of two real translation parameters $x_j$ and $t_j$ by setting
\[
\chi_{j,2}\vartheta_{j,2}-\chi_{j,1}\vartheta_{j,1}
=-\alpha\left[x_j+\left(\chi_{j,1}+\frac{\alpha}{2}\right)t_j\right].
\]
For the nonreal AB roots, the imaginary part of this relation determines
$t_j$, and the real part then fixes $x_j$. Since
$\chi_{j,2}=\chi_{j,1}+\alpha$, it follows that
\[
\omega_{j,2}-\omega_{j,1}
=\mathrm{i}\alpha\left[(x-x_j)
+\left(\chi_{j,1}+\frac{\alpha}{2}\right)(t-t_j)\right].
\]
Thus, after removing the irrelevant common factors, the phase data are
represented by the four real constants $(x_1,t_1,x_2,t_2)$, which translate
the two constituents in space and along the evolution coordinate.

To match the pre-formation background, we use the invariance of the Manakov
system under constant component phase rotations and define the
phase-normalized, slowly modulated two-AB ansatz by
\begin{equation}
q_{AB,i}^{[2]}(x,t)=\frac{\chi _{2}+b_{i}}{\chi _{2}^{\ast }+b_{i}}\frac{%
\chi _{4}+b_{i}}{\chi _{4}^{\ast }+b_{i}}%
q_{i}^{[2]}(x,t;x_{1}(x),t_{1}(x),x_{2}(x),t_{2}(x)).  \label{eq:two-AB-gene}
\end{equation}%
Here the four constant translations have been replaced by slowly varying
functions $(x_1(x),t_1(x),x_2(x),t_2(x))$. The resulting field is an
asymptotic ansatz and is not, in general, an exact solution of the Manakov system.
We next consider the joint pre-formation asymptotic limit of the two-AB solution
\eqref{eq:two-AB-gene}, in which
$t-t_{1}(x)\rightarrow -\infty$ and
$t-t_{2}(x)\rightarrow -\infty$.
{
Write $\chi _m=\chi _{mR}+\mathrm{i}\chi _{mI}$. In the AB spectral
variables, the two positive tail rates are
\[
G_1=\alpha\chi_{1I}>0,\qquad G_2=\alpha\chi_{3I}>0.
\]
Their identification with the MI gains follows from the matching below. Hence
$\mathrm{e}^{G_1t}$ and $\mathrm{e}^{G_2t}$ are exponentially small as
$t\to-\infty$. We organize the pre-formation expansion according to whether
these rates are equal or unequal. Define
$D_{m,l}\equiv \chi _{l}-\chi _{m}^{\ast }$,
$C_{0}\equiv 1/(D_{2,2}D_{4,4})-1/(D_{2,4}D_{4,2})$, and

\[
A_{j}\equiv \mathrm{i}\alpha
\left[x-x_{j}(x)+\left(\chi _{(2j-1)R}+\frac{\alpha}{2}\right)
(t-t_{j}(x))\right].
\]
With the phase convention above,
\[
\omega_{j,2}-\omega_{j,1}
=-G_j\bigl[t-t_j(x)\bigr]+A_j,
\]
which makes the translation dependence of the pre-formation exponentials
explicit.
The coefficients $C_i^{(r)}$, $r=1,\ldots,4$, are then given by
\begin{equation}
\begin{aligned} C_{i}^{(1)} &=
\frac{\alpha}{\chi_1+b_i}\,\frac{1}{C_0}\left( \frac{1}{D_{2,1}D_{4,4}} -
\frac{1}{D_{4,1}D_{2,4}} \right),\\ C_{i}^{(2)} &=
-\frac{\alpha}{\chi_2^*+b_i}\,\frac{1}{C_0}\left( \frac{1}{D_{1,2}D_{4,4}} -
\frac{1}{D_{1,4}D_{4,2}} \right),\\ C_{i}^{(3)} &=
\frac{\alpha}{\chi_3+b_i}\,\frac{1}{C_0}\left( \frac{1}{D_{4,3}D_{2,2}} -
\frac{1}{D_{4,2}D_{2,3}} \right),\\ C_{i}^{(4)} &=
-\frac{\alpha}{\chi_4^*+b_i}\,\frac{1}{C_0}\left( \frac{1}{D_{3,4}D_{2,2}} -
\frac{1}{D_{3,2}D_{2,4}} \right). \end{aligned}
\end{equation}%
Finally, define
$P_{i,1}\equiv C_i^{(1)}\mathrm{e}^{-A_1}+C_i^{(2)}\mathrm{e}^{A_1}$ and
$P_{i,2}\equiv C_i^{(3)}\mathrm{e}^{-A_2}+C_i^{(4)}\mathrm{e}^{A_2}$.
At each fixed slow position, hold the translations fixed and write
$z_j=\mathrm{e}^{G_j[t-t_j(x)]}$. For a nondegenerate state with $C_0\ne0$,
expanding the determinant ratio in these two tail amplitudes gives
\[
q_{AB,i}^{[2]}=a_i\mathrm{e}^{\theta_i}
\left[1+z_1P_{i,1}+z_2P_{i,2}
+\mathcal O(z_1^2+z_1z_2+z_2^2)\right].
\]
The coefficients of $z_1$ and $z_2$ will be used for matching. Their
ordering along the limit $t\to-\infty$ depends on the two rates, as follows.

\textbf{Equal-growth case, $G_{1}=G_{2}\equiv G_{\rm eq}$.}
The two fundamental modes contribute at the same asymptotic order, so we
match their coefficients simultaneously in the expansion
\begin{equation}
	q_{AB,i}^{[2]}(x,t)=a_i \mathrm{e}^{\theta_i}
	\left[1+\mathrm{e}^{G_{\rm eq}(t-t_1(x))}P_{i,1}
	+\mathrm{e}^{G_{\rm eq}(t-t_2(x))}P_{i,2}
	+\mathcal O(\mathrm{e}^{2G_{\rm eq}t})\right].
	\label{eq:two-AB-1}
\end{equation}
This is the natural two-mode first-order limit used in the equal-growth
numerical tests.

\textbf{Unequal-growth case, $G_{1}\neq G_{2}$.}
Let $G_{\mathrm{w}}=\min(G_1,G_2)$ and
$G_{\mathrm{s}}=\max(G_1,G_2)$. For the mode with gain $G_{\mathrm{w}}$, write
$P_{i,\mathrm{w}}$ for the coefficient $P_{i,j}$ and $t_{\mathrm{w}}(x)$ for the
translation $t_j(x)$. The strict
leading pre-formation asymptotic expansion is
\begin{equation}
	q_{AB,i}^{[2]}(x,t)=a_i \mathrm{e}^{\theta_i}
	\left[1+\mathrm{e}^{G_{\mathrm{w}}(t-t_{\mathrm{w}}(x))}P_{i,\mathrm{w}}
	+\mathcal O(\mathrm{e}^{G_*t})\right],\qquad
	G_*=\min\{G_{\mathrm{s}},2G_{\mathrm{w}}\}>G_{\mathrm{w}}.
	\label{eq:two-AB-2}
\end{equation}
As $t\to-\infty$, the smaller-gain exponential decays more slowly and
is the strict leading contribution. The larger-gain term is subleading;
it precedes the quadratic smaller-gain terms only when
$G_{\mathrm{s}}<2G_{\mathrm{w}}$. Equation~\eqref{eq:two-AB-2} states this
temporal ordering, while the two-amplitude expansion above retains the
coefficients of both fundamental modes. This tail ordering does not
determine which breather forms first in forward evolution: both gains and
initial modal projections affect the formation coordinates. Higher-order
content, including an unstable $2k$ mode, may also grow during this interval;
its significance cannot be inferred from the gain map alone. We therefore
use equal gains as the principal benchmark for comparing the two fundamental
matching predictions on a common formation-coordinate scale.}

Let $f_j(x)$ denote the slowly varying envelope of the growing mode
associated with constituent $j$. The local matching requires the spatial
variation scale of each envelope to be much larger than the carrier period
$2\pi/|\alpha|$; away from zeros of $f_j$, this is expressed as
$|\partial_x\ln f_j|/|\alpha|\ll1$. The matching below determines
$(x_j(x),t_j(x))$ from $f_j(x)$. As in the single-AB analysis,
the modal envelopes have amplitude $f_j=\mathcal O(\epsilon)$ and vary on
the scale $\mu^{-1}$, so that
$\partial_xf_j=\mathcal O(\epsilon\mu)$ and
$\partial_x^2f_j=\mathcal O(\epsilon\mu^2)$.
Applying the same local approximation to each envelope, we write the
leading-order two-mode gMI field as
\begin{equation}
q_{i}(x,t)=a_{i}\mathrm{e}^{\theta _{i}}\left(
1+\sum_{j=1}^{2}\mathcal P_i[\Omega^{(2j-1)},f_j](x,t)\right).
\label{eq:two-Fourier}
\end{equation}%
We match the coefficients of $z_1$ and $z_2$ in the two-amplitude
expansion above, before reducing it to the strict leading temporal order.
For equal gains, both terms occur at that order. For unequal gains, this
is a formal coefficient matching of the two fundamental modes; it does
not assert that both dominate the higher-order terms.
For each selected MI mode, comparing the exponential dependence and the coefficients
of $\mathrm{e}^{\pm\mathrm{i}kx}$ with the corresponding gMI terms yields
$\alpha =k$,
$\chi_{2j-1}^{\ast }+\frac{k}{2}=\Omega ^{(2j-1)}$ and
\begin{equation}
\begin{aligned}
\frac{f_j(x)}{k+2b_i+2\Omega^{(2j-1)}}=C_i^{(2j)}\,\mathrm{e}^{-\mathrm{i}k%
\bigl(x_j(x)+\Omega^{(2j-1)} t_j(x)\bigr)}. \end{aligned}
\end{equation}%
Define
$\Delta_1=1/(D_{1,2}D_{4,4})-1/(D_{1,4}D_{4,2})$,
$\Delta_2=1/(D_{3,4}D_{2,2})-1/(D_{3,2}D_{2,4})$, and
$\mathcal K_j=2|k\Delta_j|/|C_0|$. The spectral matching identifies the
tail rates with the selected MI gains, $G_j=-k\Omega_I^{(2j-1)}$.
For nonzero $f_j$, the translations are
\begin{equation}
\begin{aligned}
t_j(x)&=\frac{1}{G_j}\ln\!\left(\frac{\mathcal K_j}{|f_j(x)|}\right),\quad x_j(x)=-\Omega_R^{(2j-1)}t_j(x)
-\frac{1}{k}\arg\!\left(\frac{f_j(x)C_0}{-2k\Delta_j}\right).
\end{aligned}  \label{eq:two-dynamic}
\end{equation}%
The argument is chosen continuously on each nonzero-envelope interval.
After this matching, we use $k$ throughout the separated-constituent
formulas; in particular, each $A_j$ below denotes its definition above with
$\alpha$ replaced by $k$.

With the translations determined by Eq.~\eqref{eq:two-dynamic},
we now examine the two-AB solution when the two breathers are
well separated. At each fixed slow position, we hold the translation
parameters constant and consider one breather near formation while
the other is in its asymptotic tail. This differs from the preceding
$t\to-\infty$ expansion, in which both breathers are in their
pre-formation tails. The centers are identified with the local
maxima of the total density. With the convention
$D_{m,l}=\chi_l-\chi_m^*$, the coefficients of the constant term
and the largest exponential factor in each nondegenerate local
denominator are negative. After normalization, the center phase
is therefore $2n\pi-\arg C_+$, where $n\in\mathbb Z$ and $C_+$
denotes the coefficient of the positive-phase harmonic in the
corresponding denominator. Using this convention, we examine
each breather under the two orderings $t_1(x)>t_2(x)$ and
$t_1(x)<t_2(x)$, giving the following four cases.

\textbf{Case 1}: Assume $t-t_{1}=\tau $ with $\tau $ constant and $%
t_{1}>t_{2}$, and define
\begin{equation}
\Delta \equiv t_{1}-t_{2}>0,\quad t-t_{2}=\tau +\Delta.
\end{equation}%
Keeping $\tau $ fixed and $\Delta $ sufficiently large, and neglecting
exponentially small terms in the separation, we obtain the approximation
\begin{equation}
\frac{\det M^{(i)}}{\det M}\approx \;\frac{\chi _{3}^{\ast }+b_{i}}{\chi
_{3}+b_{i}}\,\frac{\widehat{\Gamma }_{i,0}+\widehat{\Gamma }_{i,+}\mathrm{e}^{k
\chi _{1I}\tau +A_{1}}+\widehat{\Gamma }_{i,-}\mathrm{e}^{k \chi _{1I}\tau
-A_{1}}+\widehat{\Gamma }_{i,11}\mathrm{e}^{2k \chi _{1I}\tau }}{\Gamma
_{0}+\Gamma _{+}\mathrm{e}^{k \chi _{1I}\tau +A_{1}}+\Gamma _{-}\mathrm{e}^{k \chi
_{1I}\tau -A_{1}}+\Gamma _{11}\mathrm{e}^{2k \chi _{1I}\tau }},
\end{equation}%
where the unhatted coefficients $\Gamma$ are determined solely by $D_{m,l}$,
whereas the hatted coefficients $\widehat{\Gamma}$ additionally contain the
component factors displayed below:
\begin{equation}
\begin{aligned}
\Gamma_0&:=\frac{1}{D_{3,3}D_{2,2}}-\frac{1}{D_{2,3}D_{3,2}},\\
\Gamma_{+}&:=\frac{1}{D_{3,3}D_{1,2}}-\frac{1}{D_{1,3}D_{3,2}},\\
\Gamma_{-}&:=\frac{1}{D_{3,3}D_{2,1}}-\frac{1}{D_{3,1}D_{2,3}},\\
\Gamma_{11}&:=\frac{1}{D_{3,3}D_{1,1}}-\frac{1}{D_{1,3}D_{3,1}},
\end{aligned}\qquad \begin{aligned}
\widehat\Gamma_{i,0}&=\left(\frac{\chi_2^*+b_i}{\chi_2+b_i}\right)\Gamma_0,%
\\
\widehat\Gamma_{i,+}&=\left(\frac{\chi_1^*+b_i}{\chi_2+b_i}\right)%
\Gamma_{+},\\
\widehat\Gamma_{i,-}&=\left(\frac{\chi_2^*+b_i}{\chi_1+b_i}\right)%
\Gamma_{-},\\ \widehat\Gamma_{i,11}&=\left(\frac{\chi_
1^*+b_i}{\chi_1+b_i}\right)\Gamma_{11}. \end{aligned}
\end{equation}%
Then, for $t-t_{1}=\tau $ with $\tau $ constant and $\Delta $ sufficiently
large, the solution \eqref{eq:two-AB-exact} can be approximated as
\begin{equation}
q_{i}^{[2]}(x,t)\approx a_{i}\mathrm{e}^{\theta _{i}}\frac{\chi _{3}^{\ast }+b_{i}}{%
\chi _{3}+b_{i}}\,\frac{\widehat{\Gamma }_{i,0}+\widehat{\Gamma }%
_{i,+}\mathrm{e}^{k \chi _{1I}\tau +A_{1}}+\widehat{\Gamma }_{i,-}\mathrm{e}^{k \chi
_{1I}\tau -A_{1}}+\widehat{\Gamma }_{i,11}\mathrm{e}^{2k \chi _{1I}\tau }}{%
\Gamma _{0}+\Gamma _{+}\mathrm{e}^{k \chi _{1I}\tau +A_{1}}+\Gamma _{-}\mathrm{e}^{k
\chi _{1I}\tau -A_{1}}+\Gamma _{11}\mathrm{e}^{2k \chi _{1I}\tau }}.
\label{eq:two-AB-case1}
\end{equation}%
In this limit, the local approximation \eqref{eq:two-AB-case1} has
the same form as the single-AB solution \eqref{eq:AB}, and its center is $%
(x,t)=(c_{n},t_{1}(c_{n})+t^{\Gamma })$, where $c_{n}-x_{1}(c_{n})=x_{n}^{%
\Gamma }$,~with $t^{\Gamma }$ and $x_{n}^{\Gamma }$ given by
\begin{equation}
t^{\Gamma }=\frac{1}{2k \chi _{1I}}\ln \left\vert \frac{\Gamma _{0}}{%
\Gamma _{11}}\right\vert ,\quad x_{n}^{\Gamma }=-(\chi _{1R}+\tfrac{k}{%
2})t^{\Gamma }+\frac{2n\pi -\arg {\Gamma _{+}}}{k},\quad n\in
\mathbb{Z}.
\end{equation}%
The expressions for $t_{1}$ and $x_{1}$ are given in Eq.~\eqref{eq:two-dynamic}.

\textbf{Case 2}: Assume $t-t_{1}=\tau $ with $\tau $ constant and $%
t_{2}>t_{1}$, and define
\begin{equation}
\Delta \equiv t_{1}-t_{2}<0,\quad t-t_{2}=\tau +\Delta.
\end{equation}%
Keeping $\tau $ fixed with $|\Delta|$ sufficiently large, and neglecting
exponentially small terms in the separation, we obtain
the approximation
\begin{equation}
q_{i}^{[2]}(x,t)\approx a_{i}\mathrm{e}^{\theta _{i}}\frac{\chi _{4}^{\ast }+b_{i}}{%
\chi _{4}+b_{i}}\,\frac{\widehat{\Lambda }_{i,0}+\widehat{\Lambda }%
_{i,+}\mathrm{e}^{k \chi _{1I}\tau +A_{1}}+\widehat{\Lambda }_{i,-}\mathrm{e}^{k
\chi _{1I}\tau -A_{1}}+\widehat{\Lambda }_{i,11}\mathrm{e}^{2k \chi _{1I}\tau }}{%
\Lambda _{0}+\Lambda _{+}\mathrm{e}^{k \chi _{1I}\tau +A_{1}}+\Lambda
_{-}\mathrm{e}^{k \chi _{1I}\tau -A_{1}}+\Lambda _{11}\mathrm{e}^{2k \chi _{1I}\tau }%
},  \label{eq:two-AB-case2}
\end{equation}%
where the unhatted coefficients $\Lambda$ are determined solely by $D_{m,l}$,
whereas the hatted coefficients $\widehat{\Lambda}$ additionally contain the
component factors displayed below:
\begin{equation}
\begin{aligned}
\Lambda_0&:=\frac{1}{D_{4,4}D_{2,2}}-\frac{1}{D_{2,4}D_{4,2}},\\
\Lambda_{+}&:=\frac{1}{D_{4,4}D_{1,2}}-\frac{1}{D_{1,4}D_{4,2}},\\
\Lambda_{-}&:=\frac{1}{D_{4,4}D_{2,1}}-\frac{1}{D_{2,4}D_{4,1}},\\
\Lambda_{11}&:=\frac{1}{D_{4,4}D_{1,1}}-\frac{1}{D_{1,4}D_{4,1}},
\end{aligned}\quad \begin{aligned}
\widehat\Lambda_{i,0}&:=\left(\frac{\chi_2^*+b_i}{\chi_2+b_i}\right)%
\Lambda_0,\\
\widehat\Lambda_{i,+}&:=\left(\frac{\chi_1^*+b_i}{\chi_2+b_i}\right)%
\Lambda_{+},\\
\widehat\Lambda_{i,-}&:=\left(\frac{\chi_2^*+b_i}{\chi_1+b_i}\right)%
\Lambda_{-},\\
\widehat\Lambda_{i,11}&:=\left(\frac{\chi_1^*+b_i}{\chi_1+b_i}\right)%
\Lambda_{11}. \end{aligned}
\end{equation}%
In this limit, the local approximation \eqref{eq:two-AB-case2} has
the same form as the single-AB solution \eqref{eq:AB}, and its center is $%
(x,t)=(c_{n},t_{1}(c_{n})+t^{\Lambda })$, where $c_{n}-x_{1}(c_{n})=x_{n}^{%
\Lambda }$,~with $t^{\Lambda }$ and $x_{n}^{\Lambda }$ given by
\begin{equation}
t^{\Lambda }=\frac{1}{2k \chi _{1I}}\ln \left\vert \frac{\Lambda _{0}}{%
\Lambda _{11}}\right\vert ,\quad x_{n}^{\Lambda }=-(\chi _{1R}+\tfrac{k
}{2})t^{\Lambda }+\frac{2n\pi -\arg {\Lambda _{+}}}{k},\quad n\in
\mathbb{Z}.
\end{equation}%
The expressions for $t_{1}$ and $x_{1}$ are given in Eq.~\eqref{eq:two-dynamic}%
.

\textbf{Case 3}: Assume $t-t_{2}=\bar{\tau}$ with $\bar{\tau}$ constant and $%
t_{2}>t_{1}$, and define
\begin{equation}
\Delta \equiv t_{1}-t_{2}<0,\quad t-t_{1}=\bar{\tau}-\Delta.
\end{equation}%
Keeping $\bar{\tau}$ fixed and $\Delta <0$ with a large absolute value, and
neglecting exponentially small terms in the separation, we obtain the approximation
\begin{equation}
q_{i}^{[2]}(x,t)\approx a_{i}\mathrm{e}^{\theta _{i}}\frac{\chi _{1}^{\ast }+b_{i}}{%
\chi _{1}+b_{i}}\,\frac{\widehat{\Pi }_{i,0}
+\widehat{\Pi }_{i,+}\mathrm{e}^{k\chi _{3I}\bar{\tau}+A_{2}}
+\widehat{\Pi }_{i,-}\mathrm{e}^{k\chi _{3I}\bar{\tau}-A_{2}}
+\widehat{\Pi }_{i,33}\mathrm{e}^{2k\chi _{3I}\bar{\tau}}}
{\Pi _{0}+\Pi _{+}\mathrm{e}^{k\chi _{3I}\bar{\tau}+A_{2}}
+\Pi _{-}\mathrm{e}^{k\chi _{3I}\bar{\tau}-A_{2}}
+\Pi _{33}\mathrm{e}^{2k\chi _{3I}\bar{\tau}}},
\label{eq:two-AB-case3}
\end{equation}%
where the unhatted coefficients $\Pi$ are determined solely by $D_{m,l}$,
whereas the hatted coefficients $\widehat{\Pi}$ additionally contain the
component factors displayed below:
\begin{equation}
\begin{aligned} \Pi_0&:=\frac{1}{D_{1,1}D_{4,4}}-\frac{1}{D_{1,4}D_{4,1}},\\
\Pi_{+}&:=\frac{1}{D_{1,1}D_{3,4}}-\frac{1}{D_{1,4}D_{3,1}},\\
\Pi_{-}&:=\frac{1}{D_{1,1}D_{4,3}}-\frac{1}{D_{1,3}D_{4,1}},\\
\Pi_{33}&:=\frac{1}{D_{1,1}D_{3,3}}-\frac{1}{D_{1,3}D_{3,1}}, \end{aligned}%
\qquad \begin{aligned}
\widehat\Pi_{i,0}&:=\left(\frac{\chi_4^*+b_i}{\chi_4+b_i}\right)\Pi_0,\\
\widehat\Pi_{i,+}&:=\left(\frac{\chi_3^*+b_i}{\chi_4+b_i}\right)\Pi_{+},\\
\widehat\Pi_{i,-}&:=\left(\frac{\chi_4^*+b_i}{\chi_3+b_i}\right)\Pi_{-},\\
\widehat\Pi_{i,33}&:=\left(\frac{\chi_3^*+b_i}{\chi_3+b_i}\right)\Pi_{33}.
\end{aligned}
\end{equation}%
In this limit, the local approximation \eqref{eq:two-AB-case3} has
the same form as the single-AB solution \eqref{eq:AB}, and its center is $%
(x,t)=(c_{n},t_{2}(c_{n})+t^{\Pi })$, where $c_{n}-x_{2}(c_{n})=x_{n}^{\Pi }$%
,~with $t^{\Pi }$ and $x_{n}^{\Pi }$ given by
\begin{equation}
t^{\Pi }=\frac{1}{2k \chi _{3I}}\ln \left\vert \frac{\Pi _{0}}{\Pi _{33}%
}\right\vert ,\quad x_{n}^{\Pi }=-(\chi _{3R}+\tfrac{k}{2})t^{\Pi }+%
\frac{2n\pi -\arg {\Pi _{+}}}{k},\quad n\in \mathbb{Z}.
\end{equation}%
The expressions for $t_{2}$ and $x_{2}$ are given by Eq.~\eqref{eq:two-dynamic}.

\textbf{Case 4}: Assume $t-t_{2}=\bar{\tau}$ with $\bar{\tau}$ constant and $%
t_{2}<t_{1}$, and define
\begin{equation}
\Delta \equiv t_{1}-t_{2}>0,\quad t-t_{1}=\bar{\tau}-\Delta.
\end{equation}%
Keeping $\bar{\tau}$ fixed and $\Delta $ sufficiently large, and neglecting
exponentially small terms in the separation, we obtain the approximation
\begin{equation}
q_{i}^{[2]}(x,t)\approx a_{i}\mathrm{e}^{\theta _{i}}\frac{\chi _{2}^{\ast }+b_{i}}{%
\chi _{2}+b_{i}}\,\frac{\widehat{\Upsilon }_{i,0}+\widehat{\Upsilon }%
_{i,+}\mathrm{e}^{k \chi _{3I}\bar{\tau}+A_{2}}+\widehat{\Upsilon }%
_{i,-}\mathrm{e}^{k \chi _{3I}\bar{\tau}-A_{2}}+\widehat{\Upsilon }%
_{i,33}\mathrm{e}^{2k \chi _{3I}\bar{\tau}}}{\Upsilon _{0}+\Upsilon
_{+}\mathrm{e}^{k \chi _{3I}\bar{\tau}+A_{2}}+\Upsilon _{-}\mathrm{e}^{k \chi _{3I}%
\bar{\tau}-A_{2}}+\Upsilon _{33}\mathrm{e}^{2k \chi _{3I}\bar{\tau}}},
\label{eq:two-AB-case4}
\end{equation}%
where the unhatted coefficients $\Upsilon$ are determined solely by $D_{m,l}$,
whereas the hatted coefficients $\widehat{\Upsilon}$ additionally contain the
component factors displayed below:
\begin{equation}
\begin{aligned}
\Upsilon_0&:=\frac{1}{D_{2,2}D_{4,4}}-\frac{1}{D_{2,4}D_{4,2}},\\
\Upsilon_{+}&:=\frac{1}{D_{2,2}D_{3,4}}-\frac{1}{D_{2,4}D_{3,2}},\\
\Upsilon_{-}&:=\frac{1}{D_{2,2}D_{4,3}}-\frac{1}{D_{2,3}D_{4,2}},\\
\Upsilon_{33}&:=\frac{1}{D_{2,2}D_{3,3}}-\frac{1}{D_{2,3}D_{3,2}},
\end{aligned}\qquad \begin{aligned}
\widehat\Upsilon_{i,0}&:=\left(\frac{\chi_4^*+b_i}{\chi_4+b_i}\right)%
\Upsilon_0,\\
\widehat\Upsilon_{i,+}&:=\left(\frac{\chi_3^*+b_i}{\chi_4+b_i}\right)%
\Upsilon_{+},\\
\widehat\Upsilon_{i,-}&:=\left(\frac{\chi_4^*+b_i}{\chi_3+b_i}\right)%
\Upsilon_{-},\\
\widehat\Upsilon_{i,33}&:=\left(\frac{\chi_3^*+b_i}{\chi_3+b_i}\right)%
\Upsilon_{33}. \end{aligned}
\end{equation}%
In this limit, the local approximation \eqref{eq:two-AB-case4} has
the same form as the single-AB solution \eqref{eq:AB}, and its center is $%
(x,t)=(c_{n},t_{2}(c_{n})+t^{\Upsilon })$, where $c_{n}-x_{2}(c_{n})=x_{n}^{%
\Upsilon }$, with $t^{\Upsilon }$ and $x_{n}^{\Upsilon }$ given by
\begin{equation}
t^{\Upsilon }=\frac{1}{2k \chi _{3I}}\ln \left\vert \frac{\Upsilon _{0}%
}{\Upsilon _{33}}\right\vert ,\quad x_{n}^{\Upsilon }=-(\chi _{3R}+\tfrac{%
k}{2})t^{\Upsilon }+\frac{2n\pi -\arg {\Upsilon _{+}}}{k}%
,\quad n\in \mathbb{Z}.
\end{equation}%
The expressions for $t_{2}$ and $x_{2}$ are given by Eq.~\eqref{eq:two-dynamic}%
.
To complement the separated two-AB test presented in the main text, we
retain here the partially overlapping configuration. The test uses the first-order two-mode
initial perturbation $q_{i}(x,0)=a_{i}\mathrm{e}^{\mathrm{i}b_{i}x}
\left[1+\sum_{j=1}^{2}\mathcal P_i[\Omega^{(2j-1)},f_j](x,0)\right]$, while the two slowly varying
envelopes control the local formation coordinates of the constituent ABs.
Figure~\ref{figS15} shows an example with complementary
exponential envelopes, in which the two predicted center trajectories
approach one another. Where the breathers overlap strongly, the
center formulas derived for well-separated breathers no longer apply. Consequently,
the predicted center curves are kept piecewise and are not continued
through the region in which a unique componentwise AB center cannot be
assigned.

Using the gains $G_j$ and matching factors $\mathcal K_j$ in
Eq.~\eqref{eq:two-dynamic}, the predicted center of each constituent in
a fixed separated asymptotic state is
\begin{equation}
T_j^{(s)}(x)=
\frac{1}{G_j}\ln\!\left[
\frac{\mathcal K_j}{|f_j(x)|}\right]
+\delta_j^{(s)},
\label{eq:two-ab-general-centers}
\end{equation}
where $s$ labels the relevant asymptotic case and
$\delta_j^{(s)}$ is the corresponding constant interaction-induced
shift, given by $t^\Gamma$, $t^\Lambda$, $t^\Pi$, or $t^\Upsilon$. Hence, with $s$ labeling the state of the second breather and $r$ that of
the first, the predicted signed separation along the evolution coordinate is
\begin{equation}
\Delta T^{(s,r)}(x):=T_2^{(s)}(x)-T_1^{(r)}(x)
=\frac{1}{G_2}\ln\!\left[\frac{\mathcal K_2}{|f_2(x)|}\right]
-\frac{1}{G_1}\ln\!\left[\frac{\mathcal K_1}{|f_1(x)|}\right]
+\delta_2^{(s)}-\delta_1^{(r)}.
\label{eq:two-ab-coordinate-gap}
\end{equation}
For the four explicit separated asymptotic cases above, the general
interaction shifts give the following piecewise center difference. Here
$T_j(x)$ denotes the center trajectory with the shift appropriate to the
local ordering:
\begin{equation}
T_2(x)-T_1(x)=
\begin{cases}
t_2(x)-t_1(x)+t^\Upsilon-t^\Gamma,
&t_1(x)>t_2(x),\\
t_2(x)-t_1(x)+t^\Pi-t^\Lambda,
&t_1(x)<t_2(x).
\end{cases}
\label{eq:two-ab-piecewise-shifted-gap}
\end{equation}
Likewise, the phase-coordinate translations are
$(x_n^\Gamma,x_\ell^\Upsilon)$ in the first ordering and
$(x_n^\Lambda,x_\ell^\Pi)$ in the second, with $n,\ell\in\mathbb Z$.
The center difference in Eq.~\eqref{eq:two-ab-piecewise-shifted-gap} is used to assess
whether the constituents are well separated; the unshifted difference $t_2-t_1$ contains only
the envelope and spectral matching contributions.
Thus the center gap is jointly determined by the two linear growth
rates, the spectral matching coefficients, the state-dependent
interaction shifts, and the two envelope magnitudes.  To identify its
leading ordering, take same-order perturbation envelopes
$f_j(x)=\epsilon f_{b,j}(\mu x)$ with fixed envelope profiles satisfying
$f_{b,j},f_{b,j}',f_{b,j}''=\mathcal O(1)$ and
$0<m_j\leq |f_{b,j}(\mu x)|\leq M_j$ throughout the observation interval,
where $m_j,M_j$ are independent of $\epsilon$ and $\mu$.
The gains, spectral matching coefficients, and selected interaction shifts
are also held fixed. The lower bound keeps the logarithmic envelope terms
uniformly bounded; it is a local condition on the observation interval,
not a global lower bound on a decaying envelope. For complex envelopes,
their phases are chosen continuously on this nonzero interval.
Equation~\eqref{eq:two-ab-coordinate-gap} then gives
\[
T_2^{(s)}(x)-T_1^{(r)}(x)=
\left(\frac{1}{G_2}-\frac{1}{G_1}\right)
\ln\frac{1}{\epsilon}+\mathcal O(1).
\]
For $G_1\neq G_2$, the two formation coordinates therefore separate
logarithmically as $\epsilon\to0$ and cannot remain in a common
$\mathcal O(1)$ formation-coordinate window.  Within this same-order
scaling, such a common finite window consequently requires equal growth,
$G_1=G_2$. This is an ordering condition for simultaneous
finite-window observation, not a condition for the spectral existence of
the two unstable branches.  Once the gains are equal, the divergent
logarithmic term cancels and the envelope-dependent part of the finite gap
reduces to
\begin{equation}
\Delta T_{\mathrm{env}}(x)
=\frac{1}{G_{\rm eq}}\ln\left|\frac{f_1(x)}{f_2(x)}\right|.
\label{eq:two-ab-amplitude-gap}
\end{equation}

The separated-state approximation assumes the magnitude of
$\Delta T^{(s,r)}$ to be large compared with the intrinsic
widths of both ABs along the evolution coordinate,
$W_j=\mathcal O(G_j^{-1})$, so that
$|\Delta T^{(s,r)}|\gg W_1+W_2$ in the asymptotic sense.
The envelopes must also remain slow on the carrier scale, for example
$(2\pi/|k|)|\partial_x\ln f_j|\ll1$, throughout the observation interval.
These assumptions specify the regime of the local construction. The center gap cannot, however, be enlarged without limit by
changing the envelope ratio while retaining a small perturbation and a
useful observation window. In the equal-growth case,
the signed relation is
$|f_1/f_2|=\exp(G_{\rm eq}\Delta T_{\mathrm{env}})$,
with either sign of $\Delta T_{\mathrm{env}}$ allowed.
Making one envelope too large violates the small-perturbation condition
$|f_j|\ll1$, whereas making the other smaller delays the corresponding
center and leaves a longer interval over which higher-order residuals,
numerical noise, and weak equation perturbations may be amplified.

\begin{figure}[!t]
	\centering
	\includegraphics[width=0.75\textwidth]{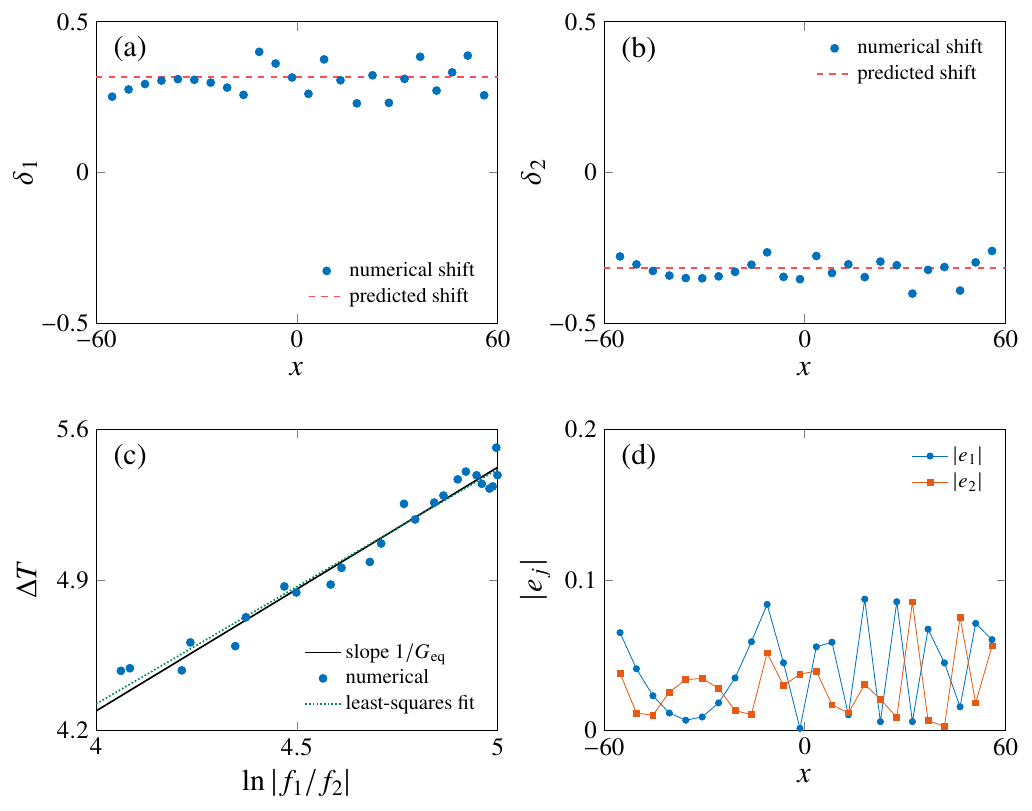}
\caption{Combined quantitative validation of the separated equal-growth
			two-AB example in Fig.~2(a) of the main text.  The perturbation envelopes are
			$f_1(x)=\epsilon\exp[2\cos(x/80)]$ and
			$f_2(x)=\epsilon \mathrm{e}^{-1}\exp[-2\cos(x/80)]$, with
			$\epsilon=10^{-2}$, $\beta=k=1.3$, and
			$G_{\rm eq}=0.882939$. All four panels use the same 24 center
			positions $\widehat c_n$. Panels (a) and (b) show the interaction
			shifts of the first ($j=1$) and second ($j=2$) breather constituents,
			respectively. Blue symbols denote the extracted values
			$\delta_{j,\mathrm{num}}(\widehat c_n)$, and red dashed lines show the
			determinant predictions: $t^{\Lambda}$ in (a) and $t^{\Pi}$ in (b).
			Their root-mean-square (RMS) differences over the retained points are
			$4.91\times10^{-2}$ and $3.57\times10^{-2}$, respectively.  Panel (c)
			tests Eq.~\eqref{eq:two-ab-gap-slope-test}.  The black line is the
			theoretical affine law and is labeled by its slope
			$G_{\rm eq}^{-1}=1.13258$; the blue
			points are the numerical values, and the green dotted line is the
			least-squares fit, whose slope is $1.09085$.  The RMS value of
			$\Delta T_{\rm num}-\Delta T_{\rm asy}$ is $6.28\times10^{-2}$.  Panel (d)
			shows the absolute evolution-coordinate errors $|e_1|$ and $|e_2|$
			for the two constituents over the same
			displayed interval; their RMS values are $4.91\times10^{-2}$ and
			$3.57\times10^{-2}$. These equal the corresponding RMS differences
			in (a) and (b), since $e_j=\delta_{j,\rm num}-\delta_j^{\rm theory}$.}
	\label{figS14}
\end{figure}

\begin{figure}[!t]
	\centering
	\includegraphics[width=0.9\textwidth]{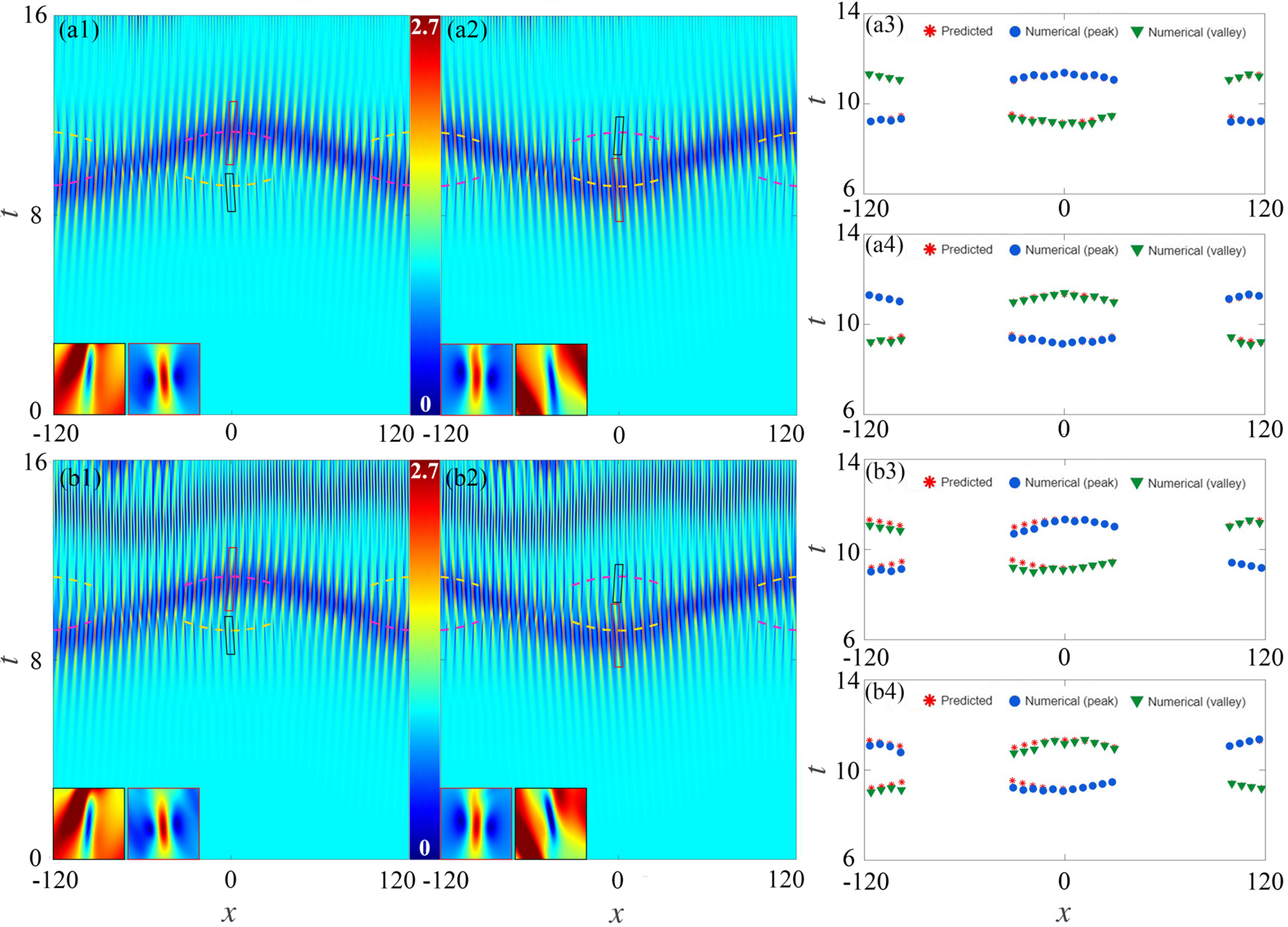}
\caption{Partially overlapping excitation and center matching of the slowly varying
			two-AB structure.
			The initial conditions are $
			q_{i}(x,0)=a_{i}\mathrm{e}^{\mathrm{i}b_{i}x}
			\left[1+\sum_{j=1}^{2}\mathcal P_i[\Omega^{(2j-1)},f_j](x,0)\right]$
			with modal envelopes $f_{1}(x)=\epsilon\exp[\cos(\mu x)]$ and
			$f_{2}(x)=\epsilon\exp[-\cos(\mu x)]$. The parameters are $\protect\epsilon %
			=10^{-2}$, $\mu=1/40$, $k=1.05$, $\protect\beta =1.1$.
			Panels (a1) and (a2) show the numerical evolutions of
			$|q_1|$ and $|q_2|$, respectively, for integrable system,
			whereas panels (b1) and (b2) show the corresponding evolutions of the
			perturbed system \eqref{eq:cnls1}, with $\epsilon_1=0.02$.
			The yellow and magenta dashed curves represent the asymptotically
			predicted center trajectories of the first and second breather constituents,
			respectively, and the insets display representative local peak and
			valley structures.
			Panels (a3) and (a4) compare the predicted and numerically extracted
			centers for components $q_1$ and $q_2$ in the integrable case,
			while panels (b3) and (b4) present the corresponding comparisons for
			the perturbed system.
			Red asterisks denote the asymptotic predictions, blue circles the
			numerical peak centers, and green downward triangles the numerical
			valley centers.}
	\label{figS15}
\end{figure}

We test the separated-state center predictions using one numerical
dataset for all four panels of Fig.~\ref{figS14}.
Each constituent's formation coordinate is extracted from the component
in which it has a local maximum, with one center retained per resolved
carrier cell. Let $\widehat c_n$ be the numerical spatial center of constituent 1;
the hat distinguishes it from an analytically predicted center $c_n$.
The formation coordinates of constituent 2 are interpolated onto these
positions using shape-preserving piecewise cubic Hermite interpolation,
and the same interpolation is used for both theoretical curves at $\widehat c_n$.
The resulting quantities are denoted by $T_{j,\mathrm{num}}(\widehat c_n)$ and
$T_{j,\mathrm{asy}}(\widehat c_n)$, with the latter defined by
Eq.~\eqref{eq:two-ab-general-centers}. We retain the same 24 samples in
$-60\leq \widehat c_n\leq60$ for every panel. The computation uses a periodic domain
of length $2\pi\times80$, 2048 spatial grid points, an evolution step
$\Delta t=1/256$, and saved fields every four steps over $0\leq t\leq12$.
The displayed case has $t_1<t_2$, so the interaction shifts are
$\delta_1^{\rm theory}=t^\Lambda$ and $\delta_2^{\rm theory}=t^\Pi$.
We define
$\delta_{j,\mathrm{num}}(\widehat c_n)=T_{j,\mathrm{num}}(\widehat c_n)-t_j(\widehat c_n)$ and
$\Delta T_{\rm num}(\widehat c_n)=T_{2,\rm num}(\widehat c_n)-T_{1,\rm num}(\widehat c_n)$.
No constant offset is fitted to the center predictions.

For equal growth, the predicted formation-coordinate gap is affine in the
logarithmic envelope ratio:
\begin{equation}
\Delta T_{\rm asy}(x)=G_{\rm eq}^{-1}
\ln\!\left|\frac{f_1(x)}{f_2(x)}\right|+C_{\rm gap},\qquad
C_{\rm gap}=G_{\rm eq}^{-1}\ln\!\left(\frac{\mathcal K_2}{\mathcal K_1}\right)
+t^{\Pi}-t^{\Lambda}.
\label{eq:two-ab-gap-slope-test}
\end{equation}
Equations~\eqref{eq:two-ab-coordinate-gap} and
\eqref{eq:two-ab-amplitude-gap} therefore predict the parametric slope
$\mathrm d\Delta T_{\rm asy}/\mathrm d\ln|f_1/f_2|=G_{\rm eq}^{-1}$.
The coordinate $x$ only parametrizes this relation;
$\ln|f_1(x)/f_2(x)|$ need not be linear in $x$.  We use
the common set of 24 retained samples described above, for which both
centers remain inside the evolution-coordinate window. The signed
formation-coordinate errors are defined by
$e_j(\widehat c_n)=T_{j,\rm num}(\widehat c_n)-T_{j,\rm asy}(\widehat c_n)$.
Because $T_{j,\rm asy}=t_j+\delta_j^{\rm theory}$, the identity
$e_j=\delta_{j,\rm num}-\delta_j^{\rm theory}$ holds on this common dataset.
Panel (d) displays the absolute values of the signed deviations in panels (a) and (b).
In panel (c), we fit $\Delta T_{\rm num}(\widehat c_n)$ as a linear function of
$\ln|f_1(\widehat c_n)/f_2(\widehat c_n)|$ using unweighted least squares.
All 24 retained points are included without smoothing or further point rejection.  Because
the chosen envelope is even, the points at $x$ and $-x$ can have nearly the
same logarithmic envelope ratio while retaining small, side-dependent center-extraction
errors; both are kept as separate retained data points, which accounts for the visible
vertical scatter.

Figure~\ref{figS14} provides numerical support
for both the constant interaction shifts and the logarithmic dependence of
the formation-coordinate gap on the envelope ratio in this equal-growth
example. The fitted slope is close to $G_{\rm eq}^{-1}$, while panel (d)
quantifies the remaining errors in the individual formation coordinates.
These complementary diagnostics test the prediction for the stated
parameters and observation window, without establishing a general
separation threshold.

For the complementary exponential envelopes in
Fig.~\ref{figS15},
$f_1(x)=\epsilon\exp[\cos(\mu x)]$ and
$f_2(x)=\epsilon\exp[-\cos(\mu x)]$. The common factor cancels in their
ratio, and Eq.~\eqref{eq:two-ab-amplitude-gap}
then gives $\Delta T_{\mathrm{env}}(x)
=\frac{2}{G_{\rm eq}}\cos(\mu x)$.
This envelope contribution changes sign and vanishes whenever
$\cos(\mu x)=0$. The full predicted separation in
Eq.~\eqref{eq:two-ab-piecewise-shifted-gap} also contains the spectral
term $G_{\rm eq}^{-1}\ln(\mathcal K_2/\mathcal K_1)$ and the
state-dependent interaction shifts. A zero of the envelope contribution
therefore does not locate a coincidence of the breather centers.
For the parameters shown, the separated left, middle, and right segments
in panels (a3)--(a4) and (b3)--(b4) reflect changes in the local ordering
of $t_1(x)$ and $t_2(x)$. In the intervening regions, the breathers overlap,
and a componentwise peak or valley cannot be uniquely assigned to one
constituent AB. We therefore apply the four asymptotic center formulas
only to the separated segments, without extrapolating them into the
interaction regions. The explicit trajectories are given below. For the first constituent with $t_1(x)>t_2(x)$,
\textbf{Case 1} applies, with
\begin{equation}
t_{1}(x)=\frac{\cos (\mu x)}{k \Omega _I^{(1)}}+\frac{1}{k \Omega
_I^{(1)}}\ln \!\left( \frac{\epsilon |C_{0}|}{2|k\Delta _{1}|}\right).
\label{eq:tx1}
\end{equation}%
The corresponding center is $(x,t)=(c_{n},\,t_{1}(c_{n})+t^{\Gamma })$%
, with $c_{n}-x_{1}(c_{n})=x_{n}^{\Gamma }$, and
\begin{equation}
t^{\Gamma }=-\frac{1}{2k \Omega _I^{(1)}}\ln \left\vert \frac{\Gamma _{0}}{%
\Gamma _{11}}\right\vert ,\quad x_{n}^{\Gamma }=-\Omega _R^{(1)}t^{\Gamma }+%
\frac{2n\pi -\arg (\Gamma _{+})}{k},\quad n\in \mathbb{Z}.
\end{equation}%
For the first constituent with $t_1(x)<t_2(x)$,
\textbf{Case 2} applies. The trajectory $t_{1}(x)$ is given by
Eq.~\eqref{eq:tx1}, and the center is $%
(x,t)=(c_{n},\,t_{1}(c_{n})+t^{\Lambda })$, with $c_{n}-x_{1}(c_{n})=x_{n}^{%
\Lambda }$, and
\begin{equation}
t^{\Lambda }=-\frac{1}{2k \Omega _I^{(1)}}\ln \left\vert \frac{\Lambda _{0}%
}{\Lambda _{11}}\right\vert ,\quad x_{n}^{\Lambda }=-\Omega _R^{(1)}t^{\Lambda
}+\frac{2n\pi -\arg (\Lambda _{+})}{k},\quad n\in \mathbb{Z}.
\end{equation}%
For the second constituent with $t_2(x)>t_1(x)$,
\textbf{Case 3} applies, with the trajectory $%
t_{2}(x)$ given by
\begin{equation}
t_{2}(x)=-\frac{\cos (\mu x)}{k \Omega _I^{(3)}}+\frac{1}{k \Omega
_I^{(3)}}\ln \!\left( \frac{\epsilon |C_{0}|}{2|k\Delta _{2}|}\right).
\label{eq:t2x}
\end{equation}%
The corresponding center is $(x,t)=(c_{n},\,t_{2}(c_{n})+t^{\Pi })$,
with $c_{n}-x_{2}(c_{n})=x_{n}^{\Pi }$, and
\begin{equation}
t^{\Pi }=-\frac{1}{2k \Omega _I^{(3)}}\ln \left\vert \frac{\Pi _{0}}{\Pi
_{33}}\right\vert ,\quad x_{n}^{\Pi }=-\Omega _R^{(3)}t^{\Pi }+\frac{2n\pi
-\arg (\Pi _{+})}{k},\quad n\in \mathbb{Z}.
\end{equation}%
For the second constituent with $t_2(x)<t_1(x)$,
\textbf{Case 4} applies. The trajectory $t_{2}(x)$ is given by
Eq.~\eqref{eq:t2x}. The center is $(x,t)=(c_{n},\,t_{2}(c_{n})+t^{\Upsilon })$,
where $c_{n}-x_{2}(c_{n})=x_{n}^{\Upsilon }$, and
\begin{equation}
t^{\Upsilon }=-\frac{1}{2k \Omega _I^{(3)}}\ln \left\vert \frac{\Upsilon
_{0}}{\Upsilon _{33}}\right\vert ,\quad x_{n}^{\Upsilon }=-\Omega
_R^{(3)}t^{\Upsilon }+\frac{2n\pi -\arg (\Upsilon _{+})}{k},\quad
n\in \mathbb{Z}.
\end{equation}%

The equal-growth partially overlapping example in Fig.~\ref{figS15}
complements the separated main-text configuration. Its two-constituent geometry
persists, but the separated-state center formulas do not describe the
strong-overlap region. The observed offsets do not separately resolve the
effects of finite overlap, slow modulation, and the assignment of
componentwise extrema to individual constituents.

\section{Residual estimate for the slowly modulated AB construction}

We evaluate the local equation residual generated by slowly varying
translation parameters and distinguish it from the initial matching defect
produced by truncating the pre-formation AB tail. This calculation concerns
the smooth explicit ansatz under the bounds stated below; it does not
assume that the initial-value evolution follows it. The residual estimates
are tested by direct numerical substitution.

Let
$\mathbf q_{AB}^{[1]}(x,t;\mathbf p)$ denote the phase-normalized exact
single-AB field with constant translations $\mathbf p=(x_1,t_1)$.
Its components are $q_{AB,i}^{[1]}=[(\chi_2+b_i)/(\chi_2^*+b_i)]q_i^{[1]}$
and approach $a_i\mathrm e^{\theta_i}$ as $t-t_1\to-\infty$.
Replacing the translations by slowly varying functions gives
\begin{equation}
 \mathbf q_{\rm asy}(x,t)=\mathbf q_{AB}^{[1]}(x,t;\mathbf p(x)).
 \label{eq:residual-ansatz}
\end{equation}
For a vector field $\mathbf u$, write
\begin{equation}
 \mathcal N[\mathbf u]
 :=\mathrm i\mathbf u_t+\frac12\mathbf u_{xx}
 +\sigma(\mathbf u^\dagger\mathbf u)\mathbf u.
 \label{eq:residual-operator}
\end{equation}
In derivatives of $\mathbf q_{AB}^{[1]}(x,t;\mathbf p)$, $\partial_x$
holds $\mathbf p$ fixed, whereas $\nabla_{\mathbf p}$ acts on the independent
translation parameters. All template derivatives in the residual below
are evaluated at $\mathbf p=\mathbf p(x)$ after differentiation.
Since $\mathcal N[\mathbf q_{AB}^{[1]}]=0$ for constant $\mathbf p$,
the chain rule gives
\begin{equation}
\begin{aligned}
 \mathbf R[\mathbf q_{\rm asy}]
 :=\mathcal N[\mathbf q_{\rm asy}]
 =&\mathbf p_x\cdot
       \nabla_{\mathbf p}\partial_x\mathbf q_{AB}^{[1]}+\frac{1}{2}\left[
       \mathbf p_{xx}\cdot\nabla_{\mathbf p}\mathbf q_{AB}^{[1]}
       +\sum_{r,s=1}^{2}(p_x)_r(p_x)_s
       \frac{\partial^2\mathbf q_{AB}^{[1]}}{\partial p_r\partial p_s}\right].
\end{aligned}
\label{eq:exact-residual}
\end{equation}
Suppose on the local observation interval that
$\|\mathbf p_x\|_{\infty}\leq C_p\mu$ and
$\|\mathbf p_{xx}\|_{\infty}\leq C_p\mu^2$, where $C_p$ is independent
of $\mu$ and $\epsilon$, and that the displayed
derivatives of the nonsingular AB are uniformly bounded independently of
these small parameters throughout
$|t-t_1(x)|\leq T$, with fixed $T>0$, where $t_1(x)$ is the matched
local formation coordinate defined above. On a fixed local interval $I$ containing a
prescribed number of carrier periods, define the normalized vector $L^2$ norm
by
\begin{equation*}
 \|\mathbf u\|_{L^2_{\rm av}(I)}
 :=\left[\frac{1}{|I|}\int_I
 \left(|u_1|^2+|u_2|^2\right)\,\mathrm dx\right]^{1/2}.
\end{equation*}
This is the usual $L^2$ norm divided by $|I|^{1/2}$ and is the measure used
in the numerical residual test below. Here $C_T$ denotes a
positive constant independent of $\mu$ and $\epsilon$, although it may depend
on the fixed observation width $T$, the interval $I$, and the fixed AB
parameters.  Equation~\eqref{eq:exact-residual}
then implies
\begin{equation}
 \|\mathbf R[\mathbf q_{\rm asy}]\|_{L^2_{\rm av}(I)}
 \leq C_T(\mu+\mu^2)=\mathcal O(\mu).
 \label{eq:center-residual-order}
\end{equation}
Thus the uncorrected slowly modulated AB has an $\mathcal O(\mu)$
equation defect in the nonlinear formation region.  An
$\mathcal O(\mu^2)$ estimate would require a first adiabatic corrector that
cancels the $\mathcal O(\mu)$ term in Eq.~\eqref{eq:exact-residual}.

The same residual admits a sharper estimate in the pre-formation linear tail.
For the single AB matched to a growing MI mode of gain $G$, define
$\zeta(t,x):=\exp\{G[t-t_1(x)]\}$. With fixed nondegenerate AB parameters,
the explicit phase-normalized formula has a regular tail expansion
$\mathbf q_{AB}^{[1]}-\mathbf q_{\rm bg}=\mathcal O(\zeta)$, whose background is
independent of $\mathbf p$. Differentiating this expansion gives
$\mathcal O(\zeta)$ bounds for $\nabla_{\mathbf p}\mathbf q_{AB}^{[1]}$,
$\nabla_{\mathbf p}\partial_x\mathbf q_{AB}^{[1]}$, and $\nabla_{\mathbf p}^{2}\mathbf q_{AB}^{[1]}$
uniformly on $I$ when $\sup_{x\in I}\zeta(t,x)\leq\zeta_0<1$, with fixed
sufficiently small $\zeta_0$. Applying Eq.~\eqref{eq:exact-residual} with
the same bounds on $\mathbf p_x$ and $\mathbf p_{xx}$ then yields
\begin{equation}
 \|\mathbf R_{\rm tail}(t)\|_{L^2_{\rm av}(I)}
  \leq C_{\rm tail}(\mu+\mu^2)
  \|\zeta(t,\cdot)\|_{L^\infty(I)}.
  \label{eq:tail-residual-order}
\end{equation}
Here $\mathbf R_{\rm tail}$ denotes the residual in this region, and
$C_{\rm tail}$ is independent of $\mu$, $\epsilon$, and $t$, but may depend
on $I$, $\zeta_0$, and the fixed AB parameters.
Equations~\eqref{eq:center-residual-order} and
\eqref{eq:tail-residual-order} therefore describe the same slow-modulation
residual in two different regions: the former is the uniform estimate near
the nonlinear center, whereas the latter uses the small tail amplitude to
refine it before formation.  They are not two additive error sources.

Let $\mathbf q_{\rm in}$ denote the initial field formed by retaining the
first-order MI perturbation at the initial matching coordinate $t_0$
($t_0=0$ in the simulations). The amplitude matching then gives
$\|\zeta(t_0,\cdot)\|_{L^\infty(I)}=\mathcal O(\epsilon)$, so the instantaneous equation residual
due to slow modulation is $\mathcal O(\epsilon\mu)$.  Separately, truncating
the exact pre-formation AB tail after its first MI term produces the initial-data
mismatch
\begin{equation}
 \|\mathbf q_{\rm in}-\mathbf q_{\rm asy}(t_0)\|_{L^2_{\rm av}(I)}
 =\mathcal O(\epsilon^2).
 \label{eq:initial-matching-defect}
\end{equation}

The initial matching defect measures agreement between the input fields at
$t_0$, whereas the equation residual measures the instantaneous defect of
the ansatz in satisfying the evolution equation. Neither estimate alone
controls the difference from the evolved field at later times.

For the two-AB construction, the same chain rule applies with
$\mathbf p=(x_1,t_1,x_2,t_2)$. On intervals where the translation derivatives
and the fixed-parameter two-AB derivatives obey the same boundedness
assumptions, its local equation residual is again $\mathcal O(\mu)$, with
the parameter sums extended to all four translations. This extension
remains conditional on those bounds.

The center and component-modulus errors reported above test evolved
observables, but do not by themselves verify the local equation defect in
Eq.~\eqref{eq:center-residual-order}.  We therefore performed a separate
substitution test in which the analytical slowly modulated AB was inserted
directly into Eq.~\eqref{eq:residual-operator}, without numerical time evolution or center extraction.  We used the focusing equal-growth background
$a_1=a_2=1$, $b_1=-b_2=1$, $k=1$, and $\epsilon=10^{-2}$, with
$f(x)=\epsilon\exp[\cos(\mu x)]$.  The normalized local vector $L^2$ norm was evaluated on
$W_x=\{x:|x-x_*|\le2L_{\rm AB}\}$, a four-period reference window
centered at $x_*=\pi/(2\mu)$, where the slow-envelope slope does not vanish.  We use the
reference formation coordinate $t_{\rm ref}:=t_1(x_*)$ and set
$t_{\rm tail}:=t_{\rm ref}-4/G$, so both residuals are evaluated on fixed
evolution-coordinate slices across $W_x$. For each tested scale, we write $\zeta_*(\mu):=\zeta(t_{\rm tail},x_*)$
for the tail amplitude at the reference position. The chosen sampling
coordinate gives $\zeta_*(\mu)=\mathrm e^{-4}$ for every scale. This is
a scalar normalization factor, not the spatially varying local amplitude
throughout $W_x$.
The residual calculation uses $x\in[-L/2,L/2)$ with $L=2\pi/\mu$.
For $\mu^{-1}=20,30,45,70,100$, the grids have
$N_x=2048,4096,4096,8192,16384$, respectively.
Spatial derivatives are evaluated spectrally after removing the background
carriers, with the carrier shifts retained in the differentiation multipliers.
The $t$-derivative uses a fourth-order centered difference with step
$h_t=2\times10^{-4}$. This is a direct evaluation of the analytical field,
so $h_t$ is a differentiation step, not a PDE integration step.

More generally, consider a family of slowly modulated fields
$\mathbf q_{\rm asy}(x,t;\mu)$ with translations $\mathbf p(x;\mu)$.
Let $W_x(\mu)$ be a local window centered at $x_*(\mu)$, with reference
formation coordinate $t_{\rm ref}(\mu)=t_1(x_*(\mu);\mu)$ and a chosen
pre-formation sampling coordinate $t_{\rm tail}(\mu)$. The window and
sampling coordinates may depend on the envelope scale.
Freezing the translations at the window center gives
$\mathbf q_{\rm frozen}(x,t;\mu)=\mathbf q_{AB}^{[1]}(x,t;\mathbf p(x_*(\mu);\mu))$.
Define $R_{\rm frozen}(\mu)$ as the normalized vector $L^2$ norm of
$\mathcal N[\mathbf q_{\rm frozen}(\cdot,\cdot;\mu)]$ on $W_x(\mu)$ at
$t=t_{\rm ref}(\mu)$. For the modulated field, write
$\mathbf R(x,t;\mu)=(R_1,R_2)^{\mathsf T}
=\mathcal N[\mathbf q_{\rm asy}(\cdot,\cdot;\mu)](x,t)$.
The local RMS residuals are
\begin{equation*}
\begin{aligned}
 R_{\rm c}(\mu)
 &=\left[\frac{1}{|W_x(\mu)|}\int_{W_x(\mu)}
 \left(|R_1(x,t_{\rm ref}(\mu);\mu)|^2+|R_2(x,t_{\rm ref}(\mu);\mu)|^2\right)\,\mathrm dx
 \right]^{1/2},\\
 R_{\rm tail}(\mu)
 &=\left[\frac{1}{|W_x(\mu)|}\int_{W_x(\mu)}
 \left(|R_1(x,t_{\rm tail}(\mu);\mu)|^2+|R_2(x,t_{\rm tail}(\mu);\mu)|^2\right)\,\mathrm dx
 \right]^{1/2}.
\end{aligned}
\end{equation*}
The amplitude-normalized tail diagnostic is
$R_{\rm tail}(\mu)/\zeta_*(\mu)$, where
$\zeta_*(\mu)=\exp\{G[t_{\rm tail}(\mu)-t_1(x_*(\mu);\mu)]\}$
is the tail amplitude at the reference position.

\begin{samepage}
Let $\mu_n$ denote the $n$th tested inverse envelope scale,
ordered so that $\mu_n<\mu_{n-1}$. Retaining the normalization factor at
each scale, the local effective orders are
\begin{equation*}
 p_{{\rm c},n}=
 \frac{\log[R_{\rm c}(\mu_{n-1})/R_{\rm c}(\mu_n)]}
 {\log(\mu_{n-1}/\mu_n)},
 \qquad
 p_{{\rm tail},n}=
 \frac{\log\!\left[\dfrac{R_{\rm tail}(\mu_{n-1})/\zeta_*(\mu_{n-1})}
 {R_{\rm tail}(\mu_n)/\zeta_*(\mu_n)}\right]}
 {\log(\mu_{n-1}/\mu_n)}.
\end{equation*}
\end{samepage}
The dashes in the first row occur because no preceding scale is available.
For the specific numerical test in Table~\ref{tab:direct-residual-scaling},
$R_{\rm frozen}(\mu)$ is at most $1.8\times10^{-10}$, well below the measured
modulated residuals. This provides a numerical check on the formula
evaluation and differentiation for the frozen case.

In addition to these successive orders, we determine a global effective
order for each residual diagnostic from the four smallest tested values of $\mu$,
indexed by $\mathcal I_4=\{n:\mu_n^{-1}\in\{30,45,70,100\}\}$, through the log--log least-squares
problems below.  In each fit, $\eta\in\mathbb R$ is the intercept (the logarithm of
the power-law prefactor) and $p$ is the candidate power-law exponent; hats
denote the minimizing fitted values:
\begin{equation*}
\begin{aligned}
 (\widehat\eta_{\rm c},\widehat p_{\rm c})
 &=\mathop{\arg\min}_{\eta,p}
 \sum_{n\in\mathcal I_4}
 \left[\log R_{\rm c}(\mu_n)-\eta-p\log\mu_n\right]^2,\\
 (\widehat\eta_{\rm tail},\widehat p_{\rm tail})
 &=\mathop{\arg\min}_{\eta,p}
 \sum_{n\in\mathcal I_4}
 \left[\log\!\left(R_{\rm tail}(\mu_n)/\zeta_*(\mu_n)\right)-
 \eta-p\log\mu_n\right]^2.
\end{aligned}
\end{equation*}
Here, $\arg\min$ denotes the pair of parameters $(\eta,p)$ at which the
displayed sum of squared logarithmic residuals attains its minimum.
The fitted independent variable is $\log\mu$,
whereas the dependent variables are $\log R_{\rm c}$ and
$\log(R_{\rm tail}(\mu)/\zeta_*(\mu))$, respectively.  Equivalently, the fitted power
laws are $R_{\rm c}\simeq \mathrm{e}^{\widehat\eta_{\rm c}}
\mu^{\widehat p_{\rm c}}$ and
$R_{\rm tail}(\mu)/\zeta_*(\mu)\simeq \mathrm{e}^{\widehat\eta_{\rm tail}}
\mu^{\widehat p_{\rm tail}}$. The fits for the four smallest tested values of $\mu$ in
Table~\ref{tab:direct-residual-scaling} give
$\widehat p_{\rm c}=0.926778\simeq0.927$ for $R_{\rm c}$ and
$\widehat p_{\rm tail}=1.013914\simeq1.014$ for
$R_{\rm tail}(\mu)/\zeta_*(\mu)$; the successive center order also approaches one as
$\mu$ decreases. Over the tested scales, these measurements are
consistent with the $\mathcal O(\mu)$ center residual and the
$\mathcal O(\mu\zeta)$ tail refinement.

\begin{table}[!tb]
\centering

\begin{ruledtabular}
\begin{tabular}{cccccc}
$\mu^{-1}$ & $R_{\rm c}$ & $p_{{\rm c},n}$ &
$R_{\rm tail}(\mu)/\zeta_*(\mu)$ & $p_{{\rm tail},n}$ & $R_{\rm frozen}$\\
\hline
$20$  &  $0.1304$ & ---    & $0.0976$ & ---    & $2.08\!\times\!10^{-11}$\\
$30$  &  $0.1006$ & $0.640$& $0.0633$ & $1.069$& $2.76\!\times\!10^{-11}$\\
$45$  &  $0.0715$ & $0.842$& $0.0419$ & $1.015$& $1.75\!\times\!10^{-10}$\\
$70$  &  $0.0477$ & $0.919$& $0.0267$ & $1.023$& $9.05\!\times\!10^{-11}$\\
$100$ &  $0.0329$ & $1.041$& $0.0187$ & $0.997$& $1.30\!\times\!10^{-10}$\\
\end{tabular}
\end{ruledtabular}
\caption{Direct residual scaling of the focusing slowly modulated
single AB.  Here, $R_{\rm c}$ is the local RMS residual on the reference
formation-coordinate slice, $R_{\rm tail}(\mu)/\zeta_*(\mu)$ removes the known
pre-formation-tail amplitude, and
$p_{{\rm c},n}$ and $p_{{\rm tail},n}$ are the corresponding successive effective
orders between adjacent slow scales.  $R_{\rm frozen}$ is obtained after freezing both translation
parameters at their values at the center of the local window.}
\label{tab:direct-residual-scaling}
\end{table}

The chain-rule calculation and direct substitution test support the
local consistency of slowly varying translation parameters in the AB
construction. Table~\ref{tab:single-ab-joint-convergence} separately shows
decreasing two-component modulus-field discrepancies along the tested
joint-limit paths. These results, together with the center comparisons, provide
numerical evidence in the tested regimes. Uniform control of the difference between the asymptotic ansatz and the initial-value solution through the nonlinear formation stage is beyond the scope of the present work.

\begin{figure}[!t]
\centering
\includegraphics[width=\textwidth]{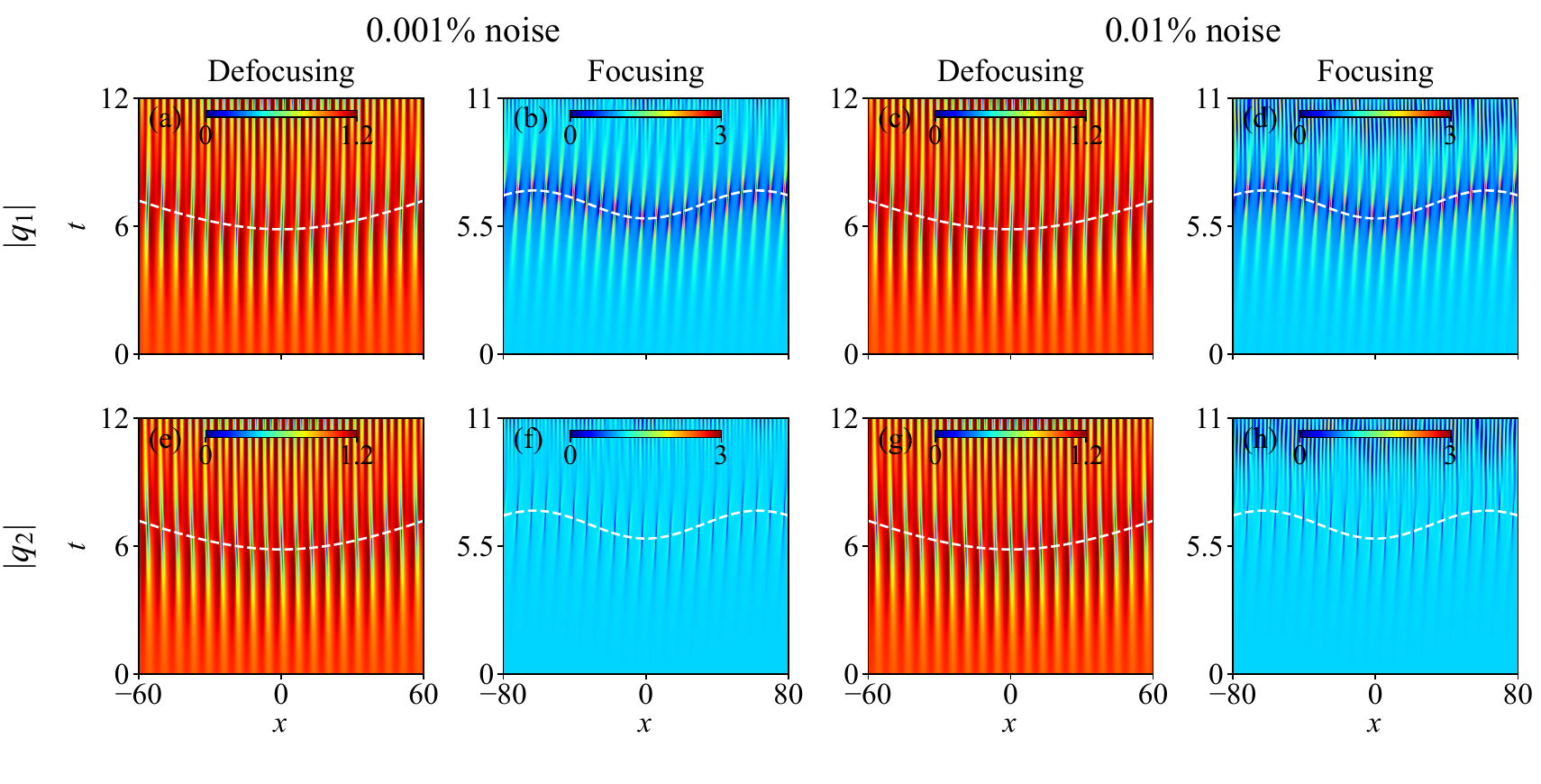}
\caption{Evolution from the noisy gMI initial
condition~\eqref{eq:noise-input}. The background amplitudes are $a_1=a_2=1$. The defocusing case uses
$\sigma=-1$, $b_1=-b_2=1.1$, $k=1$,
$f(x)=0.1\,\mathrm{sech}(x/50)$, and
$\Omega=-0.441796\mathrm{i}$.
The focusing case uses $\sigma=+1$, $b_1=-b_2=1$, $k=0.8$,
$f(x)=0.02\exp[0.4\cos(x/20)]$, and
$\Omega=-0.923184-0.832027\mathrm{i}$.
The upper and lower rows show
$|q_1|$ and $|q_2|$, respectively. The first and second columns show
the defocusing and focusing cases at $s=10^{-5}$ ($0.001\%$); the
third and fourth columns show the corresponding cases at $s=10^{-4}$
($0.01\%$). White dashed curves mark the formation time $t_1(x)$
predicted by the unperturbed asymptotic matching. The color scales
represent field moduli.}
\label{fig:noise-examples}
\end{figure}

\section{Numerical tests with weak initial-field noise}
\label{sec:noise-numerics}

We test the sensitivity of breather formation to weak random noise
in the initial field with a prescribed gMI modulation. The initial
fields are prescribed as
\begin{equation}
\begin{gathered}
q_i(x,0)=a_i\mathrm{e}^{\mathrm{i}b_i x}
\left[1+\frac{f(x)\mathrm{e}^{\mathrm{i}kx}}{k+2b_i+2\Omega}
+\frac{f^*(x)\mathrm{e}^{-\mathrm{i}kx}}{(k-2b_i-2\Omega)^*}\right]
\left[1+s\eta_i(x)\right],~\eta_i(x)\sim\mathcal N(0,1),\qquad i=1,2.
\end{gathered}
\label{eq:noise-input}
\end{equation}
The real noise $\eta_i(x)$ is independent across spatial grid points
and components, and $s\ll1$ denotes the standard deviation of the
relative amplitude noise. Noise is applied only to the initial field;
subsequent evolution follows the original Manakov
equations~\eqref{eq:cnls}. 
Within each regime, the same noise realization is used at both noise levels. The simulations employ a fourth-order symmetric split-step Fourier
scheme in the carrier-removed fields. The time steps for the defocusing and focusing cases are $2\times10^{-3}$ and
$5\times10^{-4}$, respectively; the defocusing envelope is periodized
by an image sum.

Figure~\ref{fig:noise-examples} displays the component moduli
$|q_1|$ and $|q_2|$. The first breather units form near the unperturbed
asymptotic formation curve. The defocusing patterns remain regular over the
displayed interval. In the focusing case, additional fluctuations
develop after the first formation stage and are more pronounced at
the higher noise level.
For the 19 defocusing and 20 focusing breather units shown, we match each component's
peak or valley at first formation to its numerical counterpart for the same
gMI modulation at $s=0$. We locate the peaks and valleys by local bicubic
spline interpolation of $|q_i|^2$ in $x$ and $t$. For each regime and noise level, we use the matched centers from
both field components to calculate the root-mean-square shift separately for
$x$ and $t$. At $s=10^{-5}$, these shifts are approximately $(0.0002,0.0009)$ for
defocusing and $(0.004,0.004)$ for focusing. At $s=10^{-4}$, the corresponding
values are $(0.002,0.007)$ and $(0.039,0.040)$, respectively.
At both tested noise levels, the first breather formation remains clearly
identifiable in both regimes, with only minor shifts in the center positions.


\begin{thebibliography}{99}
\bibitem{MI1} N. N. Akhmediev and V. I. Korneev, Modulation instability and
periodic solutions of the nonlinear Schr\"{o}dinger equation, \newblock\href{https://doi.org/10.1007/BF01037866}%
{Theor. Math. Phys. \textbf{69}, 1089 (1986).}

\bibitem{MI2} V. E. Zakharov and L. A. Ostrovsky, Modulation instability:
The beginning, \newblock\href{https://doi.org/10.1016/j.physd.2008.12.002}{%
Physica D \textbf{238}, 540 (2009).}

\bibitem{MI3} K. Hammani, B. Wetzel, B. Kibler, J. Fatome, C. Finot, G.
Millot, N. Akhmediev, and J. M. Dudley, Spectral dynamics of modulation
instability described using Akhmediev breather theory, \newblock\href{https://doi.org/10.1364/OL.36.002140}%
{Opt. Lett. \textbf{36}, 2140 (2011).}

\bibitem{MI4} G. Vanderhaegen, C. Naveau, P. Szriftgiser, A. Kudlinski, M.
Conforti, A. Mussot, M. Onorato, S. Trillo, A. Chabchoub, and N. Akhmediev,
``Extraordinary'' modulation instability in optics and hydrodynamics, %
\newblock\href{https://doi.org/10.1073/pnas.2019348118}{Proc. Natl. Acad.
Sci. U.S.A. \textbf{118}, e2019348118 (2021).}

\bibitem{MI5} V. E. Zakharov, and A. A. Gelash, Nonlinear Stage of
Modulation Instability, \newblock\href{https://doi.org/10.1103/PhysRevLett.111.054101}%
{Phys. Rev. Lett. \textbf{111}, 054101 (2013).}

\bibitem{Zhao2016} L.-C. Zhao and L. Ling, Quantitative relations between
modulational instability and several well-known nonlinear excitations, %
\newblock\href{https://doi.org/10.1364/JOSAB.33.000850}{J. Opt. Soc. Am. B
\textbf{33}, 850 (2016).}

\bibitem{Sun2023} L. Liu, W.-R. Sun, and B. A. Malomed, Formation of rogue
waves and modulational instability with zero-wavenumber gain in
multicomponent systems with coherent coupling, \newblock\href{https://doi.org/10.1103/PhysRevLett.131.093801}%
{Phys. Rev. Lett. \textbf{131}, 093801 (2023).}

\bibitem{Sun2025} W.-R. Sun and B. A. Malomed, Subharmonic modulational
instabilities, \newblock\href{https://doi.org/10.1016/j.physrep.2025.07.001}{%
Phys. Rep. \textbf{1139}, 1 (2025).}

\bibitem{Dudley1} J. M. Dudley, F. Dias, M. Erkintalo, and G. Genty,
Instabilities, breathers and rogue waves in optics, \newblock\href{https://doi.org/10.1038/nphoton.2014.220}%
{Nat. Photonics \textbf{8}, 755 (2014).}

\bibitem{Dudley2} J. M. Dudley, G. Genty, A. Mussot, A. Chabchoub, and F.
Dias, Rogue waves and analogies in optics and oceanography, \newblock\href{https://doi.org/10.1038/s42254-019-0100-0}%
{Nat. Rev. Phys. 1, 675 (2019).}

\bibitem{Kevrekidis-09} P. G. Kevrekidis, D. J. Frantzeskakis, and R.
Carretero-Gonz\'alez, eds., \textit{Emergent Nonlinear Phenomena in Bose--Einstein
Condensates: Theory and Experiment} (Springer, Berlin, 2008).

\bibitem{fiber} J. M. Dudley, G. Genty, and S. Coen, Supercontinuum
generation in photonic crystal fiber, \newblock\href{https://doi.org/10.1103/RevModPhys.78.1135}%
{Rev. Mod. Phys. \textbf{78}, 1135 (2006).}

\bibitem{fiber2} R. Schiek and F. Baronio, Spatial Akhmediev breathers and
modulation instability growth-decay cycles in a quadratic optical medium, %
\newblock\href{https://doi.org/10.1103/PhysRevResearch.1.032036}{Phys. Rev.
Research \textbf{1}, 032036(R) (2019).}

\bibitem{fiber3} R. Schiek, Excitation of nonlinear beams: From the linear
Talbot effect through modulation instability to Akhmediev breathers, %
\newblock\href{https://doi.org/10.1364/OE.425626}{Opt. Express \textbf{29},
15830 (2021).}

\bibitem{water} A. Chabchoub, N. Hoffmann, M. Onorato, and N. Akhmediev,
Super rogue waves: Observation of a higher-order breather in water waves, %
\newblock\href{https://doi.org/10.1103/PhysRevX.2.011015}{Phys. Rev. X
\textbf{2}, 011015 (2012).}

\bibitem{water2} A. Gomel, A. Chabchoub, M. Brunetti, S. Trillo, J.
Kasparian, and A. Armaroli, Stabilization of unsteady nonlinear waves by
phase-space manipulation, \newblock\href{https://doi.org/10.1103/PhysRevLett.126.174501}%
{Phys. Rev. Lett. \textbf{126}, 174501 (2021).}

\bibitem{BEC} K. Manikandan, P. Muruganandam, M. Senthilvelan, and M.
Lakshmanan, Manipulating matter rogue waves and breathers in Bose-Einstein
condensates, \newblock\href{https://doi.org/10.1103/PhysRevE.90.062905}{%
Phys. Rev. E \textbf{90}, 062905 (2014).}

\bibitem{Vishnu} N. Vishnu Priya, M. Senthilvelan, and M. Lakshmanan,
Akhmediev breathers, Ma solitons, and general breathers from rogue waves: A
case study in the Manakov system, \newblock\href{https://doi.org/10.1103/PhysRevE.88.022918}%
{Phys. Rev. E \textbf{88}, 022918 (2013).}

\bibitem{Guo2011} B.-L. Guo and L.-M. Ling, Rogue wave, breathers and
bright-dark rogue solutions for the coupled Schr\"{o}dinger equations, %
\newblock\href{https://doi.org/10.1088/0256-307X/28/11/110202}{Chin. Phys.
Lett. \textbf{28}, 110202 (2011).}

\bibitem{Ling2019} L. Ling and L.-C. Zhao, Modulational instability and
homoclinic orbit solutions in vector nonlinear Schr\"{o}dinger equation, %
\newblock\href{https://doi.org/10.1016/j.cnsns.2019.01.008}{Commun.
Nonlinear Sci. Numer. Simul. \textbf{72}, 449 (2019).}

\bibitem{Chen2021} S.-C. Chen, C. Liu, X. Yao, L.-C. Zhao, and N. Akhmediev,
Extreme spectral asymmetry of Akhmediev breathers and Fermi-Pasta-Ulam
recurrence in a Manakov system, \newblock\href{https://doi.org/10.1103/PhysRevE.104.024215}%
{Phys. Rev. E \textbf{104}, 024215 (2021).}

\bibitem{Gelash} A. Gelash and A. Raskovalov, Vector breathers in the
Manakov system, \newblock\href{https://doi.org/10.1111/sapm.12558}{Stud.
Appl. Math. \textbf{150}, 841 (2023).}

\bibitem{Kraus} D. Kraus, G. Biondini, and G. Kova\v{c}i\v{c}, The focusing
Manakov system with nonzero boundary conditions, \newblock\href{https://doi.org/10.1088/0951-7715/28/9/3101}%
{Nonlinearity \textbf{28}, 3101 (2015).}

\bibitem{DSWang} J.-J. Qi, D.-S. Wang, X.-Y. Wen, H. Zhu, Breathers,
rational solutions, and their exact physical spectra in $F=1$ spinor
Bose-Einstein condensates, \newblock\href{https://doi.org/10.1063/5.0323804}{%
Chaos \textbf{36}, 063142 (2026).}

\bibitem{Wright2000} O. C. Wright and M. G. Forest, On the B\"{a}%
cklund-gauge transformation and homoclinic orbits of a coupled nonlinear Schr%
\"{o}dinger system, \newblock\href{https://doi.org/10.1016/S0167-2789(00)00021-X}%
{Physica D \textbf{141}, 104 (2000).}

\bibitem{Kibler2015} B. Kibler, A. Chabchoub, A. Gelash, N. Akhmediev, and
V. E. Zakharov, Superregular Breathers in Optics and Hydrodynamics:
Omnipresent Modulation Instability beyond Simple Periodicity, \newblock\href{https://doi.org/10.1103/PhysRevX.5.041026}%
{Phys. Rev. X \textbf{5}, 041026 (2015).}

\bibitem{Gelash2022} A. Gelash, G. Xu, and B. Kibler, Management of breather interactions, \newblock\href{https://doi.org/10.1103/PhysRevResearch.4.033197}{Phys. Rev. Research \textbf{4}, 033197 (2022).}


\bibitem{Grinevich2018} P. G. Grinevich and P. M. Santini, The exact rogue
wave recurrence in the NLS periodic setting via matched asymptotic
expansions, for 1 and 2 unstable modes, \newblock\href{https://doi.org/10.1016/j.physleta.2018.02.014}%
{Phys. Lett. A \textbf{382}, 973 (2018).}

\bibitem{Trillo2019} S. Trillo and M. Conforti, Quantitative approach to
breather pair appearance in nonlinear modulational instability, \newblock
\href{https://doi.org/10.1364/OL.44.004275}{Opt. Lett. \textbf{44}, 4275
(2019).}

%

\bibitem{Frisquet2015} B. Frisquet, B. Kibler, J. Fatome, P. Morin, F.
Baronio, M. Conforti, G. Millot, and S. Wabnitz, Polarization modulation
instability in a Manakov fiber system, \newblock\href{https://doi.org/10.1103/PhysRevA.92.053854}%
{Phys. Rev. A \textbf{92}, 053854 (2015).}

\bibitem{Forest2000} M. G. Forest, D. W. McLaughlin, D. J. Muraki, and O. C.
Wright, Nonfocusing instabilities in coupled, integrable nonlinear Schr\"{o}%
dinger PDEs, \newblock\href{https://doi.org/10.1007/s003329910012}{J.
Nonlinear Sci. \textbf{10}, 291 (2000).}

\bibitem{Ling2017} L. Ling, L.-C. Zhao, Z.-Y. Yang, and B. Guo, Generation
mechanisms of fundamental rogue wave spatial-temporal structure, \newblock
\href{https://doi.org/10.1103/PhysRevE.96.022211}{Phys. Rev. E \textbf{96},
022211 (2017).}

\bibitem{Qin2023} Y.-H. Qin, L. Ling, and L.-C. Zhao, Optical rogue-wave
patterns in coupled defocusing systems, \newblock\href{https://doi.org/10.1103/PhysRevA.108.023519}%
{Phys. Rev. A \textbf{108}, 023519 (2023).}

\bibitem{Zhao2013} L.-C. Zhao and J. Liu, Rogue-wave solutions of a
three-component coupled nonlinear Schr\"{o}dinger equation, \newblock
\href{https://doi.org/10.1103/PhysRevE.87.013201}{Phys. Rev. E \textbf{87},
013201 (2013).}

\bibitem{Liu2024} C. Liu, S.-C. Chen, and N. Akhmediev, Fundamental and
second-order superregular breathers in vector fields, \newblock\href{https://doi.org/10.1103/PhysRevLett.132.027201}%
{Phys. Rev. Lett. \textbf{132}, 027201 (2024).}

\bibitem{Manakov-74} S. V. Manakov, On the theory of two-dimensional
stationary self-focusing of electromagnetic waves, \newblock\href{https://www.jetp.ras.ru/cgi-bin/e/index/e/38/2/p248?a=list}%
{Sov. Phys. JETP \textbf{38}, 248 (1974).}

\bibitem{Baronio2012} F. Baronio, A. Degasperis, M. Conforti, and S.
Wabnitz, Solutions of the vector nonlinear Schr\"{o}dinger equations:
Evidence for deterministic rogue waves, \newblock\href{https://doi.org/10.1103/PhysRevLett.109.044102}%
{Phys. Rev. Lett. \textbf{109}, 044102 (2012).}

\bibitem{Ling2014} L. Ling, B. Guo, and L.-C. Zhao, High-order rogue waves
in vector nonlinear Schr\"{o}dinger equations, \newblock\href{https://doi.org/10.1103/PhysRevE.89.041201}%
{Phys. Rev. E \textbf{89}, 041201(R) (2014).}

\bibitem{Mossman-MI} S. Mossman, S. I. Mistakidis, G. C. Katsimiga, A.
Romero-Ros, G. Biondini, P. Schmelcher, P. Engels, and P. G. Kevrekidis,
Nonlinear stage of modulational instability in repulsive two-component
Bose-Einstein condensates, \newblock\href{https://doi.org/10.1103/6jsr-f8q1}{%
Phys. Rev. Lett. \textbf{135}, 113401 (2025).}

\bibitem{Baronio2014} F. Baronio, M. Conforti, A. Degasperis, S. Lombardo,
M. Onorato, and S. Wabnitz, Vector Rogue Waves and Baseband Modulation
Instability in the Defocusing Regime,\newblock\href{https://doi.org/10.1103/PhysRevLett.113.034101}%
{Phys. Rev. Lett. \textbf{113}, 034101 (2014).}

\bibitem{Agrawal-12} G. P. Agrawal, \textit{Nonlinear Fiber Optics}, 5th ed.
(Academic Press, San Diego, 2012).

\bibitem{model3} C. Becker, S. Stellmer, P. Soltan-Panahi, S. D\"{o}rscher,
M. Baumert, E.-M. Richter, J. Kronj\"{a}ger, K. Bongs, and K. Sengstock,
Oscillations and interactions of dark and dark-bright solitons in
Bose-Einstein condensates, \newblock\href{https://doi.org/10.1038/nphys962}{%
Nat. Phys. \textbf{4}, 496 (2008).}

\bibitem{Kang-96} J. U. Kang, G. I. Stegeman, J. S. Aitchison, and N.
Akhmediev, Observation of Manakov spatial solitons in AlGaAs planar
waveguides, \newblock\href{https://doi.org/10.1103/PhysRevLett.76.3699}{%
Phys. Rev. Lett. \textbf{76}, 3699 (1996).}

\bibitem{model2} J. He, L. Guo, Y. Zhang, and A. Chabchoub, Theoretical and
experimental evidence of non-symmetric doubly localized rogue waves,
\newblock
\href{https://doi.org/10.1098/rspa.2014.0318}{Proc. R. Soc. A \textbf{470},
20140318 (2014).}

\bibitem{Agrawal} G. P. Agrawal, Modulation instability induced by
cross-phase modulation, \newblock\href{https://doi.org/10.1103/PhysRevLett.59.880}%
{{Phys. Rev. Lett. \textbf{59}, 880 (1987).}}

\bibitem{supp} See Supplemental Material for MI spectra, 
breather pattern classification, the analysis details of asymptotic two-AB solutions, 
additional simulations of integrable and weakly nonintegrable systems, 
numerical error analysis, and local residual estimates.

\bibitem{Frisquet2014}
B. Frisquet, A. Chabchoub, J. Fatome, C. Finot, B. Kibler, and G. Millot,
Two-stage linear-nonlinear shaping of an optical frequency comb as rogue
nonlinear-Schr\"{o}dinger-equation-solution generator,
\href{https://doi.org/10.1103/PhysRevA.89.023821}
{Phys. Rev. A \textbf{89}, 023821 (2014).}

\bibitem{OF2018} F. Baronio, B. Frisquet, S. Chen, G. Millot, S. Wabnitz, 
and B. Kibler, Observation of a group of dark rogue waves in a
telecommunication optical fiber, \newblock\href{https://doi.org/10.1103/PhysRevA.97.013852}%
{Phys. Rev. A \textbf{97}, 013852 (2018).}



\bibitem{OF2016} B. Frisquet, B. Kibler, P. Morin, F. Baronio, M. Conforti,
G. Millot, and S. Wabnitz, Optical dark rogue wave, \newblock\href{https://doi.org/10.1038/srep20785}%
{Sci. Rep. \textbf{6}, 20785 (2016).}



\bibitem{OFAB} C. Liu, L. Li, S.-C. Chen, X. Yao, W.-L. Yang, and N.
Akhmediev, Experimental observation of recurrence and spectral asymmetry of
the two-component Akhmediev breathers in a single mode optical fibre, %
\newblock\href{https://doi.org/10.48550/arXiv.2503.08513}{arXiv:2503.08513
(2025).}

\end{thebibliography}
\end{document}